\documentclass[%
twocolumn,
 amsmath,amssymb,
 longbibliography,
]{revtex4-2}
\pdfoutput=1
\usepackage{hyperref}

\usepackage{graphicx}
\usepackage{dcolumn}
\usepackage{bm}
\usepackage[normalem]{ulem}
\usepackage{color,soul}

\newcommand{\blue}[1]{\color{black} #1}

\begin{document}


\title[Micromagnetic simulations of clusters of nanoparticles]{Micromagnetic simulations of clusters of nanoparticles with internal structure: Application to magnetic hyperthermia}







\author{Razyeh Behbahani$^{1, 2}$, Martin L.~Plumer$^1$ and Ivan Saika-Voivod$^{1,*}$}
\address{$^1$ Department of Physics and Physical Oceanography, Memorial University of Newfoundland, St. John's, Newfoundland and Labrador, Canada, A1B 3X7}

\email{saika@mun.ca}
 
\address{$^2$ Department of Physics, University of Western Ontario, London, Ontario, Canada, N6A 3K7}









\begin{abstract}
Micromagnetic simulation results on dynamic hysteresis loops of clusters of iron oxide nanoparticles (NPs) with internal structure composed of nanorods are compared with the widely used macrospin approximation. Such calculations allowing for nanorod-composed NPs is facilitated by a previously developed coarse-graining method based on the renormalization group approach. With a focus on applications to magnetic hyperthermia, we show that magnetostatic interactions improve the heating performance of NPs in chains and triangles, and reduce heating performance in fcc arrangements.  Hysteresis loops of triangular and fcc systems of complex NPs are not recovered within the macrospin approximation, especially at smaller interparticle distances. For triangular arrangements, the macrospin approximation predicts that magnetostatic interactions reduce loop area, in contrast to the complex NP case. An investigation of the local hysteresis loops of individual NPs and macrospins  in clusters reveals the impact of the geometry of their neighbours on  individual versus collective magnetic response, inhomogenous heating within clusters, and further differences between simulating NPs with internal structure and the use of the macrospin approximation. Capturing the internal physical and magnetic structure of NPs is thus important for some applications.

\end{abstract}
\maketitle
\section{Introduction}\label{sec:RB3_intro}


Magnetic nanoparticles (NPs) have attracted much attention due to their wide range of 
applications~\cite{LeliaertMagnetic2022,plumer2010micromagnetic, labaye2002surface}. For biomedical purposes, magnetic NP hyperthermia is a developing method that uses NPs for cancer treatment by taking advantage of their heating upon exposure to an alternating external magnetic field~\cite{kolen2014large, surowiec2017investigation, villanueva2010hyperthermia, hergt2005magnetic, shi2019enhanced}. Along with pre-clinical experiments~\cite{dennis2009nearly, lee2011exchange, albarqi2019biocompatible, chang2018biologically}, computer simulations of magnetic NPs are used to better understand the details of the heating process~\cite{usov2009hysteresis,usov2013properties,fu2015study,usov2017interaction,rosensweig2002heating,bakuzis2013chain,serantes2014multiplying,torche2020thermodynamics,munoz2020disentangling,anand2020hysteresis,dejardin2020specific,kim_dynamical_2018, valdes2020modeling,valdes_2021}, which guides further experiments and more efficient cancer treatment. Magnetic heating of immobile NPs through N\'{e}el relaxation is quantified by their magnetization-field (MH) hysteresis loop area.  

Micromagnetic simulation is a numerical method that uses the Landau-Lifshitz-Gilbert (LLG) equation for describing the magnetization dynamics of NPs~\cite{gilbert2004phenomenological, cullity2011introduction, brown1963thermal}.
When the particle size decreases to the nanometer range, thermal fluctuations play a key role in the magnetization dynamics of NPs. Using uniformly magnetized cells with the same size as atomic unit cells can be computationally very expensive. To solve the problem of expensive calculations, using fewer but larger simulation cells for describing NPs is desirable. Different methods have been prescribed for scaling the magnetic parameters to enable the use of larger simulation cell sizes~\cite{CoarseGrainingFengVisscher, kirschner2005cell, kirschner2006relaxation, grinstein2003coarse, schrefl2021}. We 
calculate here  hysteresis loops of multiple ``$6z4y$'' magnetite NPs composed of ten nanorods ordered with six along the $z$ axis and four along the $y$ axis  arranged in a variety of configurations. The simulation model is motivated by a successful pre-clinical study of magnetite NPs for breast cancer treatment in mice~\cite{dennis2009nearly}.

In two recent works~\cite{behbahani2020coarse, behbahani2021multiscale} (hereafter referred to as I and II), we implemented, amended and extended a renormalization group (RG) scaling approach introduced by Grinstein and Koch~\cite{grinstein2003coarse} for simulating dynamic hysteresis loops at 310~K to include magnetostatic interactions in addition to 
exchange  interactions and 
single-ion anisotropy.  In these works we showed that the RG scaling works properly for simulating fixed-volume nanorods over a range of cell sizes ($a=ba_0$), from $b=1$ corresponding to the atomic unit cell size when a nanorod consists of 10752 cells to $b=22$, where a single block represents a nanorod. We also tested the scaling method for stacks of nanorods and 
demonstrated its validity for simulating multiple nanorods with cell sizes as big as $a=8a_0$. Employing the RG scaling method for simulating complex NPs enabled further investigations of a macrospin (MS) model to find the effective magnetization and uniaxial anisotropy of a same-volume MS with the same coercivity $H_c$ and remanent magnetization $M_r$ in MH hysteresis loops as a complex NP. 
{\blue Modelling a system as a MS assumes that the 
behavior of the magnetization is fully coherent; this does not require uniaxial anisotropy.
However, in the present study, what we refer to as a MS model is an approximation in which the collective effects of exchange interactions, magnetocrystalline anisotropy and self-demagnetization in a complex NP are captured with a single effective magnetization vector subject to uniaxial anisotropy, speeding up calculations greatly.}
The next step in evaluating the MS model is to compare the heating performance of a system of multiple NPs with a corresponding system of MSs. We showed in II that for two NPs at different separations, the hysteresis loops match those of the MS's beyond a centre-to-centre distance of 1.5 times the particle diameter. Our work also makes use of an approximate invariance of hysteresis loops under an increase in AC field sweep rate (SR) when the damping constant $\alpha$ in the LLG equation increases commensurately, which further reduces computational time {\blue and which we refer to as SR-scaling.}

{\blue Here, we employ our scaling methods to study the response of a single complex NP to varying external field frequencies and amplitudes that are clinically relevant, and then extend investigations of the applicability of the MS model to the case of clusters of NPs.}
The primary goal is to simulate clusters of NPs and compare the heating performance of complex NPs with the MS model at different particle separations. Another objective is to study the local loops of individual MSs in clusters as a step to understand how their heating mechanism is influenced by interparticle interaction, which varies in different sites of the cluster. We also explore {\blue the validity of SR-scaling}
for multiple-NP simulations with OOMMF~\cite{OOMMF} and Vinamax~\cite{leliaert2015vinamax} software used in this study.    

Different studies have used the MS approximation to investigate NP clustering in a variety of arrangements such as chains, rings, cubes, face-centered cubic (fcc), 2D hexagonal lattice, spheres or disordered structures, and reported productive or destructive effects of dipole interactions on the collective heating performance of NPs depending on the particle arrangement~\cite{anand2020hysteresis, valdes2020modeling, anand2016spin, torche2020thermodynamics, serantes2014multiplying, mehdaoui2013increase, haase2012role, cabrera2018dynamical, landi2014role, fu2015study, usov2017interaction}.
For example, Anand~\cite{anand2020hysteresis} used Monte Carlo simulations to study the effect of dipole interactions on the heating efficiency of a chain of NPs when their uniaxial anisotropy axes make an angle $\theta$ with respect to the chain axis. He found that strong dipole interactions tend to align the NPs' magnetization parallel to the chain axis, even when $\theta=90^{\circ}$. Valdes et al.~\cite{valdes2020modeling} examined the effect of dipolar interactions on the heating efficiency of NP chains with different lengths when each particle's effective anisotropy axis is along the chain axis. They found that the chain formation of NPs could improve the heating performance even if the chains are not aligned with respect to one another. Anand et al.~\cite{anand2016spin} explored the heating behavior of micron-sized spherical clusters of NPs by changing the amplitude and frequency of the applied field. They found different heating behavior for the core and surface NPs, which highly depends on the field parameters used. Serantes et al.~\cite{serantes2014multiplying} studied the effect of dipole interactions in clusters of NPs in the form of chains, 2D hexagonal lattices, cubes and rings. They reported improved heating performance when NPs formed chains, in contrast to other studied assemblies, although the beneficial effect tapered off once chains exceeded eight NPs in length.
All of these studies assumed the MS model for the NPs. One goal of the present work is to examine the impact of internal NP structure on the relevant hysteresis loops.

While evaluating the heating efficiency of NP clusters is commonly done based on their collective heating, experimental studies reported cases of effective magnetic NP therapies without a significant global rise in temperature~\cite{creixell2011egfr, villanueva2010hyperthermia}. This can be attributed to the fast temperature drop of the surrounding tissue within a short distance (~$\sim 10$~nm) from the NP surface, i.e., local heating can be effective therapeutically while occurring on a scale too small to significantly warm the surroundings~\cite{riedinger2013subnanometer}. Although there are many unsolved questions on the ultimate reasons and mechanism of death in cancerous cells through magnetic nanoparticle hyperthermia, different studies investigated the local heating of individual NPs in clusters to try to understand the process~\cite{torche2020thermodynamics, munoz2020disentangling}. Here too, we look into the local hysteresis loops for a number of different cluster geometries. 

This paper is organized as follows. 
Our model is described in section~\ref{sec:RB3_model}. 
{\blue Section~\ref{sec:singleNP} reports on clinically relevant MH loops for a single 6$z$4$y$ NP for the cases where field frequency is varied at constant maximum field strength and where maximum field strength is varied at constant frequency.} 
In section~\ref{sec:multipleNP} we simulate three NPs in chain and triangular arrangements, and 13 NPs in an fcc structure. 
In section~\ref{sec:local_loops}, local loops for individual NPs and MSs within clusters are studied, and we report our conclusions in section~\ref{sec:conclusions}. Moreover, in Appendix~\ref{sec:Eff_K}, we explore the use of two anisotropy axes in the MS model in describing complex NPs, and quantify the effect of varying the distribution of (single) anisotropy directions on hysteresis loops. Appendix~\ref{sec:SR_alpha} tests the equivalence of simulation results when the ratio of the AC field sweep rate SR to damping constant $\alpha$ is held fixed for multiple particles. 
{\blue Appendix~\ref{sec:additional} reports additional results which
serve to clarify some discussion found in the main body of the paper.}

\section{The model}\label{sec:RB3_model}

Our goal is to simulate clusters of iron oxide NPs composed of magnetite (Fe$_3$O$_4$) 
nanorods corresponding to the magnetic NPs used in an experimental study by Dennis et al.~\cite{dennis2009nearly}. Among different possible assemblies of nanorods to make up NPs, we studied three combinations of parallel and perpendicular arrangements of increasing orientational order labelled $6z4y$, $8z2y$ and $10z$ in II, and here we use the $6z4y$ structure to explore clusters of NPs. {\blue We choose the $6z4y$ NP as a test case because it is the most disordered NP considered in II, the disorder having the most impact on single-NP loops (see Fig.~5 in II).} 
{\blue Magnetite is most common candidate for magnetic nanoparticle hyperthermia, an iron oxide that the US Food and Drug Administration and European Medicine Agency approved for medical usage~\cite{colombo2012biological}. 
It has cubic magnetocrystalline anisotropy.}

For simulating complex NPs composed of ten nanorods with dimensions 6.7~nm$\times$ 20~nm$\times$ 47~nm, we use the OOMMF~\cite{OOMMF} software package and its Theta Evolve module~\cite{theta_evolve} for finite temperature calculations. In micromagnetics, instead of simulating individual atomic spins, a magnetization vector represents the collective behavior of the spins in a simulation cell of size $a$, which is usually larger than the atomic unit cell size $a_0$. In I and II, we implemented an RG-based scaling approach for simulating fixed-volume nanorods using fewer but larger cells, with size $a=ba_0$ ($b\geq1$), and here we employ the same approach for simulating clusters of multiple NPs. The LLG equation describes the magnetization dynamics of simulation cells and involves a phenomenological damping constant $\alpha$ that quantifies the energy dissipated as a magnetic moment precesses about the local effective field, which is the sum of different contributions: the external magnetic field (Zeeman) and the effective fields arising from magnetocrystalline anisotropy, exchange, magnetostatics and temperature (stochastic thermal field). In magnetic hyperthermia, to control the unwanted heating of healthy tissue due to eddy currents, the frequency $f$ and amplitude $H_{\rm max}$ of the applied magnetic field should be chosen so that the sweep rate of the AC field be less than a threshold, i.e., ${\rm SR}=4fH_{\rm max} \leq$0.25~Oe/ns~\cite{hergt2007magnetic, dutz2013magnetic}. As we explored in I, equivalent hysteresis loops are achievable using faster SR when simulating nanorods, provided that the ratio ${\rm SR}/\alpha$ remains constant. For the clinically relevant SR of 0.25~Oe/ns and an $\alpha$ of 0.1 for magnetite nanorods, the target ratio is ${\rm SR}/\alpha = 2.5$.
In this study, the NP simulations are performed in OOMMF {\blue with ${\rm SR^{sim}=25}$~Oe/ns and $\alpha^{\rm sim}=10$, preserving ${\rm SR^{sim}}/\alpha^{\rm sim}=2.5$. }In Appendix~\ref{sec:SR_alpha} we confirm that these parameter choices fall within  the range of validity of this scaling.

As in I and II, for simulating NPs with micromagnetic cells of the same size as the crystalline unit cell ($a_1=a_0=0.839$~nm), we use magnetic parameters of bulk magnetite: saturation magnetization $M_{\rm s}=480$~kA/m~\cite{dutz2013magnetic}, and exchange constant $A_0=0.98\times10^{-11}$~J/m~\cite{heider1988note}, which reproduces the experimental critical temperature of $T_c=858$~K~\cite{bercoff1997exchange} by LLG simulations. {\blue Magnetite has cubic crystalline anisotropy with constant $K_c=-10$~kJ/m$^3$, which we omit since it does not contribute significantly to hysteresis, especially given the relatively large shape anisotropy of the nanorods.  
In Appendix~\ref{sec:Eff_K}, we show that including cubic anisotropy reduces loop areas for nanorods by approximately 4\%, and for NPs by approximately 10\%.}
A $6z4y$ NP is made up of ten nanorods each with dimensions 8$a_0\times$24$a_0\times$56$a_0$ and volume $V=6350.0$~nm$^3$, with six of them lying along the $z$ direction and four lying along the $y$ direction. Using the modified RG-based coarse-graining procedure explained in I and II, we simulate NPs with micromagnetic cells of side length $a_4=4 a_0$, i.e., we set the scaling parameter to $b=4$, and with scaled exchange, anisotropy, and magnetostatic interactions, at $T=310$~K. 
At this level of coarse-graining, each nanorod consists of $2\times 6 \times 14 =168$ micromagnetic cells, and so each NP, containing 10 nanorods, consists of 1680 cells.
{\blue Briefly, scaling depends on the function $\zeta(b)=t/b+1-t$, with $t=T/T_c=310/858=0.361$, yielding $\zeta(4)=0.729$ and scaled exchange constant $A_4=\zeta(4)A_0$. 
Any anisotropy energy density, cubic or uniaxial, is scaled within simulations to be $K_4=\zeta(4)^3K_0$, although in the present study, unless otherwise noted, $K_0=0$. In Appendix~\ref{sec:Eff_K}, we consider the case of non-zero uniaxial anisotropy.}
Magnetostatic interactions need to be scaled by a factor $\zeta(4)^3$, and this is accomplished by using the following input parameters for OOMMF {\blue simulations:
$M_{\rm s}^{\rm sim} = \zeta^{3/2}(4) M_{\rm s}= 298.78$~kA/m;
$A^{\rm sim}= \zeta^{4}(4) A_4/2= 0.138\times 10^{-11}$~J/m (the extra factor of 1/2 arises from the definition of the exchange interaction used by OOMMF);
$K^{\rm sim}= \zeta^{9/2}(4) K_4= 0$ for $K_0=0$ (or $2.412$~kJ/m$^3$ when $K_0=10$~kJ/m$^3$); and
$T^{\rm sim}= \zeta^{3/2}(4) T= 193$~K.}
In these coarse-grained simulations, the field is applied along the $z$ axis with {\blue component $H^{\rm sim} \equiv H(b)=H_{\rm max}^{\rm sim}\sin (2\pi\omega t)$.} The $z$ component of the resulting magnetization for the coarse-grained system, {\blue  i.e., the output from the simulations, is $M_H^{\rm sim} \equiv M_H(b)$,} which we report in normalized form as $m_H(b)=M_H(b)/M_{\rm s}$.  The quantities of interest in determining hysteresis loops, however, are the corresponding quantities in the original, non-coarse-grained system, and are given by $H_0 = H(b)/\zeta(b)$ and $m_0=(\delta\zeta(b)+1-\delta)m_H(b)$, where $\delta=0.511$ is a phenomenological parameter. 
{\blue $H_0$ and $m_0$ are the quantities corresponding to experiments.}
Neighboring cells on different nanorods interact via exchange with {\blue half the strength as cells within the same nanorod, $A_0/2$, (appropriately scaled in the coarse-grained simulations).  This choice of parameter value reflects a likely reduced inter-rod exchange strength~\cite{johan}, and the impact of its precise value was studied in II.} We simulate using full magnetostatic interactions, including ``self-demag'' within cells, in addition to magnetostatic interactions between cells.  Including magnetostatics means that it is not necessary to include a non-zero value of $K_0$ for magnetite to approximate the effects of shape anisotropy.
Hysteresis loops are calculated by averaging over at least 100 independent runs. 



{\blue 
As described above, in this study we employ two scaling techniques to simulate experimental outcomes for complex nanoparticles, and we wish to clarify here how simulation results are related to experiments.
For example, when employing SR-scaling with no spatial coarse-graining, if the simulations are carried out using damping parameter $\alpha^{\rm sim}=1$, field frequency $f^{\rm sim}=625$~kHz and field amplitude $H_{\rm max}^{\rm sim}=1000$~Oe, the results are equivalent to experiments carried out with $\alpha=0.1$, frequency $f=62.5$~kHz and amplitude $H_{\rm max}=H_{\rm max}^{\rm sim}=1000$~Oe.
If both SR-scaling and coarse-graining are employed, as they are here with $b=4$, then coarse-grained simulations carried out with $\alpha^{\rm sim}=1$, field frequency $f^{\rm sim}=625$~kHz and field amplitude $H_{\rm max}^{\rm sim}=1000$~Oe are equivalent to coarse-grained simulations carried out with $\alpha^{\rm sim}=0.1$, frequency $f^{\rm sim}=62.5$~kHz and amplitude $H_{\rm max}^{\rm sim}=1000$~Oe, such that ${\rm SR}^{\rm sim}/\alpha^{\rm sim}=2.5$ is held constant; the simulation results are equivalent to experiments carried out with $\alpha=0.1$, frequency $f=62.5$~kHz and amplitude $H_{\rm max}=H_{\rm max}^{\rm sim}/\zeta(4)=1372$~Oe, so long of course as $m_0$ is reported as the (normalized) magnetization.  
This also means that as ${\rm SR}^{\rm sim}/\alpha^{\rm sim}=2.5$ for the coarse-grained system, the corresponding experimental results are for $\alpha=0.1$ and ${\rm SR}=0.25/\zeta(4)=0.34$~Oe/ns.   
When reporting values of specific loss power, SLP $=A_{\rm loop}(1000/4\pi)M_{\rm s}\mu_0f/\rho$, 
where $A_{\rm loop}$ is the area of an $m_0$-$H_0$ loop in units of Oe, $\mu_0$ is the permeability of free space, $f$ is the experimental frequency (in Hz) to which the results correspond (assuming $\alpha=0.1$), and $\rho=5.17\times 10^6$~g/m$^3$ is the mass density of magnetite. With the input quantities given in the above units, the units of SLP are W/g.  Unless otherwise stated, our simulation results correspond to experimental parameters $f=62.5$~kHz, $H_{\rm max}=1.37$~kOe, and SR=0.34~Oe/ns, assuming magnetite NPs have a damping parameter of $\alpha=0.1$.

}

We also use the MS model introduced in II, which refers to
single-moment macrospins equivalent to NPs with complex internal structure. A MS with tailored effective magnetization and uniaxial anisotropy exhibits dynamic hysteresis loops similar to that of a complex NP.
In this study, we {\blue first simulate a single $6z4y$ NP, using OOMMF, and the equivalent MS, using Vinamax~\cite{leliaert2015vinamax}, to study and compare their responses to varying field frequency and amplitude.  We then} simulate clusters of $6z4y$ magnetite NPs and compare their hysteresis loops with clusters of equivalent MSs, again using Vinamax as it gives more flexibility in assigning the position of MSs in clusters. Owing to the smaller number of calculations for MSs that interact just with dipole interactions, these hysteresis loop simulations can be performed efficiently with the clinically-relevant SR of 0.25~Oe/ns and with $\alpha=0.1$.
We use 
{\blue $M_{\rm s}^{\rm eff}=381.6$~kA/m and $K_{\rm eff}=3.50$~kJ/m$^3$, 
values determined following the methods in} II to yield MH loops equivalent to $6z4y$ NP when the field is applied along the $z$ axis for a single particle.  {\blue (The $K_{\rm eff}$ used here is slightly smaller than the value of 3.64~kJ/m$^3$ reported in II, where we used a value of $H_{\rm max}=1.00$~kOe for the MS calculation instead of the proper value of  1.37~kOe.)} Note that since temperature affects the MS less on account of its large volume, the magnetization of the MS saturates to nearly $M_{\rm s}^{\rm eff}$ at $H_{\rm max}$.  Thus, when reporting normalized magnetization, we report the quantity $M_H/M_{\rm s}$, rather than $M_H/M_{\rm s}^{\rm eff}$. The volume of the spherical MS is the same as that of a complex NP, $V_{\rm NP}=63500$~nm$^3$.

The importance of coarse-graining is particularly pronounced when we model multiple NPs. For example, the simulation time for a cycle of an AC field for 13 complex NPs in an fcc configuration, at the smallest nearest-neighbour separation, using OOMMF with $a=4a_0$ and {\blue ${\rm SR^{sim}}=25$~Oe/ns,} takes more than 16 days, using a typical workstation, whereas it takes around 11~minutes to simulate 13 MSs with ${\rm SR}=0.25$~Oe/ns, using Vinamax {\blue (for the MS model, no scaling of any kind is applied)}. This huge difference is due to various factors: the higher number of interactions required for modeling complex NPs compared to MSs; simulating the empty cells between NPs takes up some of the OOMMF simulation time, whereas this is not a consideration in Vinamax; and the small integration timestep of 10~fs required for the small cells used to model NPs compared to the time step of 1~ps for MSs.  Without spatial coarse-graining, the simulation of complex NPs with OOMMF would take approximately $10\times 4^3=640$ times as long (since a smaller cell requires a shorter time step).


\section{Properties of a single NP}\label{sec:singleNP}

\begin{figure*}
    \centering
    \includegraphics[width = 0.49 \textwidth]{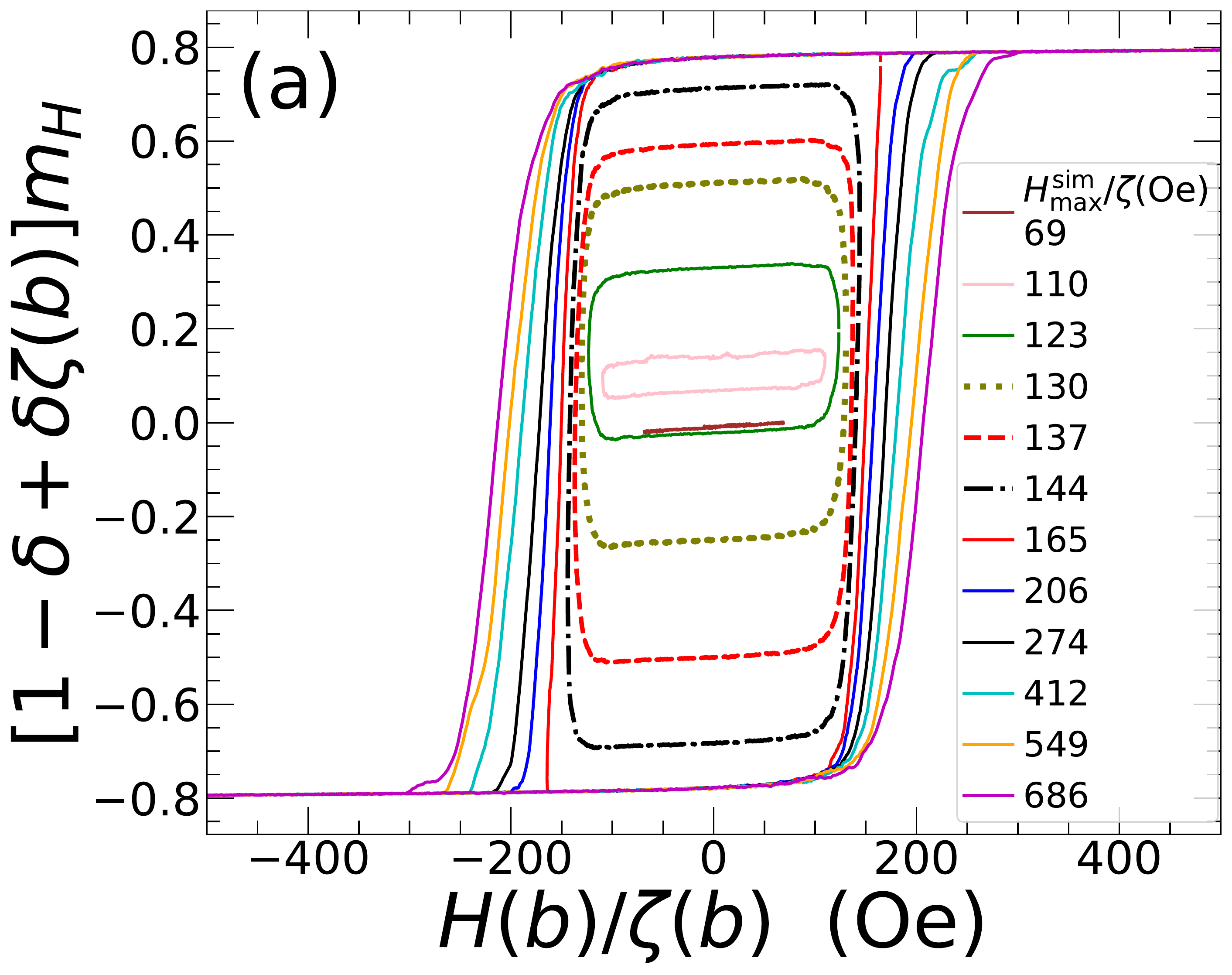}
    \includegraphics[width = 0.49 \textwidth]{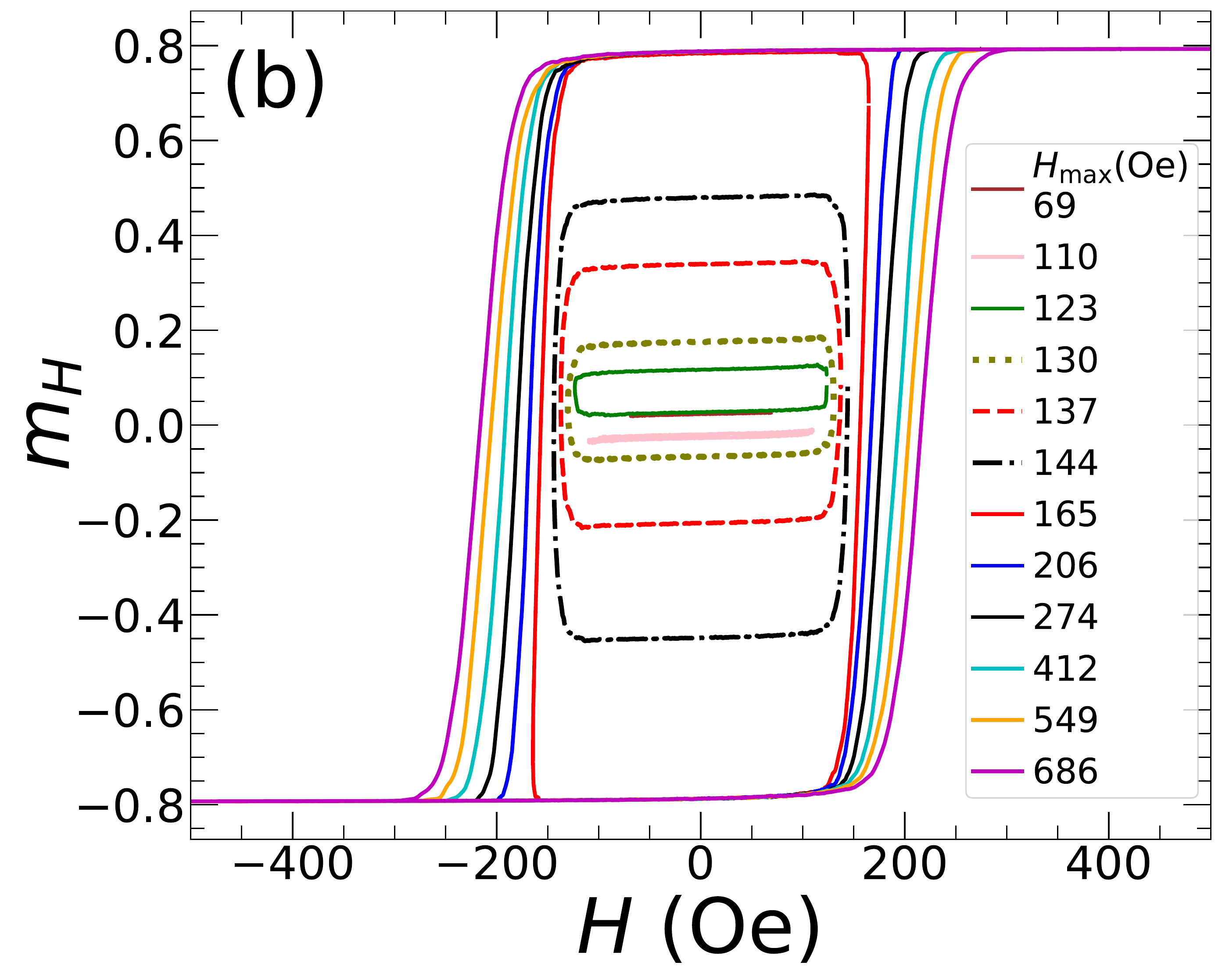}
        \includegraphics[width = 0.53 \textwidth]{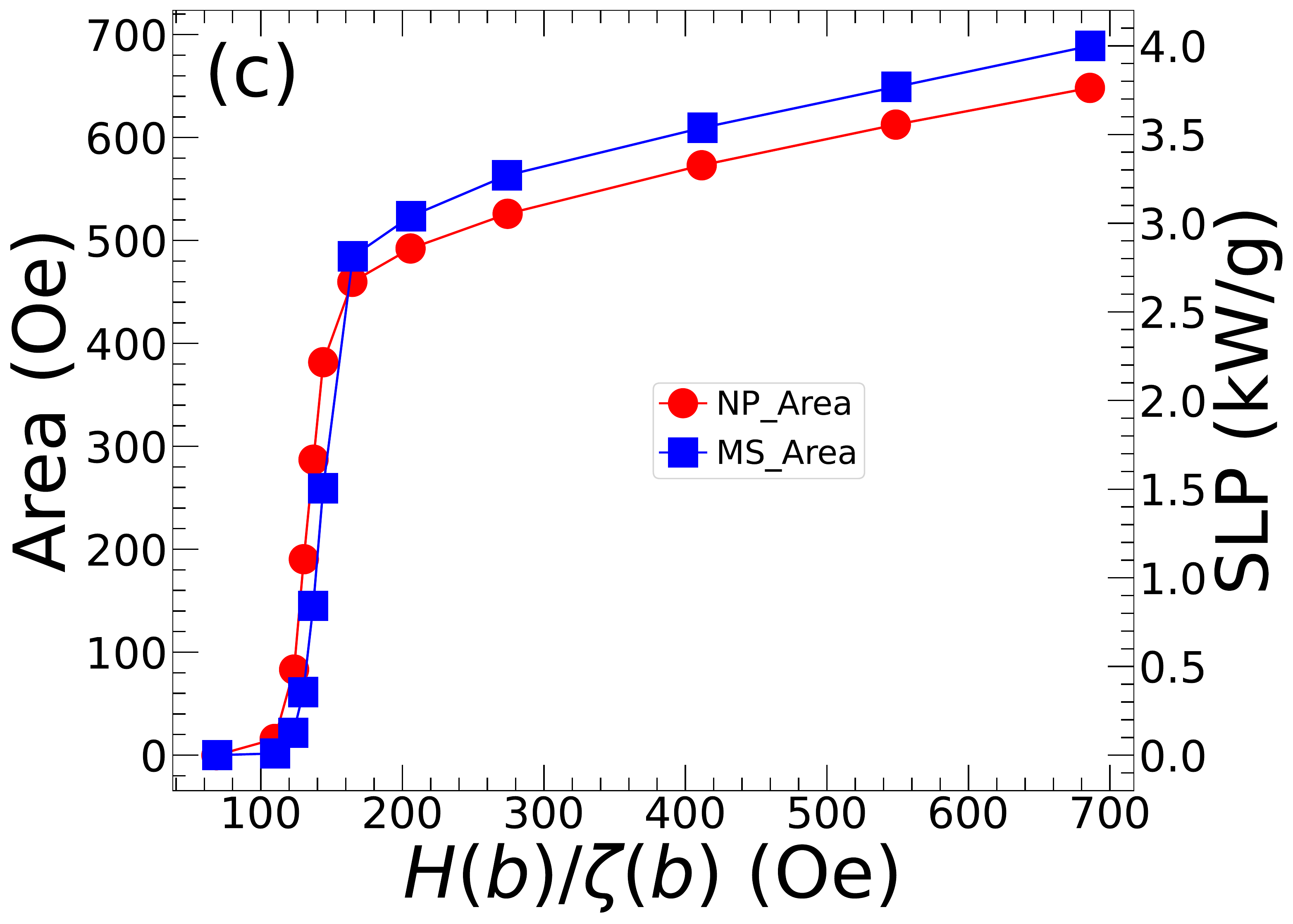}
    \caption{
    {\blue Hysteresis loops as a function of the field amplitude for (a) $6z4y$ magnetite NPs, simulations with  $\alpha^{\rm sim}=1$, $f^{\rm sim}=6.25$~MHz equivalent to the clinical parameters of $\alpha=0.1$, $f=625$~kHz, (b) MSs, with $K_u=3.50$~kJ/m$^3$, $M_{\rm s}=381.6$~kA/m, $\alpha=0.1$, $f=625$~kHz. (c) Loop area and SLP as a function of the field amplitude for NPs are shown with red circles and for MSs in blue squares, respectively.}}
    \label{fig:H_slp}
\end{figure*}

\begin{figure*}
    \centering
    \includegraphics[width = 0.495 \textwidth]{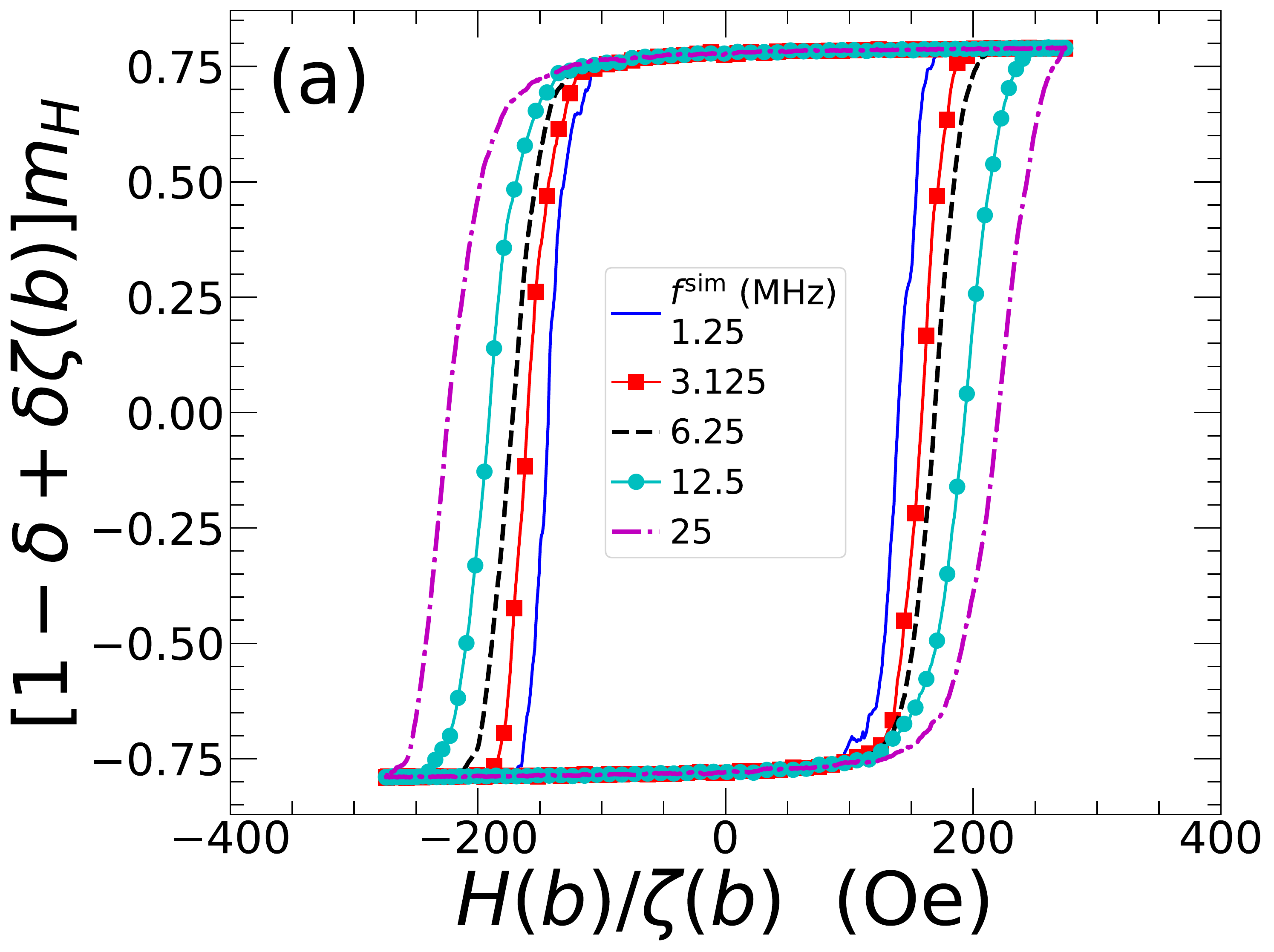}
    \includegraphics[width = 0.49 \textwidth]{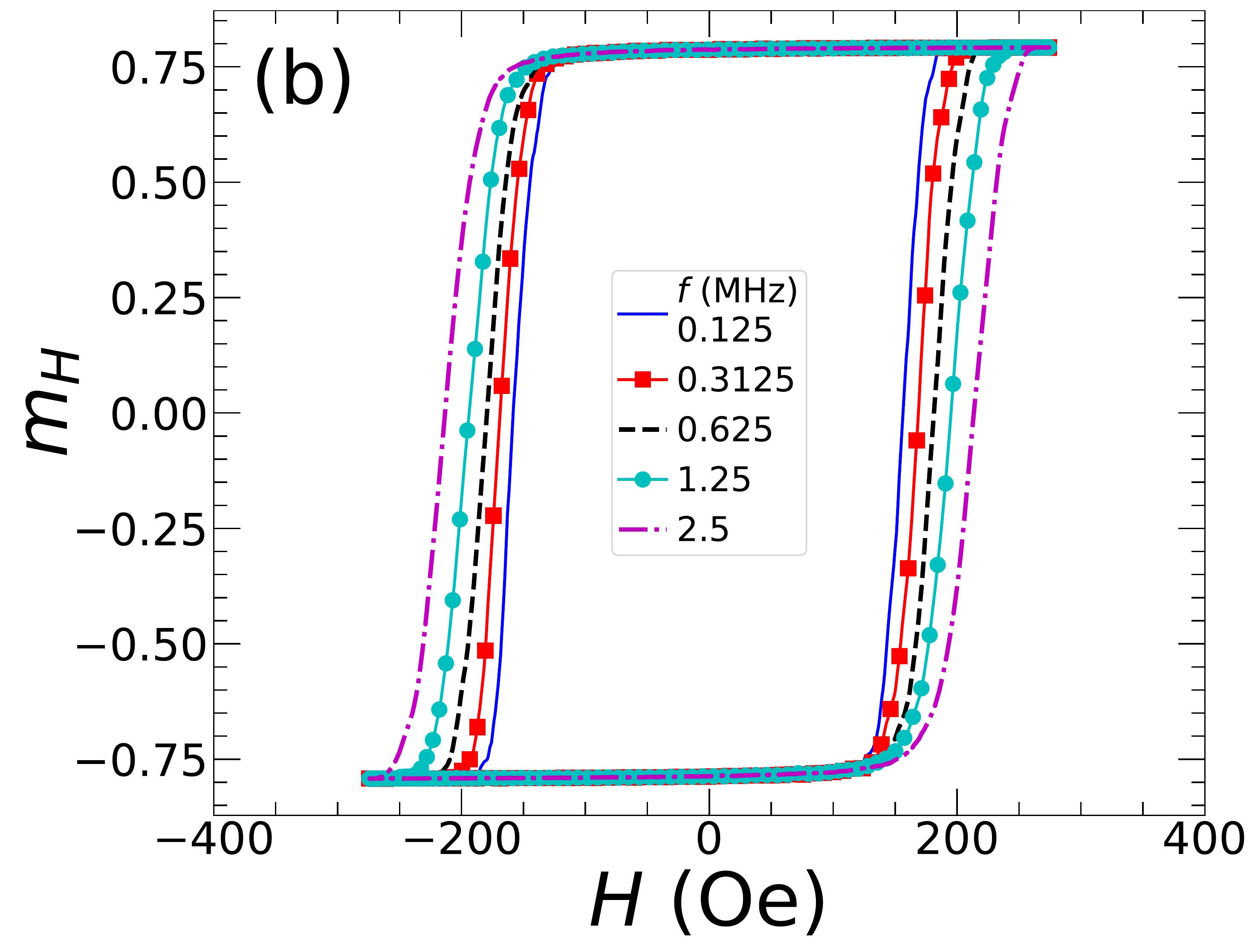}
    \includegraphics[width = 0.51 \textwidth]{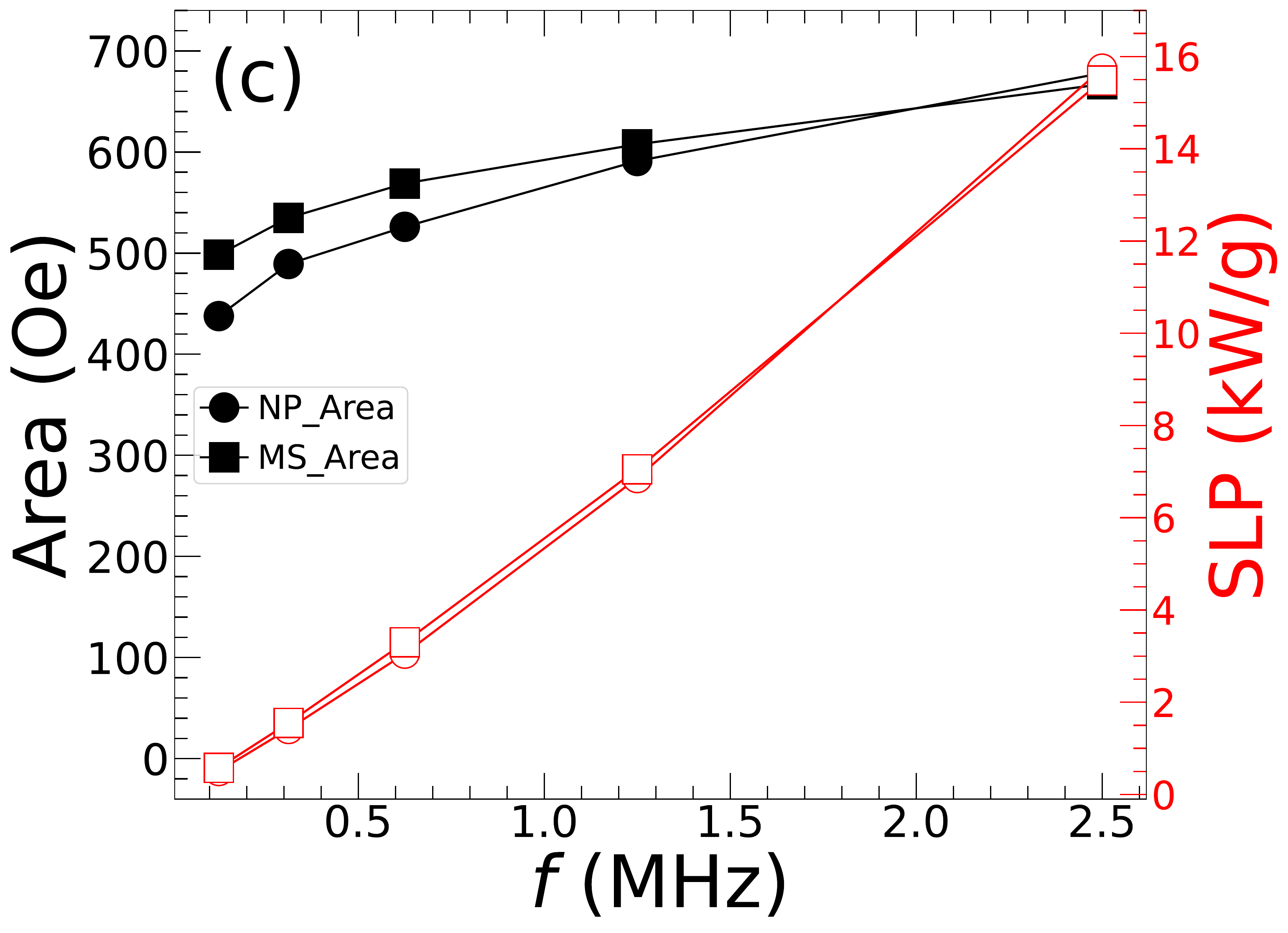}
    \caption{
    {\blue Hysteresis loops as a function of the field frequency for (a) a $6z4y$ magnetite NP with $H_{\rm max}^{\rm sim}=200$~Oe,  $\alpha^{\rm sim}=1$ and simulation frequency $f^{\rm sim}$ equivalent to laboratory parameters  $H_{\rm max}=200/\zeta=274$~Oe, $\alpha=0.1$, and frequency $f=f^{\rm sim}/10$, (b) an MS with $K_u=3.50$~kJ/m$^3$, $M_{\rm s}=381.6$~kA/m, $\alpha=0.1$, $H_{\rm max}=274$~Oe and clinical frequencies of 125, 312.5, 625, 1250, 2500~kHz corresponding to SRs of 0.137, 0.343, 0.686, 1.372, 2.743~Oe/ns. (c) Loop area and SLP as a function of the clinical field frequency for NPs are shown with red closed and open circles, and for MSs in black closed and open squares, respectively.}}
    \label{fig:f_slp}
\end{figure*}

{\blue How a magnetic nanoparticle's response to a magnetic field varies with frequency and amplitude is important to know for clinical applications.  The coarse-graining and SR-scaling approaches we have developed allow us to calculate loops and associated values of SLP for a complex NP and for the corresponding MS.  In Fig.~\ref{fig:H_slp}a we plot the MH loops for a $6z4y$ magnetite NP resulting from a field with $f=625$~kHz and amplitude varying from $H_{\rm max}=69$ to 686~Oe.  Note that the NP simulations are carried out at higher frequency $f^{\rm sim}=6.25$~MHz and lower amplitudes $H_{\rm max}^{\rm sim}$, with $H_{\rm max}=H_{\rm max}^{\rm sim}/\zeta(4)$, but the scaling methods allow us to report experimentally relevant quantities. In Fig.~\ref{fig:H_slp}b, we show the results for the equivalent MS.  For the MS simulations, no scaling is required and experimentally relevant quantities are used directly.  We see that in both cases, the transition from minor to major  loops occurs between 144 and 165~Oe.  At $H_{\rm max}=165$~Oe and $f=625$~Hz, ${\rm SR}\approx 0.41$~Oe/ns, which is near, but just above, the SR threshold for collateral heating. Fig.~\ref{fig:H_slp}c plots the loop areas from panels (a) and (b) (left axis) and the corresponding SLP (right axis), confirming that reducing $H_{\rm max}$ below $165$~Oe to meet the SR tolerance results in a dramatic drop in SLP.  The above results are examples of the kind of analysis that simulations enable for  designing and characterizing magnetic NPs.  On a more minor note, the MS overestimates the SLP for major loops while transitioning to minor loops at slightly higher field values.

In Fig.~\ref{fig:f_slp} we address the effect of varying frequency at fixed field amplitude for the same NP-MS pair.  While both the NP (Fig.~\ref{fig:f_slp}a) and MS (Fig.~\ref{fig:f_slp}b) maintain major loops, even at lower $f$, the dependence of SLP on $f$ (Fig.~\ref{fig:f_slp}c) means that a reducing $f$ to yield an acceptable SR results in an SLP of approximately 1~kW/g.
The following section describes the impact of clustering on SLP for field amplitude and frequency near the SR safety threshold.
}

\section{Multiple NP heating efficiency}\label{sec:multipleNP}

We explore and contrast in this section hyperthermia-relevant hysteresis loops of clusters of NPs and MSs arranged in chains, triangles and fcc configurations.


\subsection{Chained particles}

Chains of NPs are the most common aggregation structure reported in simulation studies~\cite{torche2020thermodynamics, anand2020hysteresis, valdes2020modeling, serantes2014multiplying}. 
\begin{figure*}
    \centering
    \includegraphics[width=0.441\textwidth]{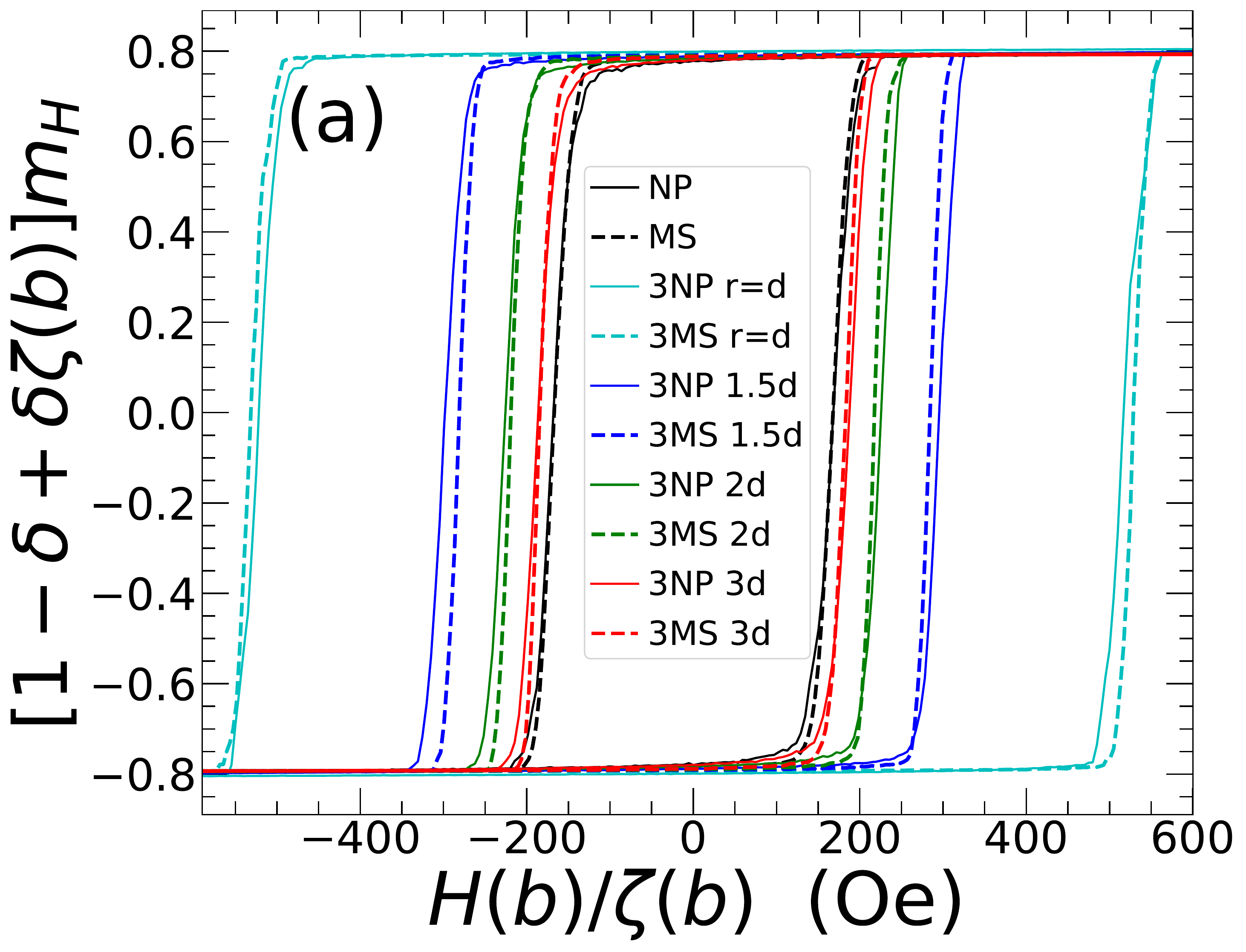}
    \includegraphics[width=0.525\textwidth]{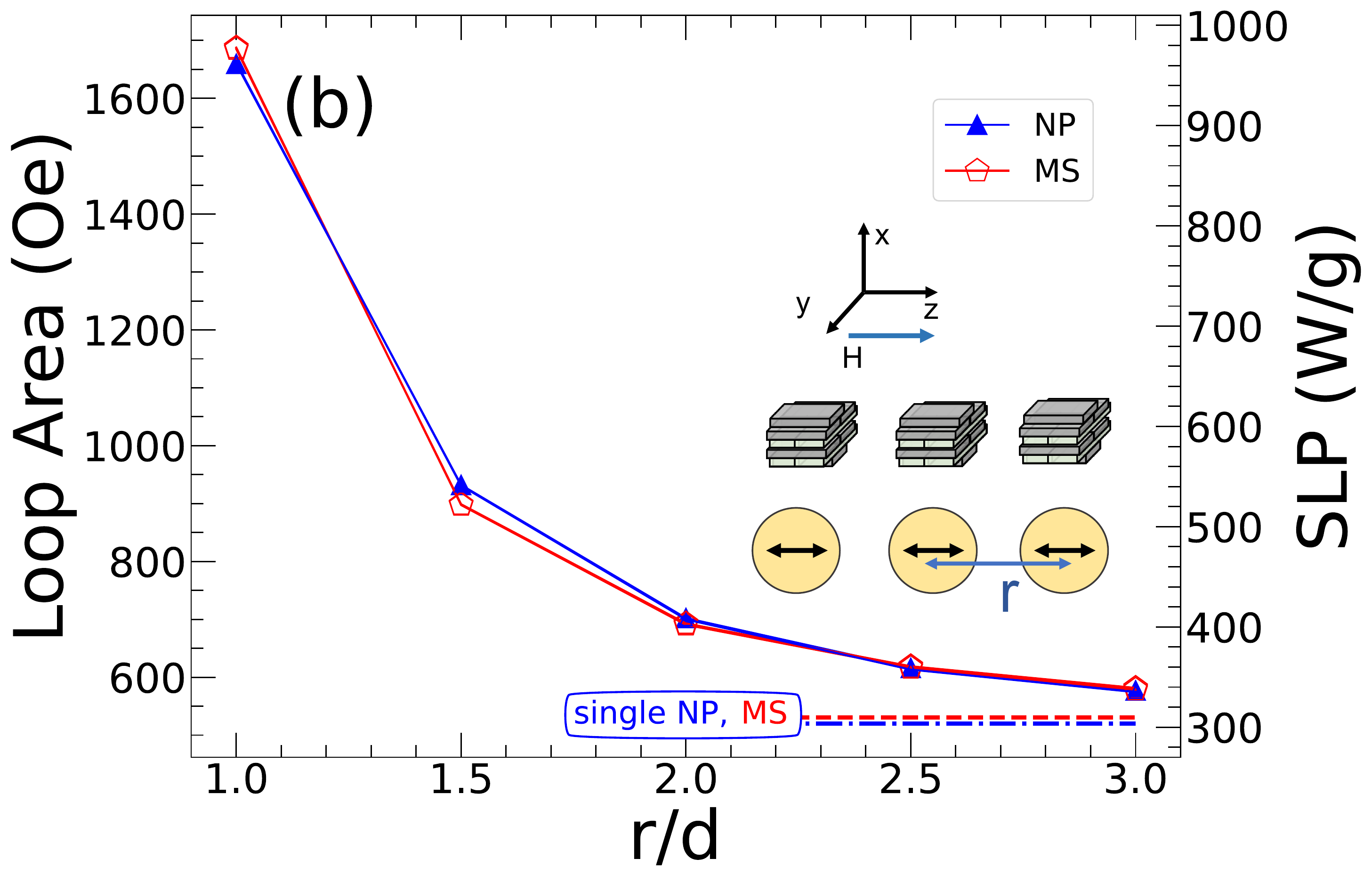}
    \caption[Hysteresis loops of three chained NPs and equivalent MSs as a function of interparticle distance]{a) Hysteresis loops of three NPs and three MSs when their center-center distance varies from 1 to 3 NP diameters (d). b) Loop areas vs center-center distance $r$ for three chained NPs and MSs as shown in the inset. The energy per loop per NP can be calculated as {\blue $E=\mu_0 M_{\rm s} V_{\rm {NP}} A_{\rm loop}$   while SLP $= A_{\rm loop} (1000/4\pi)M_{\mathrm {s}}\mu_0f/\rho$, with density  in g/m$^3$.
    The SLP is calculated using the clinical frequency of $f=62.5$~kHz, $H_{\mathrm{max}}= 1000/\zeta = 1372$~Oe, $\rho=$5.17~g/cm$^3$ and $M_{\mathrm{s}}=480$~kA/m.
    }} \label{fig:chain_omfvmx}
\end{figure*}
We start with a chain of three particles and compare hysteresis loops of three complex magnetite NPs with three equivalent MSs. In addition to the average hysteresis loop of particles at different center-to-center distances, loops for a single NP and a single MS are shown as benchmarks representing the limit of independent particles, i.e., when interactions between particles have a negligible effect on the particles' magnetic response to the field. Corresponding loop areas are also calculated as a simpler metric of comparison. As shown in Fig.~\ref{fig:chain_omfvmx}a, the hysteresis loops corresponding to MSs are in  good agreement with those from NPs. The wider loops for closer particles are a result of the effect of dipole interactions aligning the chain particles' magnetizations, in agreement with the results reported by Torche et al.~\cite{torche2020thermodynamics}, Anand~\cite{anand2020hysteresis}, Valdes~\cite{valdes2020modeling} and Serantes et al.~\cite{serantes2014multiplying}.
Convergence of the loop area to the single NP or MS case implies that particles are approximately independent when the center-to-center distance between neighbours $r$ exceeds 3 NP diameters ($d=47.0$~nm), i.e., when $r>3d$;
see Fig.~\ref{fig:chain_omfvmx}b, where the loop area follows an approximately $1/r^3$ dependence. 
Thus, for magnetite NPs in a chain, despite the internal structure of $6z4y$, the MS approximation is a good one.
 
\subsection{Triangular order: when the internal structure matters}\label{sec:triangleMS}

\begin{figure*}
    \centering
    \includegraphics[width=0.495\textwidth]{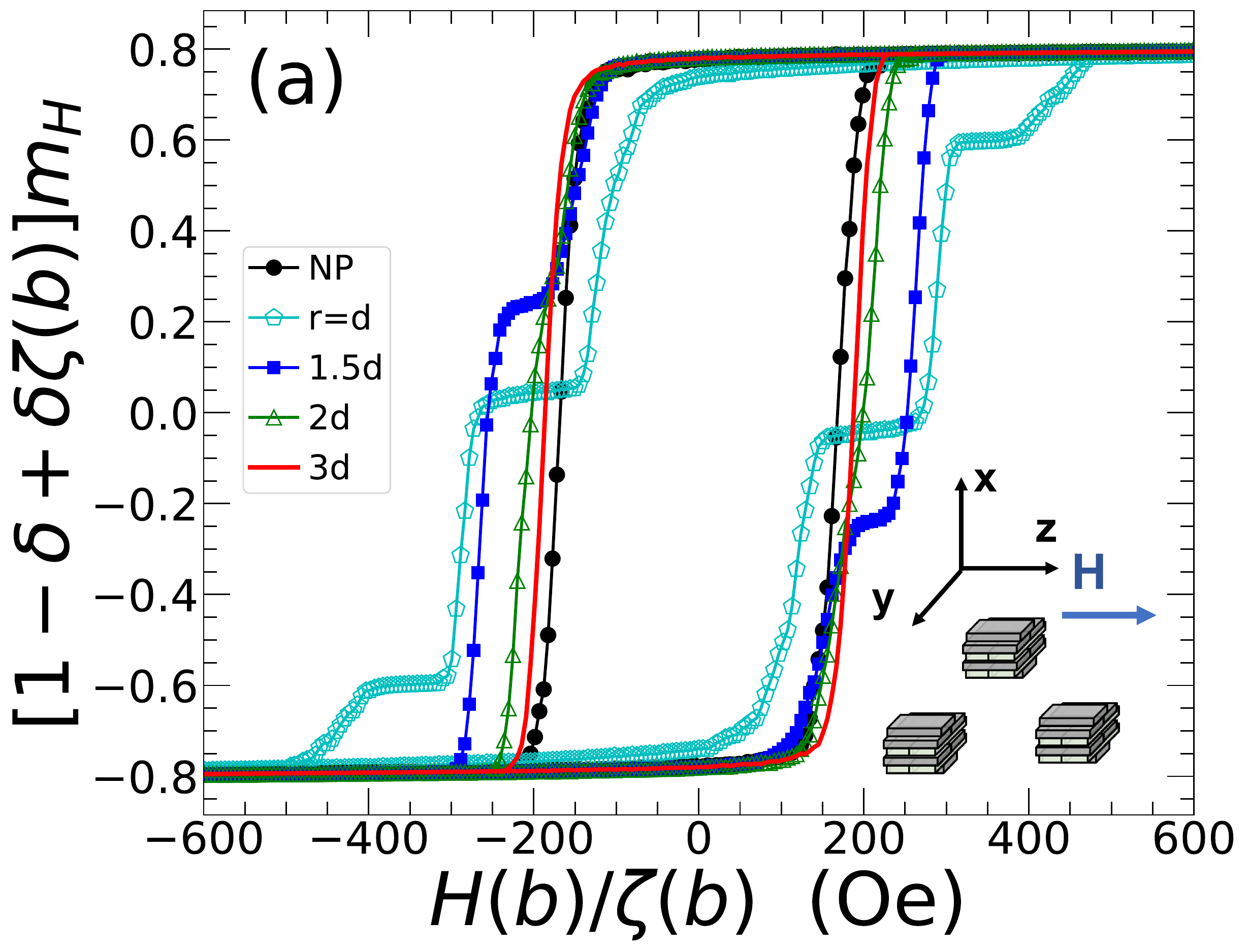}
    \includegraphics[width=0.495\textwidth]{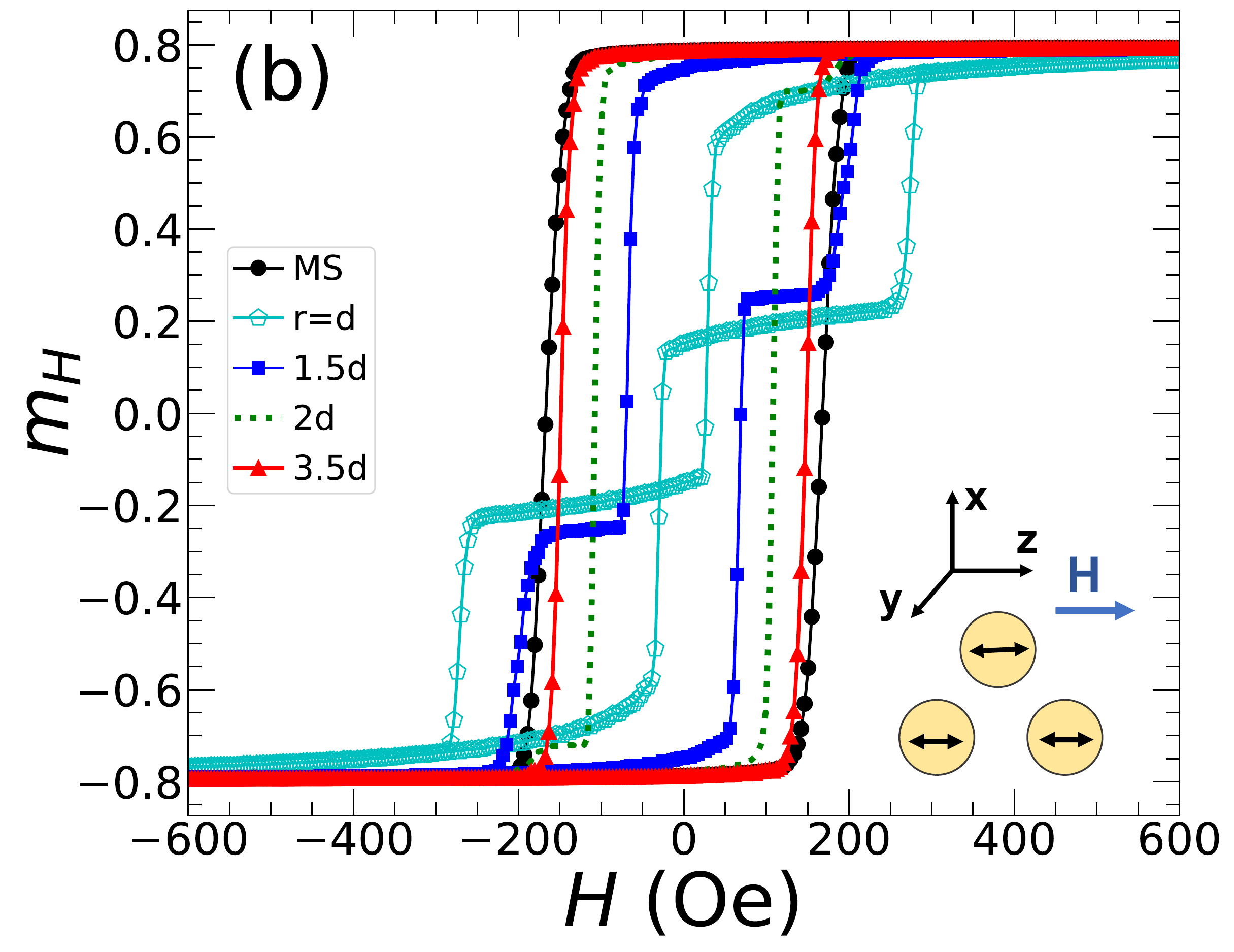}
    \caption[Hysteresis loops of three NPs and equivalent MSs in triangular order as a function of interparticle distance]{Hysteresis loops as a function of particle distances for systems of three interacting a) NPs b) MSs, on vertices of an equilateral triangle. $d$ is a NP diameter and $r$ is the particles center-center distance.\label{fig:triangle_omfvmx}}
\end{figure*}

Serantes et al.~\cite{serantes2014multiplying} studied the heating efficiency of eight NPs in a hexagonal structure, which can be considered as an extended triangular cluster, and demonstrated that dipole interactions diminish the hysteresis loop area compared to non-interacting particles. 
Figs.~\ref{fig:triangle_omfvmx}a and b show hysteresis loops corresponding to systems of three NPs on vertices of an equilateral triangle with $r$ varying from $d$ to $3d$ and similarly for three MSs with $r$ from $d$ to $3.5d$. As shown in the insets, the applied field is along $z$, parallel to the MS's uniaxial anisotropy and the $6z4y$ NP's shape anisotropy axis (see Appendix~\ref{sec:Eff_K} {\blue for further discussion}). 
The remarkable difference between the hysteresis loops of NPs and MSs for small $r$ reveals that, for the triangular configuration, the combined effect of exchange and magnetostatics on spin alignments of complex NPs is qualitatively different from that of dipole interactions for MSs, and results in significantly larger hysteresis loop areas for NPs compared to MSs.
{\blue We note in Fig.~\ref{fig:triangle_omfvmx}b that for the loop at $r=d$, there is a central portion showing inversion, where the magnetization is lower on decreasing the field than on increasing the field.  This portion has  a negative area (taking area enclosed in a regular loop as positive), reducing the heating capability of the triangular cluster.}
{\blue Having an inverted portion of the loop appears specific to the MS model. In Appendix~\ref{sec:additional} (Fig.~\ref{fig:Triangle_10z}), we show that using $10z$ NPs (with all nanorods in the field direction), and additionally by adding artificial magnetocrystalline anisotropy ($K_{u0}=10$~kJ/m$^3$) to the nanorods, does not yield inverted loops. 
The global loop for $10z$ NPs in a triangle has a different shape, but its area is approximately the same as for $6z4y$ NPs; adding magnetocrystalline anisotropy increases the loop area without qualitative changes.}

\begin{figure}
    \centering
    \includegraphics[width = 0.495\textwidth]{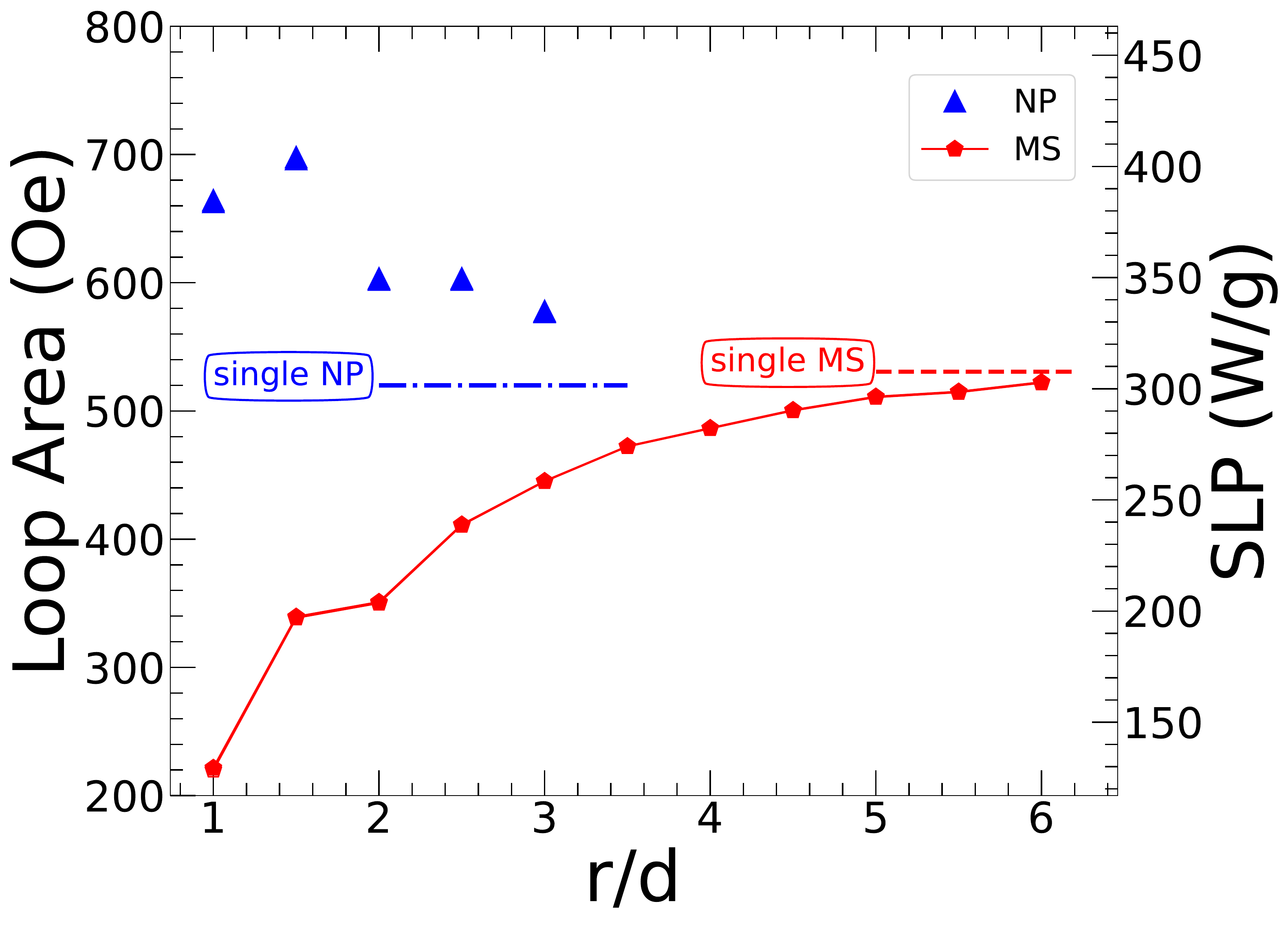}
    \caption[Loop area as a function of inter-particle distance for three NPs and MSs in triangular order]{Loop area as a function of inter-particle distance ($r$) normalized with a NP diameter ($d$), for three NPs and MSs on vertices of an equilateral triangle. {\blue The right axis gives the associated SLP for clinical conditions with $f=62.5$~kHz and $H_{\rm max}=1.37$~kOe.}
    }\label{fig:triangle_area}
\end{figure}

Fig.~\ref{fig:triangle_area} shows the loop area as a function of $r$, for the same arrangements considered in  Figs.~\ref{fig:triangle_omfvmx}a and b. This serves to quantify the poor quality of the MS approximation, especially at small $r$: 
for the MS case, interactions between particles reduce heating efficiency, while for more detailed simulations of NPs, interactions increase heating efficiency.
As expected, the loops converge to the single particle limit as $r$ increases, although more quickly for the NPs.

\begin{figure}
    \centering
    \includegraphics[width= 0.495\textwidth]{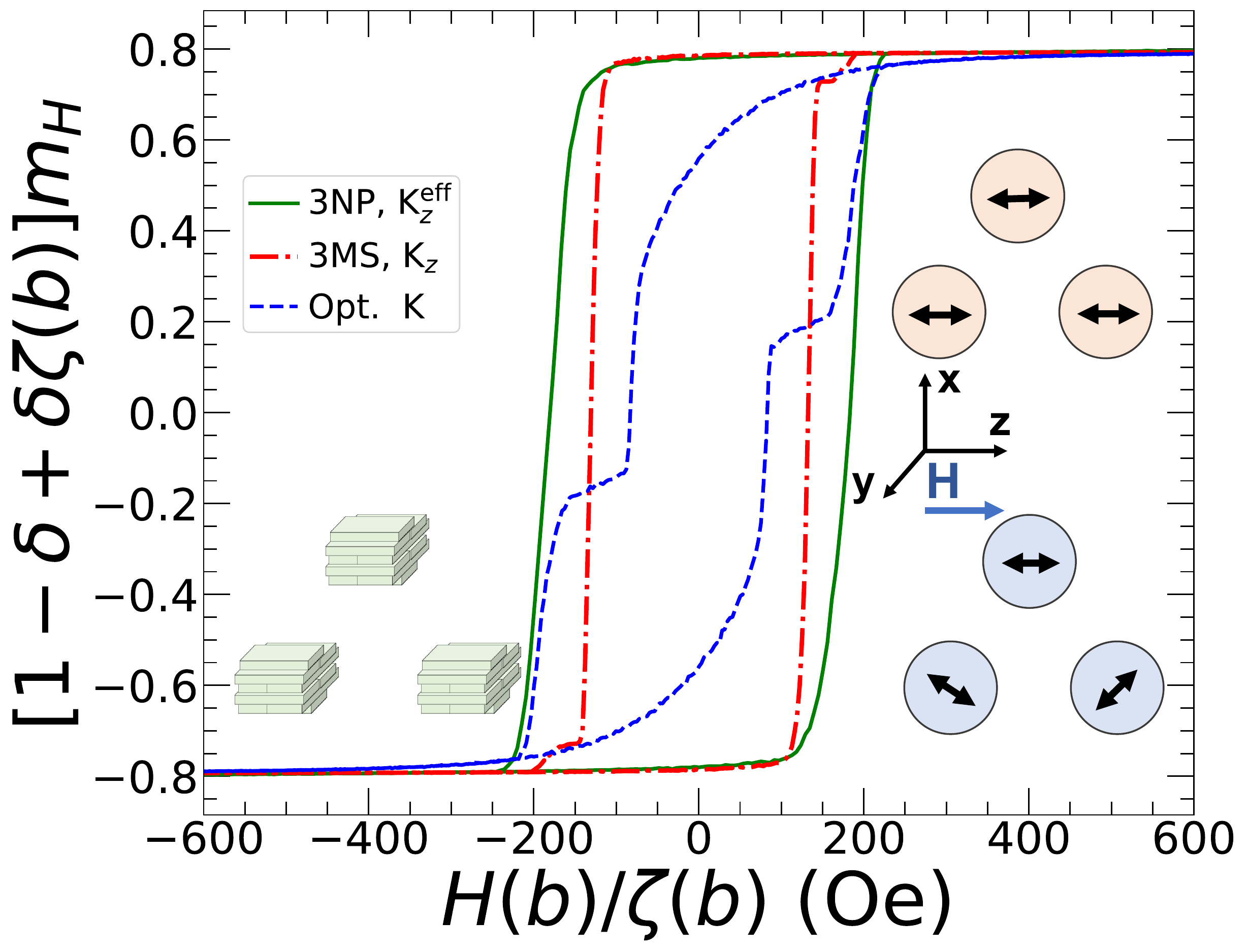}
    \caption[Loop comparison of three NPs and MSs, with different anisotropy orders, arranged in a triangle of side length $2.5d$.]{Comparison of NPs and MSs arranged in a triangle of side length $2.5d$. The green curve is the hysteresis loop for three $6z4y$ NPs, oriented as in the left inset. Black arrows represent the anisotropy axes in MSs. The dashed blue loop corresponds to assigning the MS anisotropy axes at 120$^\circ$ relative to each other (the lowest dipole energy arrangement at $H=0$). The red dot-dashed loop corresponds to the case where the MS anisotropy axes are aligned with the field.}
    \label{fig:optAnis}
\end{figure}

We now briefly consider the case when MSs are initially oriented to minimize dipolar energy, and 
make 120 degrees with respect to each other~\cite{holden2015monte}. 
Fig.~\ref{fig:optAnis} compares the hysteresis loops of three NPs and MSs in triangular order, when the MSs' anisotropy axes are aligned 120 degrees with respect to each other and when they lie along the field, with $r=2.5d$.
The biggest loop area corresponds to three complex NPs with anisotropy aligned with the field,
followed by the MSs with anisotropy axes along the external field, and finally the $120^{\circ}$ alignment case exhibits the smallest area.


\subsection{NPs in an ``fcc'' structure}

Arrays of NPs packed in different arrangements such as spheres, cubes and fcc structures have been studied and reported in the literature~\cite{serantes2014multiplying, anand2016spin, fu2015study, usov2017interaction}. Fu et al.~\cite{fu2015study} investigated the dipole interaction effects on the heating performance of a cluster of 64 and 63 superparamagnets in simple cubic and fcc structures, respectively. They introduced a concept called morphology anisotropy, which is defined in terms of the aspect ratio of the semi-axes of an ellipsoid which is equivalent to the cluster, and concluded that in the structures without morphology anisotropy the effect of dipole interaction is minimized and the cluster's loop area is almost the same as non-interacting particles. This can be interpreted to mean that the effect of dipole interactions is considerable when the cluster is extended in one direction, similar to a chain of NPs.
Serantes et al.~\cite{serantes2014multiplying} studied eight nanoparticles in a cubic structure and observed a negative effect of dipole interactions on their collective heating compared to non-interacting nanoparticles, unlike chained particles.
In the present study, by exploring the effect of dipolar interactions on the hysteresis loop at different interparticle distances, we are comparing the response of NPs and MSs in situations of the kind reported in Ref.~\cite{serantes2014multiplying}; however, they focus on eight MSs in various geometries, and at fixed interparticle distance.

\begin{figure*}
    \centering
    \includegraphics[width=0.495\textwidth]{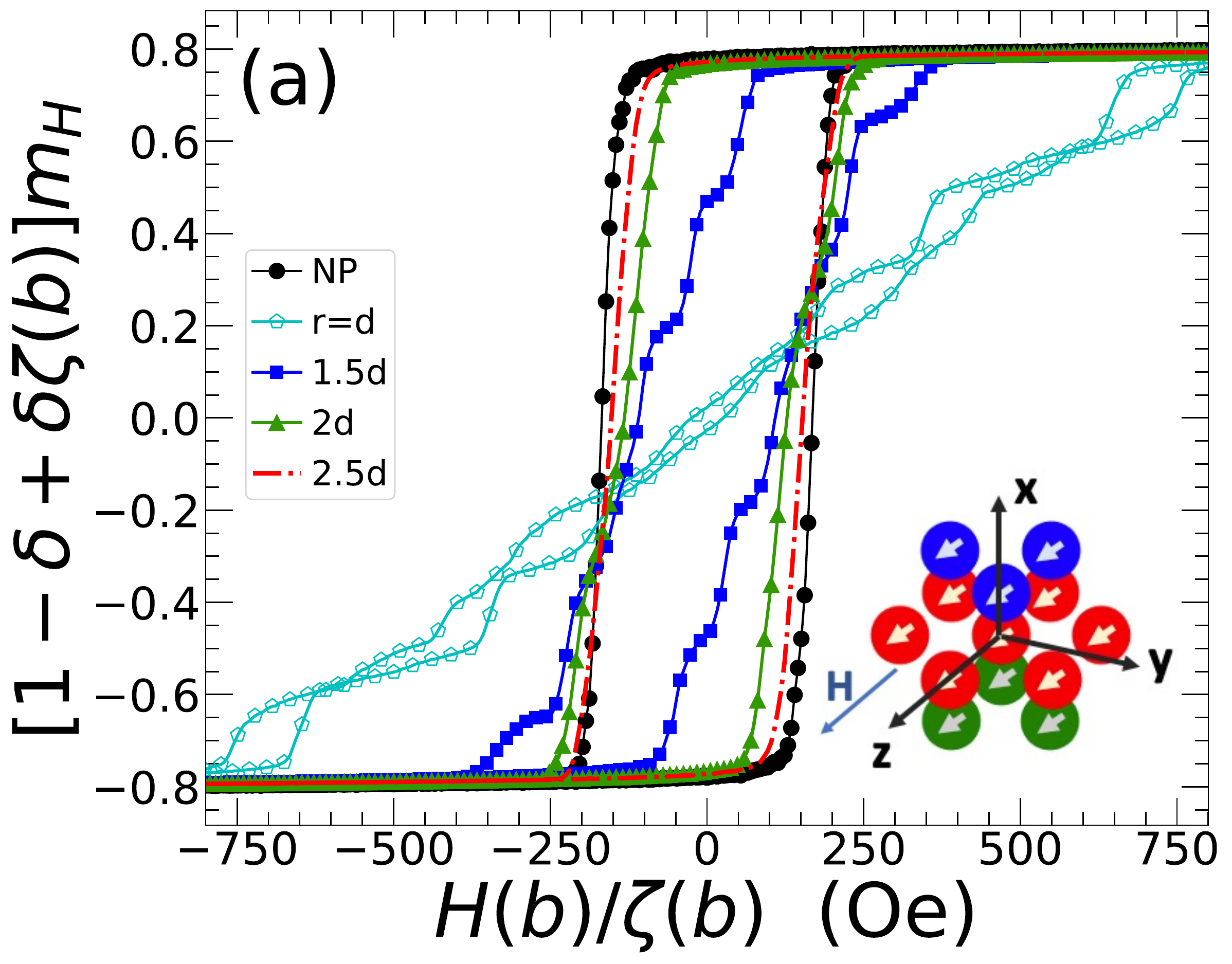}
    \includegraphics[width=0.495\textwidth]{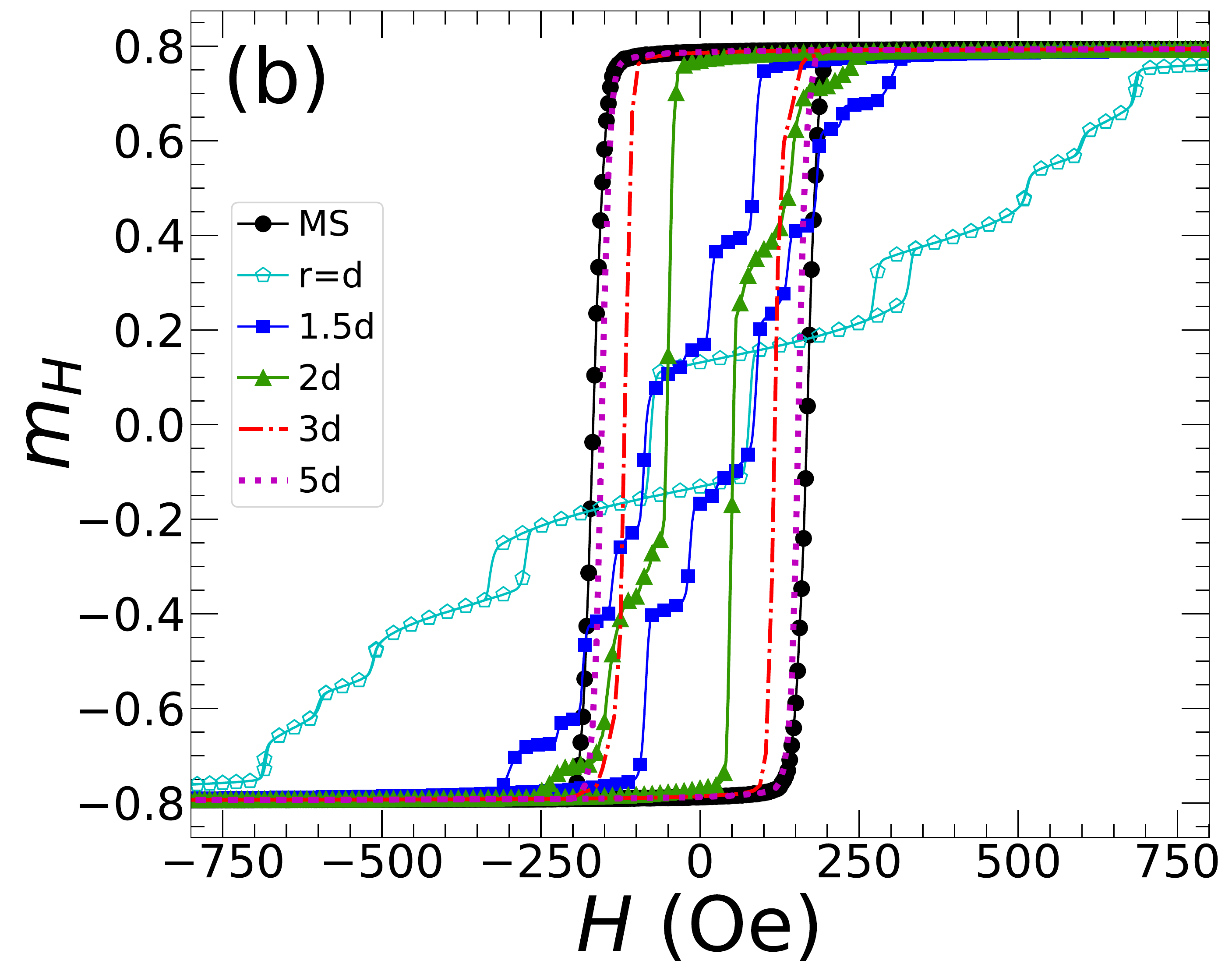}
    \caption[Hysteresis loops of NPs and MSs in an fcc structure as a function of interparticle distance]{Hysteresis loops of particles on an fcc structure (as shown in the inset) containing 13 a) NPs, b) MSs, at different center-center distances. The effective anisotropy of particles is aligned to the applied field. White arrows show the effective or explicit anisotropy axes in magnetic particles.}
    \label{fig:fcc_omfvmx}
\end{figure*}

Figs.~\ref{fig:fcc_omfvmx}a and b show the hysteresis loops of thirteen NPs and thirteen MSs in the fcc structure shown in the inset of panel (a), for which morphology anisotropy might be reasonably assumed to be small. Similar to particles arranged in a triangle, the internal structure of the NP plays a role in determining the loop shape that is not accounted for by the equivalent MSs.
Comparing the hysteresis loops in Fig.~\ref{fig:fcc_omfvmx} shows that the biggest loop differences occur at smaller $r$. For example, for $r=d$ and $1.5d$, the steeper slope of the loop in a cluster of MSs originates from bigger jumps in the total magnetization than in a cluster of NPs.
This can be attributed to the fact that the smallest contribution due to each MS flip is more pronounced in the total magnetization of thirteen MSs compared to the smallest contributions from each simulation cell among 21840 cells in a cluster of NPs. 

\begin{figure}
    \centering
    \includegraphics[width = 0.495 \textwidth]{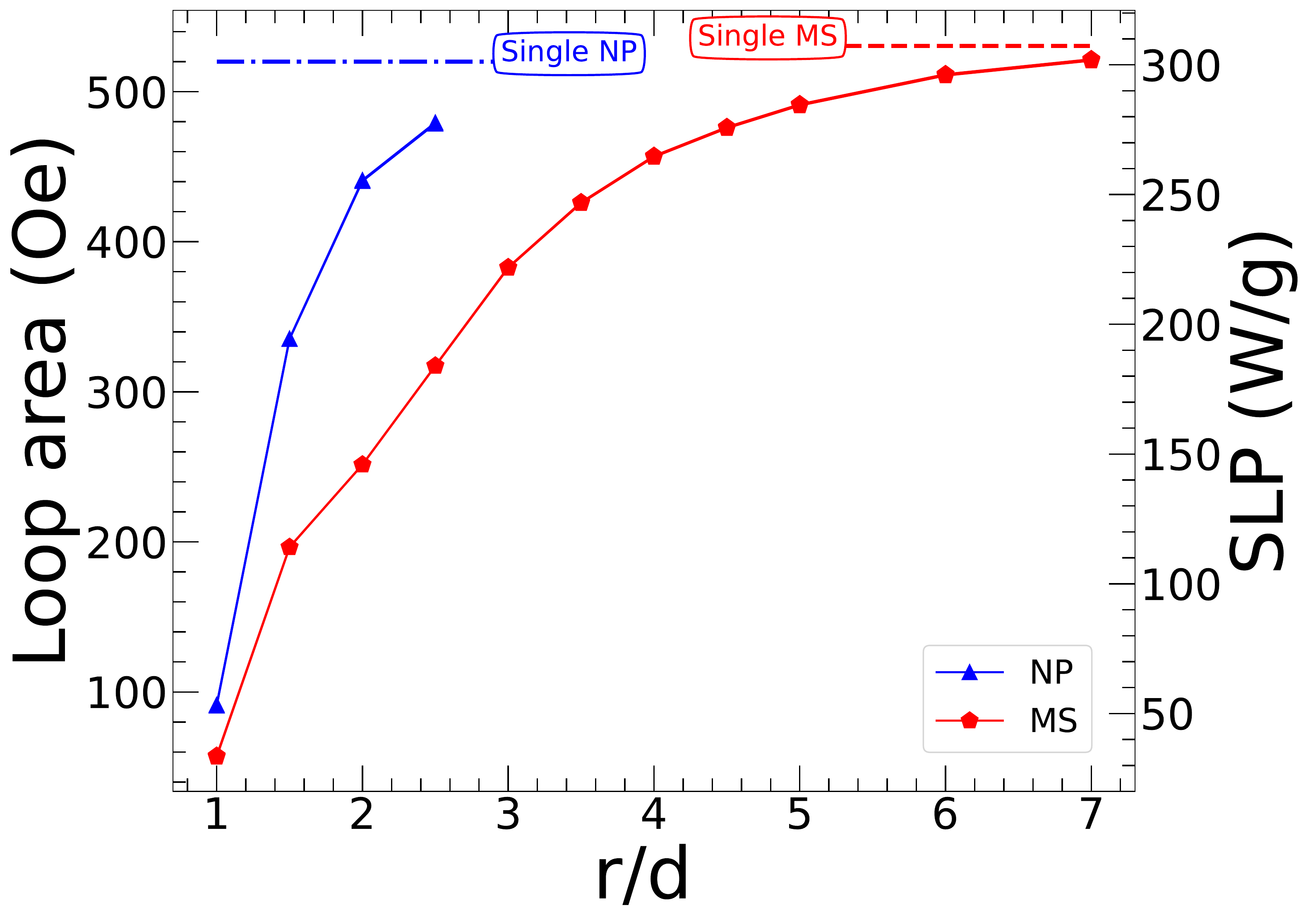}
    \caption{Loop area as a function of particle distance for clusters of complex NPs and MSs in an fcc structure. {\blue The right axis gives the associated SLP for clinical conditions with $f=62.5$~kHz and $H_{\rm max}=1.37$~kOe.}}
    \label{fig:fcc_area}
\end{figure}

As shown in Fig.~\ref{fig:fcc_area}, the $r$ dependence of the hysteresis loop area  implies that the intraparticle exchange and self-demag counteract some effects of the interparticle interactions in the case of NPs, so that NPs further apart than $2.5d$ have a loop area close to that of a single NP, whereas the role of dipole interaction between the MSs can not be ignored for $r<5d$. 
With regard to morphology anisotropy, despite the roughly spherical shape of the fcc cluster, the effects of inter-particle interactions are significant at smaller $r$.

\section{Local vs global loops}\label{sec:local_loops}

Recent studies emphasize the importance of local heating of NPs in clusters rather than their collective (global) heating, as the temperature of the surrounding tissue was found to be different in the vicinity of individual NPs~\cite{creixell2011egfr, villanueva2010hyperthermia}. 
In this section, we explore the local hysteresis loops, i.e., loops of individual NPs and MSs, in collections of particles, when they are closest to each other ($r=d$).
\subsection{Chains and triangles}

\begin{figure*}
    \centering
    \includegraphics[width= 0.45\textwidth]{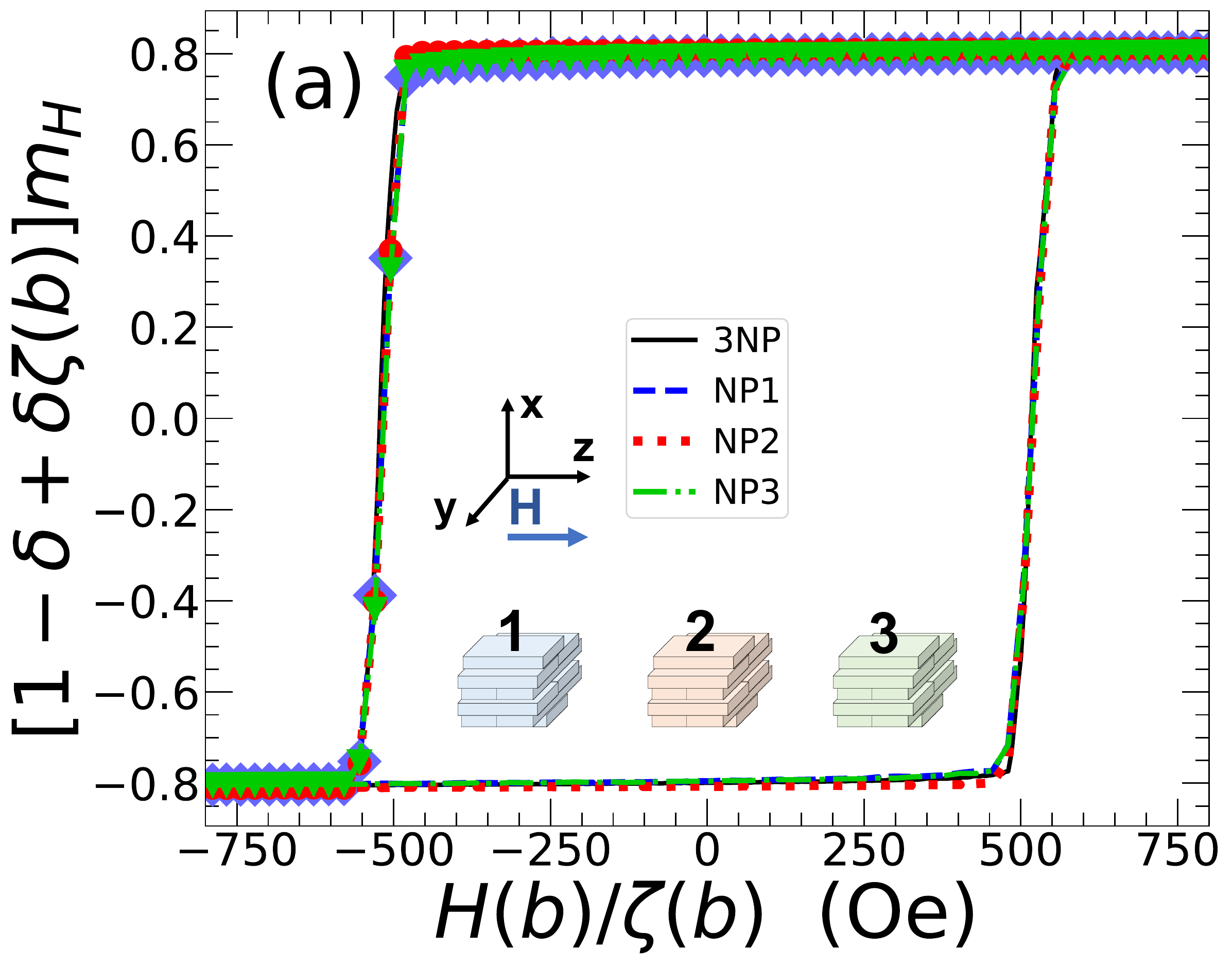}
    \includegraphics[width= 0.45\textwidth]{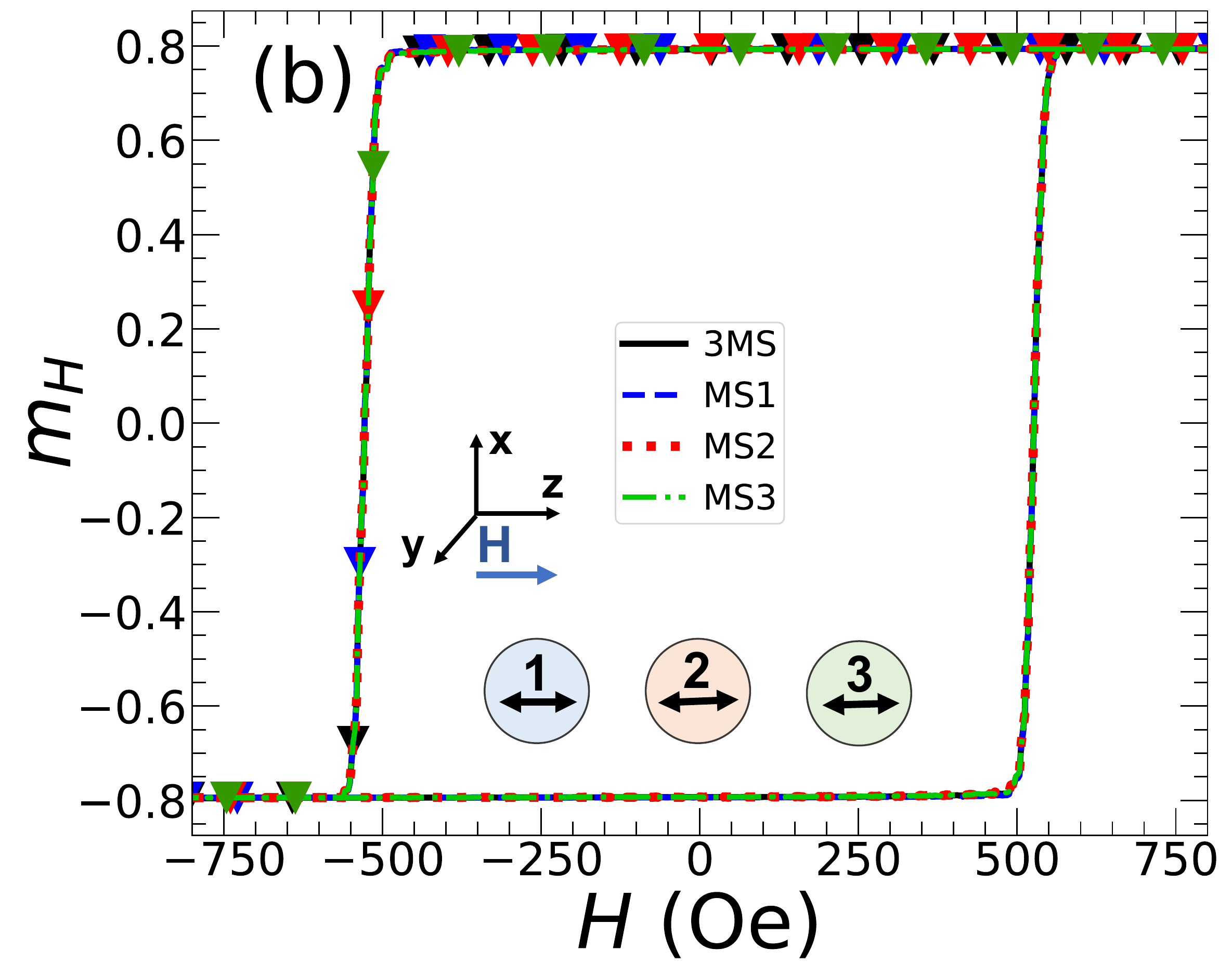}
    \includegraphics[width= 0.45\textwidth]{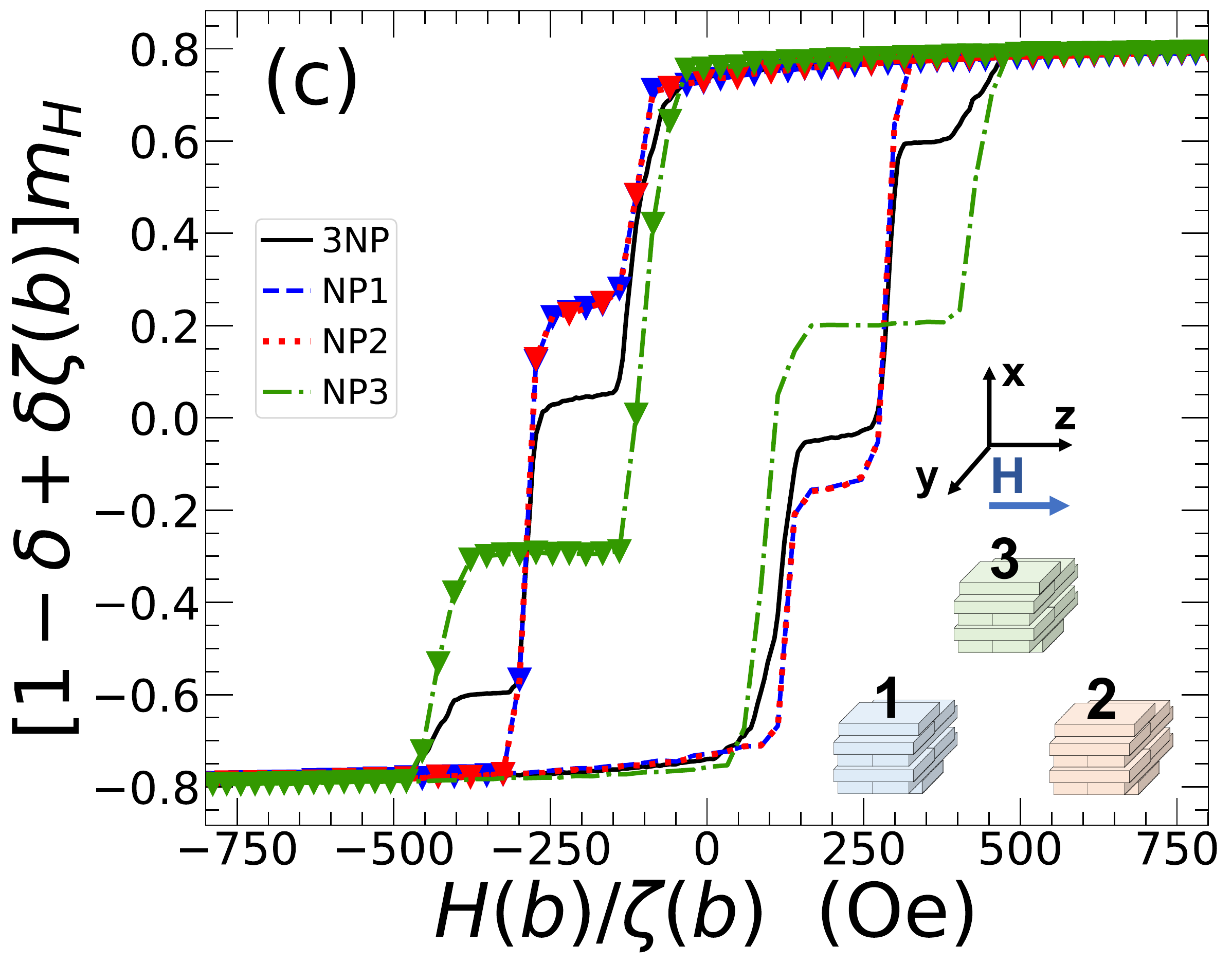}
    \includegraphics[width= 0.45\textwidth]{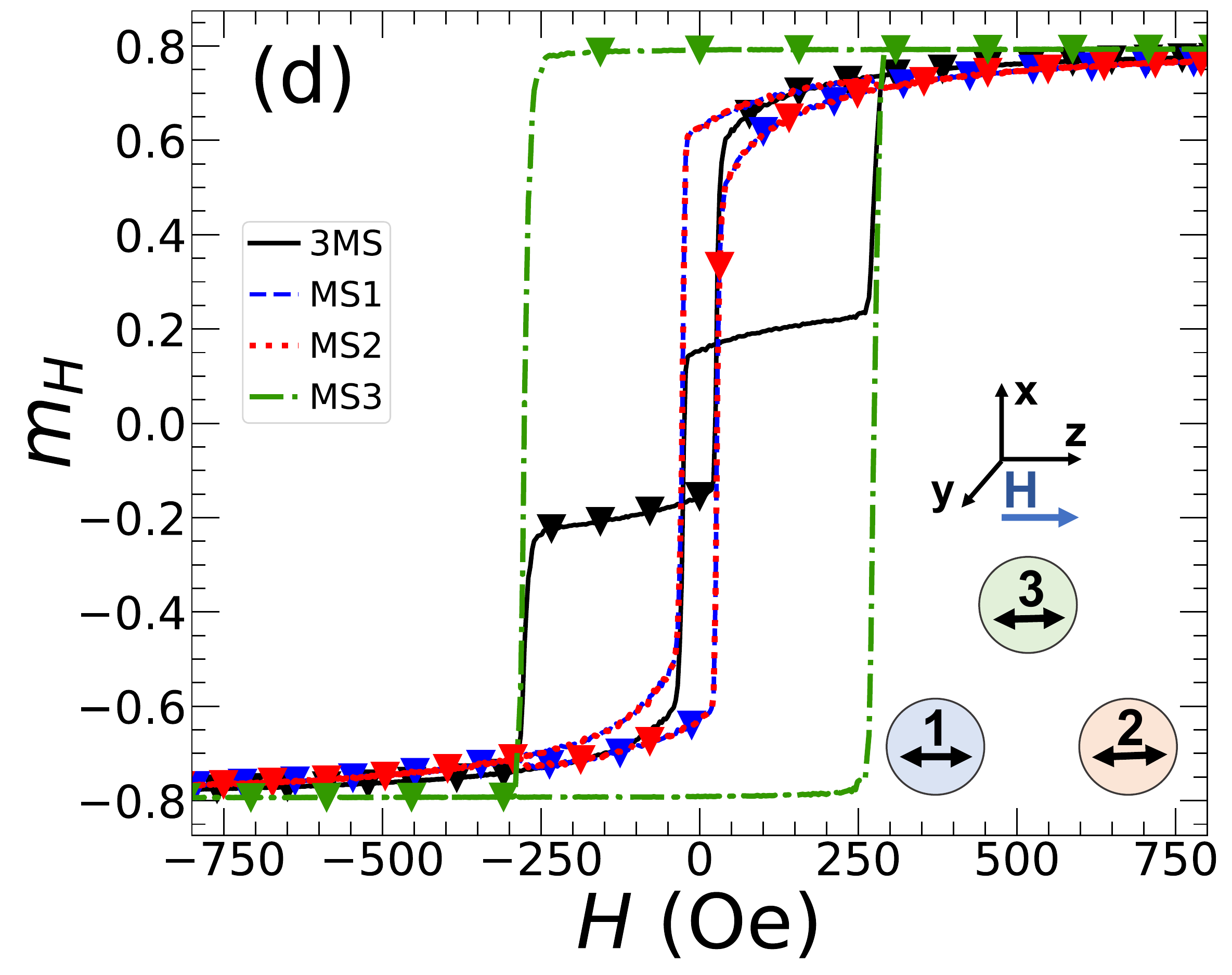}
    \includegraphics[width= 0.45\textwidth]{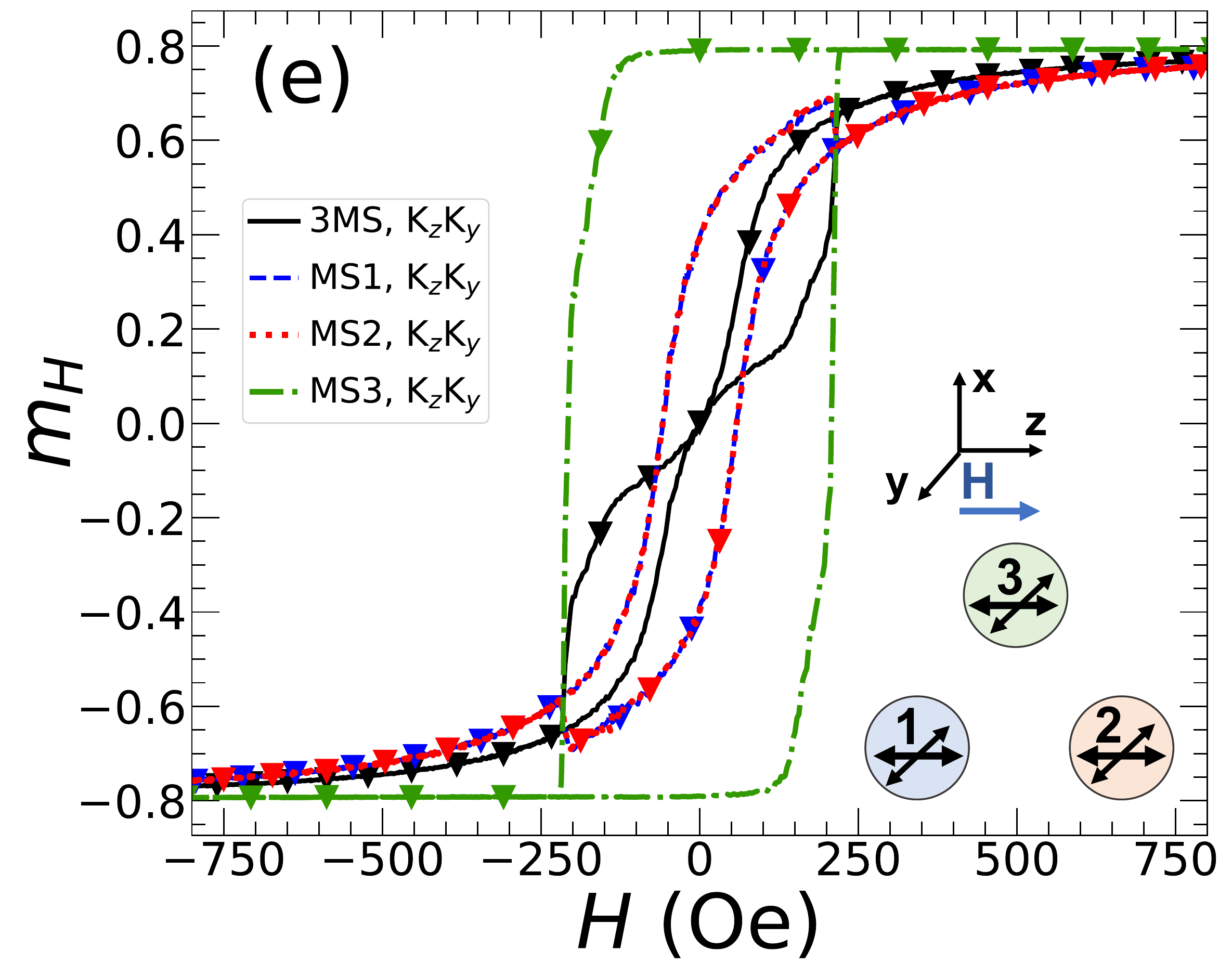}
    \includegraphics[width= 0.45\textwidth]{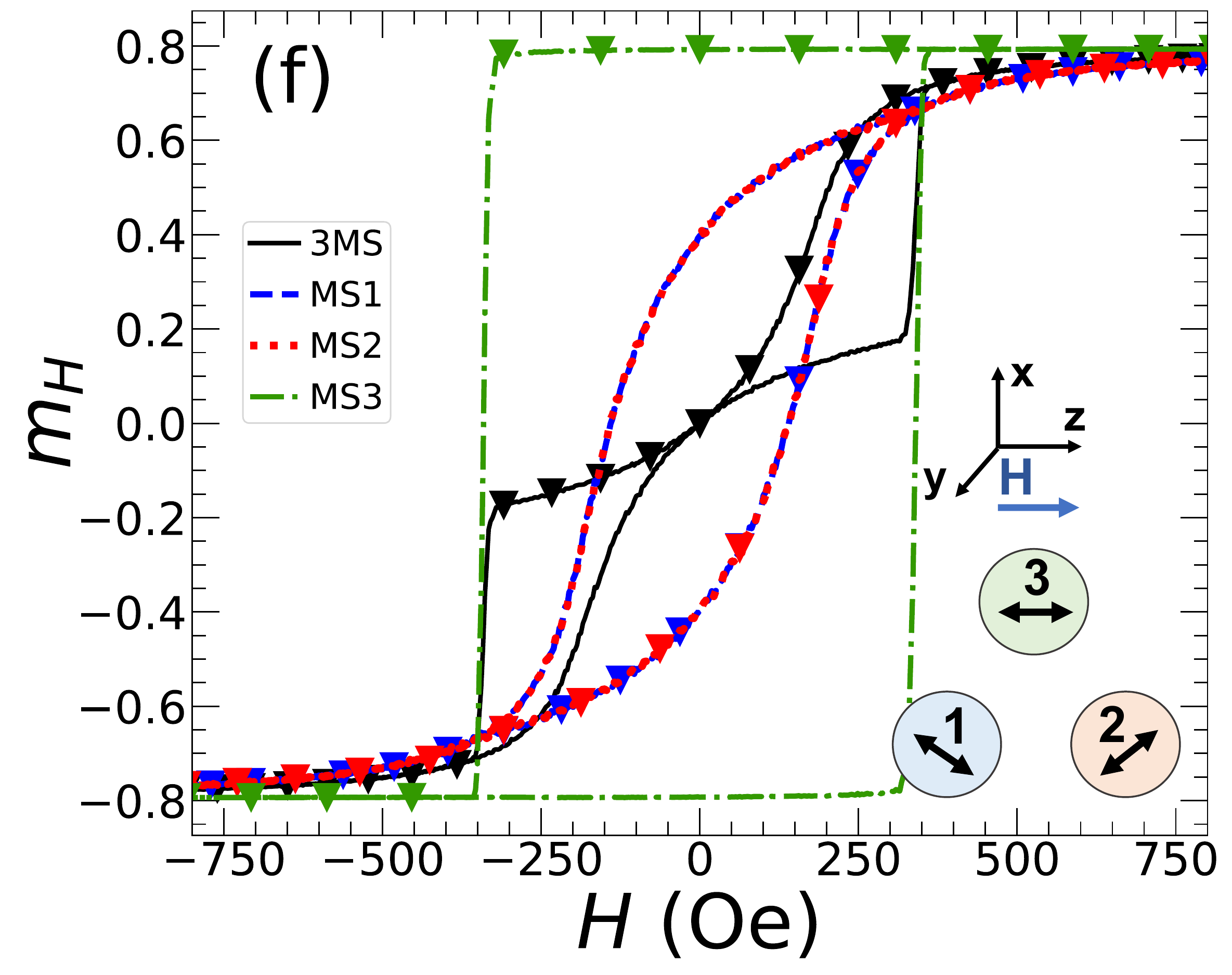}
    \caption{Individual hysteresis loops of particles within three-particle chains of (a) NPs and (b) MSs, and triangles of (c) NPs and (d, e, f) MSs. Chains lie along the field direction ($z$), and triangles lie in the $x$-$z$ plane. For NPs, the majority of nanorod long axes lie along $z$, with the minority along $y$.  For (b) and (d) MS anisotropy is along $z$; in panel (e) MSs include two anisotropy axes with associated strengths {\blue $K_z=12.15$~kJ/m$^3$} and $K_y=2K_z/3$; in panel (f) (single) anisotropies are aligned 120$^\circ$ with respect to each other, reflecting the dipolar ground state.  
    Filled triangles mark the first half-cycle when the field decreases from $H_{\rm max}$ to $-H_{\rm max}$.}
    \label{fig:localLoop_3MS}
\end{figure*}

We first examine the chain and triangular clusters considered in section~\ref{sec:multipleNP} and the local loops for NPs and MSs are shown in Fig.~\ref{fig:localLoop_3MS}a, b, respectively. 
While we use different line styles and colors to differentiate the local loops of different particles,
to make it clear when we have inverted loops, the portion of the magnetization curve corresponding to the first half of an AC field cycle, when the field decreases from $H_{\rm {max}}$ to $-H_{\rm {max}}$, is plotted with symbols; the magnetization curve for the second half of an AC field cycle, when the field increases from $-H_{\rm {max}}$ to $H_{\rm {max}}$, appears with no symbols.


It is well known that, as long as dipoles are parallel to the joining line (chain axis), the dipole energy favors their head-to-tail alignment. Similarly for NPs and MSs in Fig.~\ref{fig:localLoop_3MS}a and b, magnetic particles at the two ends of the chain, i.e., NP1, NP3 and MS1, MS3, feel the same magnetostatic or dipole interaction, which is different from the net dipole field from other particles acting on the particle in the middle of the chain (NP2 and MS2). 
Despite this difference, all three of the particles flip in unison for both NPs and MSs, although for smaller particles, this need not be the case~\cite{torche2020thermodynamics}.

In contrast, if the magnetic particles are in a triangular arrangement, as in Figs.~\ref{fig:localLoop_3MS}c and \ref{fig:localLoop_3MS}d, the loops for the particles at the base (particles 1 and 2) are the same, but very different from that of the third particle. 
Moreover, the loops for individual NPs are markedly different from those of the corresponding MSs.  {\blue In Appendix~\ref{sec:additional} (Fig.~\ref{fig:Triangle_10z}), we show that when using $10z$ NPs, individual loops are more similar to those of MSs, but no inversion occurs.} The loops for particles 1 and 2 in the NP case are large and show magnetization reversal occurring in two steps, while in the MS case, the loops are narrow and inverted.  Another difference is that for NPs, loop areas of all three particles are comparable, while for MSs, loops for different particles can be very different. Clearly, the internal structure of the NP plays a significant role in the magnetization dynamics of individual NPs, resulting in features that are not captured within the MS approximation.

To understand the inversion of the MS loops for particles 1 and 2 in Fig.~\ref{fig:localLoop_3MS}d, it is useful to recall that the lowest energy arrangement at $H=0$ and $K=0$ for three dipoles in a triangle is the 120$^\circ$ structure, as noted above.
Starting from the highest spin alignment to the high external field in the $z$ direction, as the external field decreases,
the dipole fields tend to tilt MS1 and MS2 towards the 120$^\circ$ state and away from the $z$ axis. 
When the field is small, but still positive, the net effect of the Zeeman, anisotropy and, most importantly, dipole interactions results in a magnetization flip of MS1 and MS2.
The hysteresis loops of MS1 and MS2 in Fig.~\ref{fig:localLoop_3MS}d  are ``inverted'' since the (positive) loop area naively implies that the MS is doing work on its surroundings and thus absorbing heat.
Considering that a normal hysteresis loop is due to the delayed alignment of the magnetization with the applied field, 
an inverted loop occurs when a spin flips in advance of the external field changing direction.
In the inverted case, a spin is ``helped along'' by the dipolar interactions of neighboring particles. 
MS3 exhibits a loop similar to that of an independent particle, but with a larger $H_c$ arising from the dipolar interactions with MS1 and MS2.
The average hysteresis loop of these MSs represents the global heating performance and is shown with a black curve in Fig.~\ref{fig:localLoop_3MS}d, with an inverted middle portion.

The step-like features in the NP loops shown in Fig.~\ref{fig:localLoop_3MS}c reveal that not all nanorods (or their constituent cells) within an NP switch simultaneously, and suggest the importance of modelling the two directions ($z$ and $y$) along which the nanorods lie.
To this end, we calculate loops for MSs with two uniaxial anisotropies, one along $z$ and one along $y$, with respective anisotropy strengths {\blue $K_z=12.15$~kJ/m$^3$} and $K_y=2K_z/3$. Details about modelling NPs with MSs that have two anisotropies are found in Appendix~\ref{sec:Eff_K}. As shown in Fig.~\ref{fig:localLoop_3MS}e, the local loops for MS1 and MS2 are still  
inverted but are less square compared to those in Fig.~\ref{fig:localLoop_3MS}d.  The loops in Fig.~\ref{fig:localLoop_3MS}e are also still significantly different from those for NPs in Fig.~\ref{fig:localLoop_3MS}c. Our inclusion of two anisotropy directions does not  improve the MS model {\blue for the triangular arrangement}.
The global loop for these two-axes MSs is 50\% smaller than for MSs with single anisotropy axes oriented along the external field.

Another triangular arrangement to study is one in which the MSs' (single) anisotropy axes follow the alignment favored by dipole interactions, as shown in the inset of Fig.~\ref{fig:localLoop_3MS}f. In this case, similar to that shown in Fig.~\ref{fig:localLoop_3MS}e, there are no sudden magnetization flips for MS1 and MS2,  implying the absence of an energetic barrier for magnetization reversal, and
the loops for MS1 and MS2 are inverted from start to finish.
Similar to the two previous cases, MS3 experiences a sudden magnetization flip. The loop for MS3, however, is much wider than expected from anisotropy alone. This can be understood from the $H=0$ and $K=0$ ground state, where there is a large (positive) $z$ component in the dipolar fields of MS1 and MS2 at the position of MS3, which counteracts the (negative) external field.
Despite the larger loop area for MS3, the average of the local loops results in a global loop area 21\% smaller compared to the case of Fig.~\ref{fig:localLoop_3MS}d, where anisotropies along the $z$-axis.

The marked differences in (global) loops for NPs and MSs in a triangular arrangement at close distances can be attributed to very different dynamics of individual particles, as quantified by their local loops.  The inverted loops of some particles, and wider loops of others, in the MS case, arise from dipolar interactions that can either enhance or counteract the external field, depending on the configuration.  By contrast, the loops for different NPs in the triangle are more similar to each other than in the MS case, exhibit no inversions, and thus show weaker effects of interparticle magnetostatic interactions. 
Ultimately, for the triangular arrangement at small separation, the simpler MS model exhibits strong dipolar effects that lower heating efficiency, while for the more complex NPs, magnetostatic interactions slightly enhance heating.  Evidently, the MS is too simple a model for this scenario.

Munoz et al.~\cite{munoz2020disentangling} discuss in detail how local hysteresis loops of interacting MSs do not represent the heating of individual particles because the particles exchange energy with their neighbours through dipole interactions.  They calculated local heating by considering the dynamical process of energy dissipation through the damping torque. 
We note that the loops we present are calculated with respect to the external field alone, and not with respect to the local field acting on an individual NP.  Calculations based on the local field may help us bridge the insights from Munoz et al.'s dynamical theory and local heating based on a thermodynamic description.
\subsection{fcc cluster}
\begin{figure*}
    \centering
    \includegraphics[width= 0.495\textwidth]{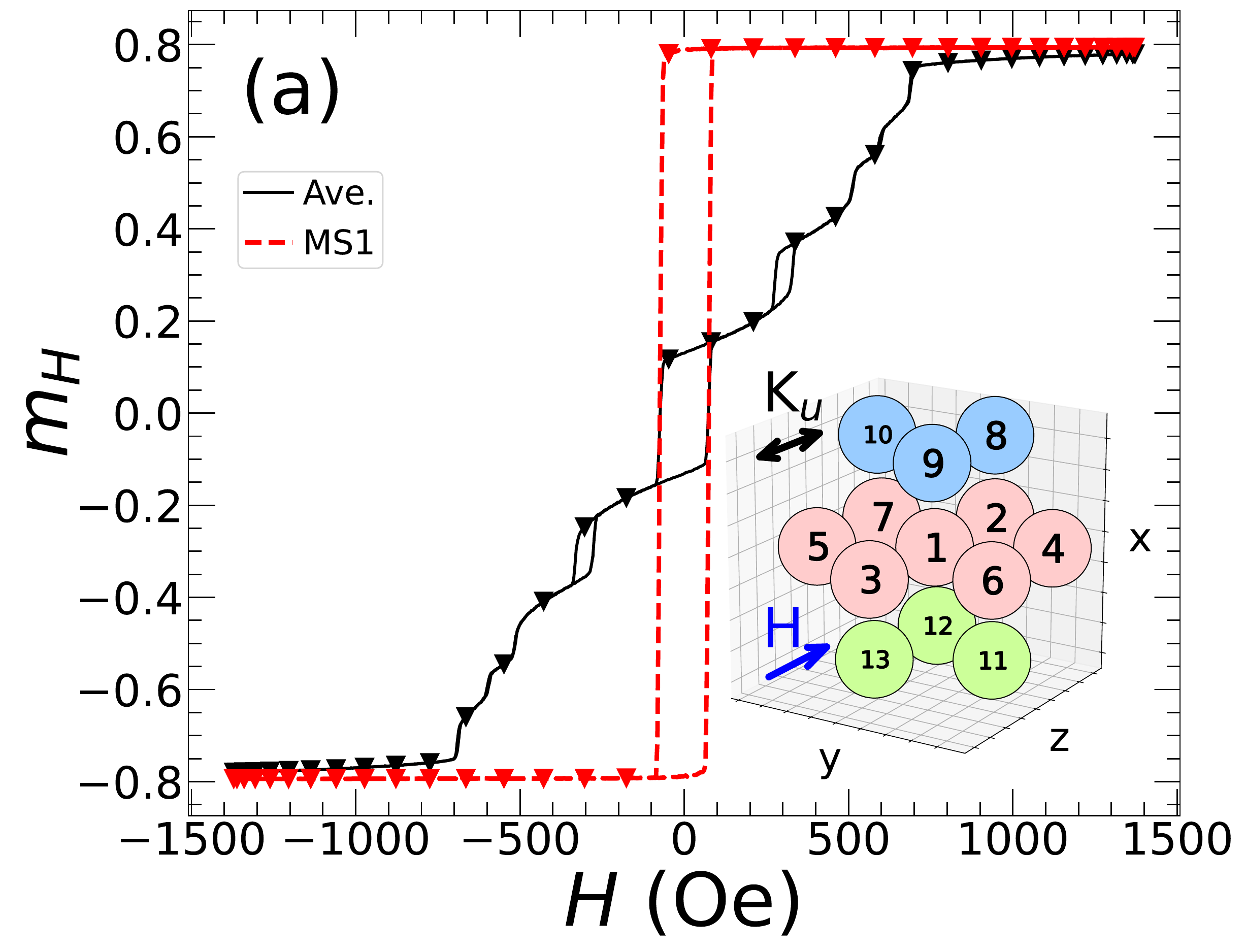}
    \includegraphics[width= 0.495\textwidth]{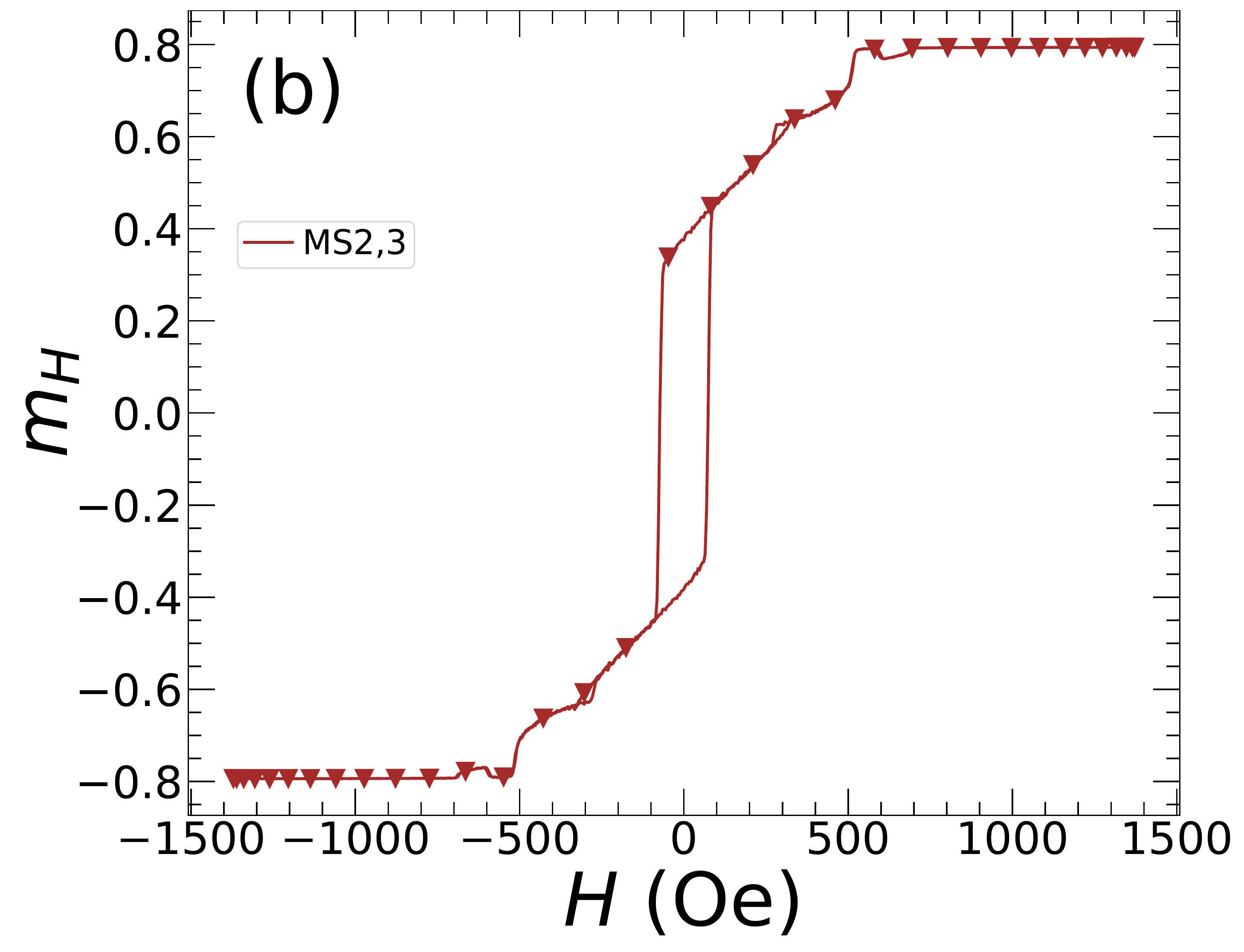}
    \includegraphics[width= 0.495\textwidth]{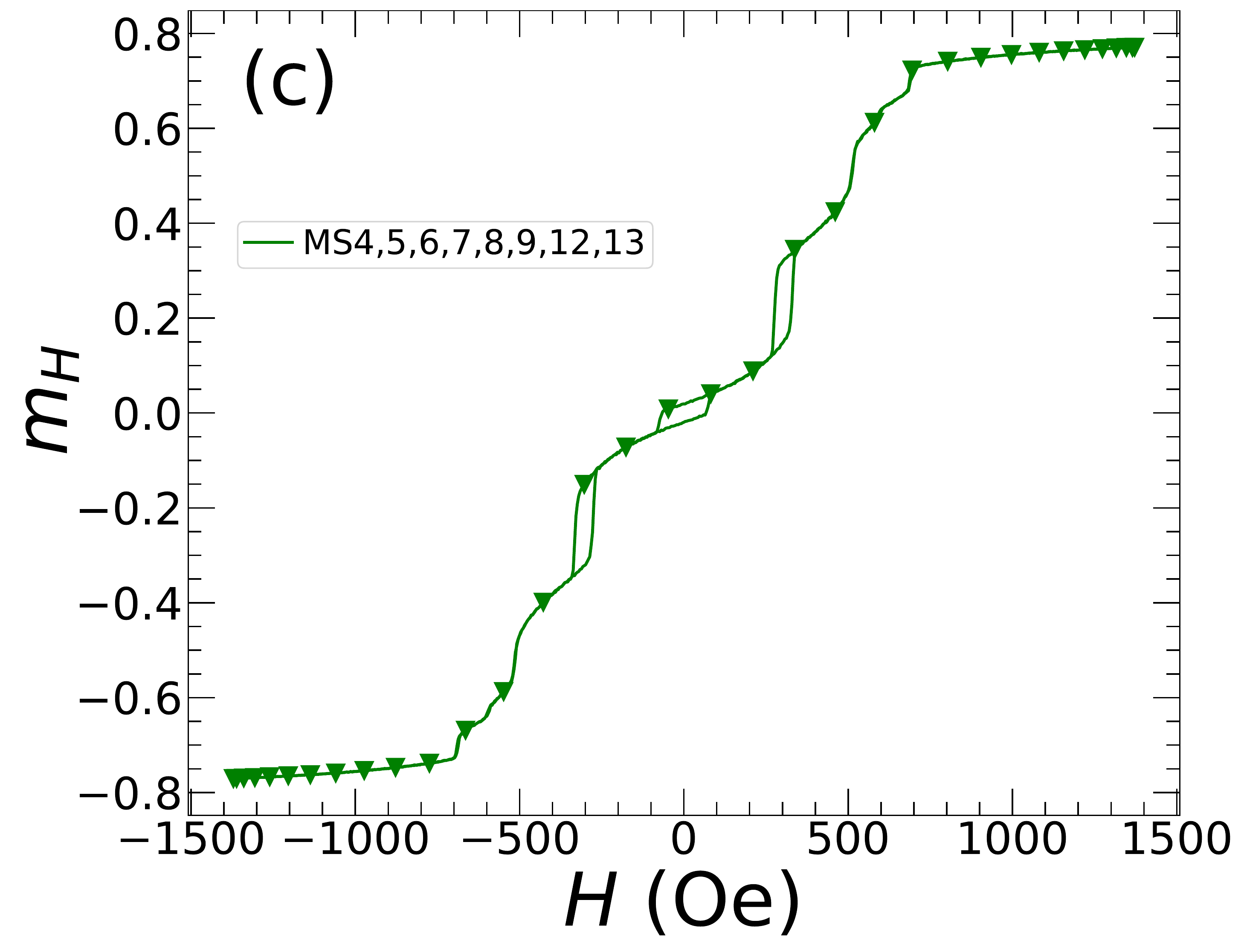}
    \includegraphics[width= 0.495\textwidth]{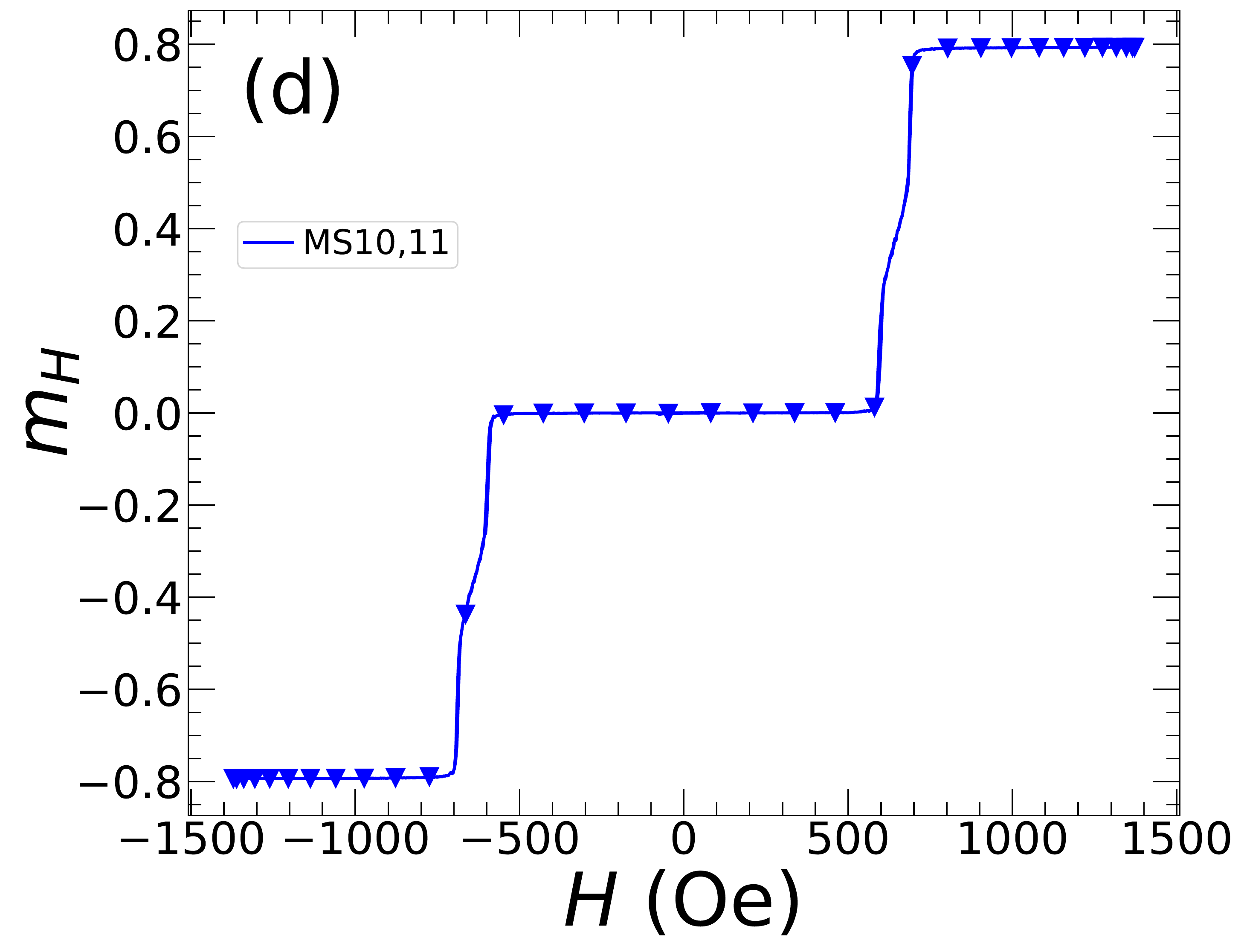}
    \caption{fcc structure made of 13 MSs, and their global and local hysteresis loops when particles are at the closest distance, $r=d$. Closed markers distinguish the first ($H_{\rm max}\rightarrow -H_{\rm max}$) half of a cycle. Inset of panel (a) shows the labeled MSs on the particle arrangement having uniaxial anisotropy along the applied field. a) Global hysteresis loop for 13 MSs shown in black and the local loop for the central particle MS1 in red. Local hysteresis loops for b) MS2 and MS3, c) MS4, MS5, MS6, MS7, MS8, MS9, MS12, MS13, and d) MS10 and MS11. Each loop is calculated via averaging over 1500 independent field cycles}
    \label{fig:fcc_Global_Local_r1d}
\end{figure*}

\begin{figure*}
    \centering
    \includegraphics[width= 0.495\textwidth]{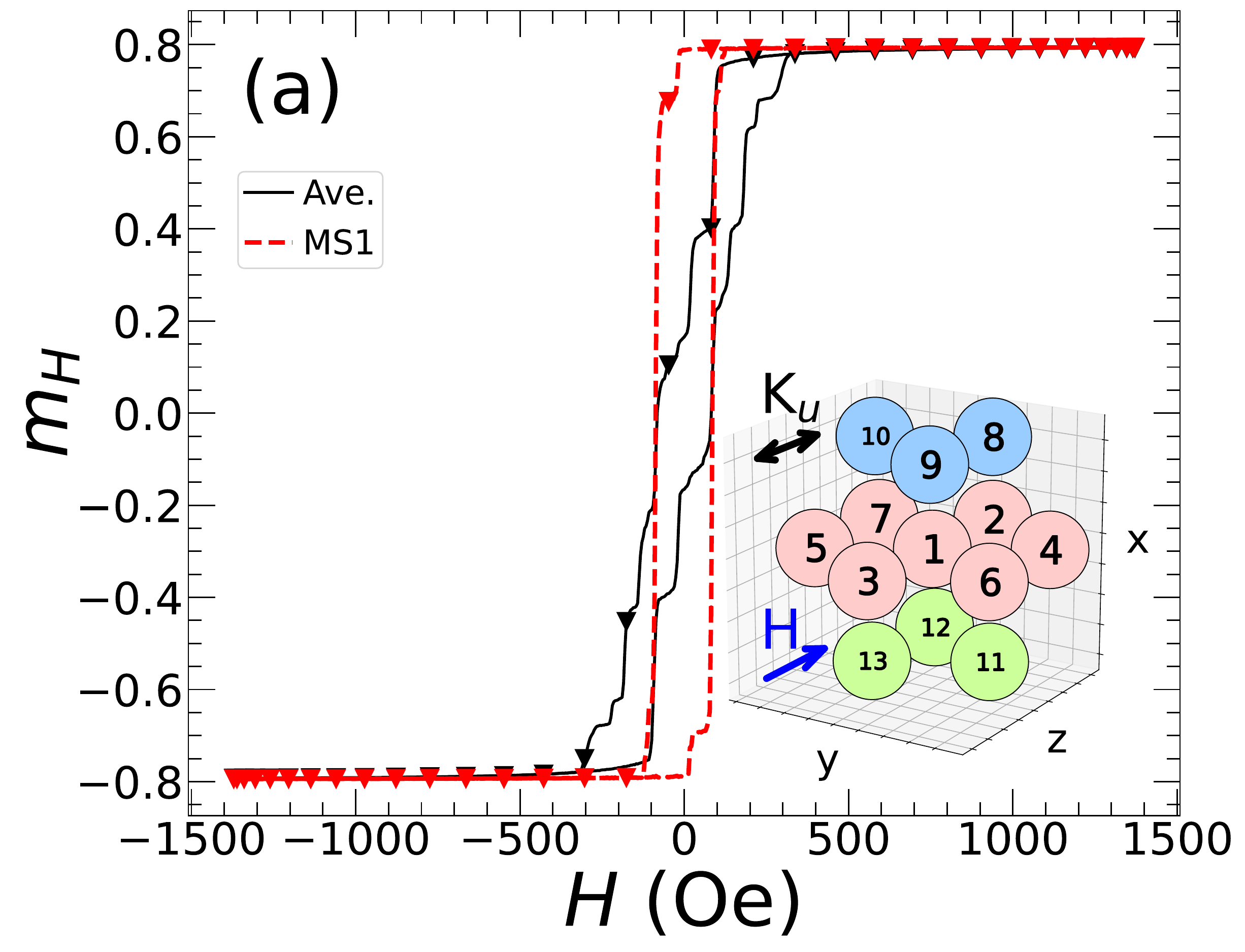}
    \includegraphics[width= 0.495\textwidth]{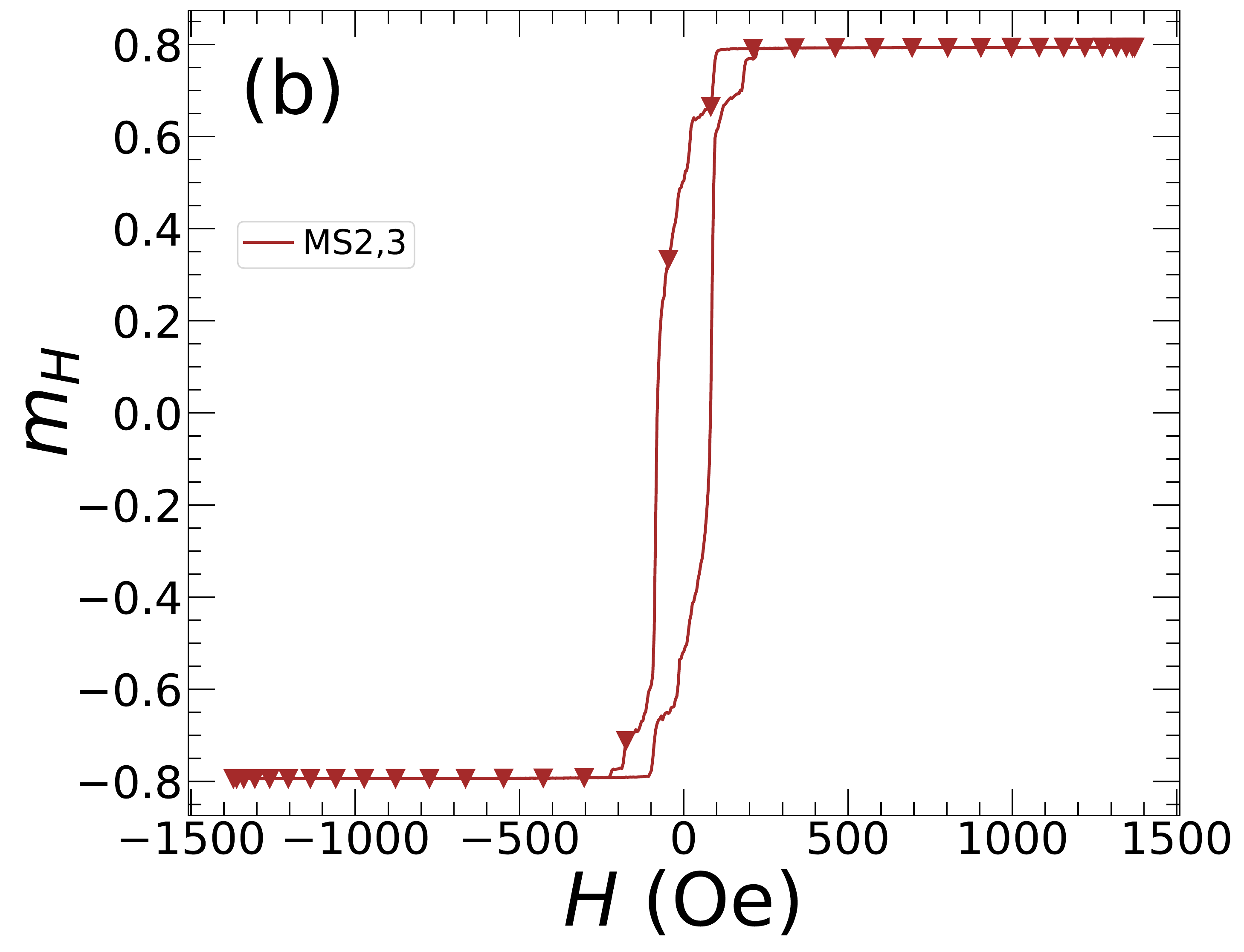}
    \includegraphics[width= 0.495\textwidth]{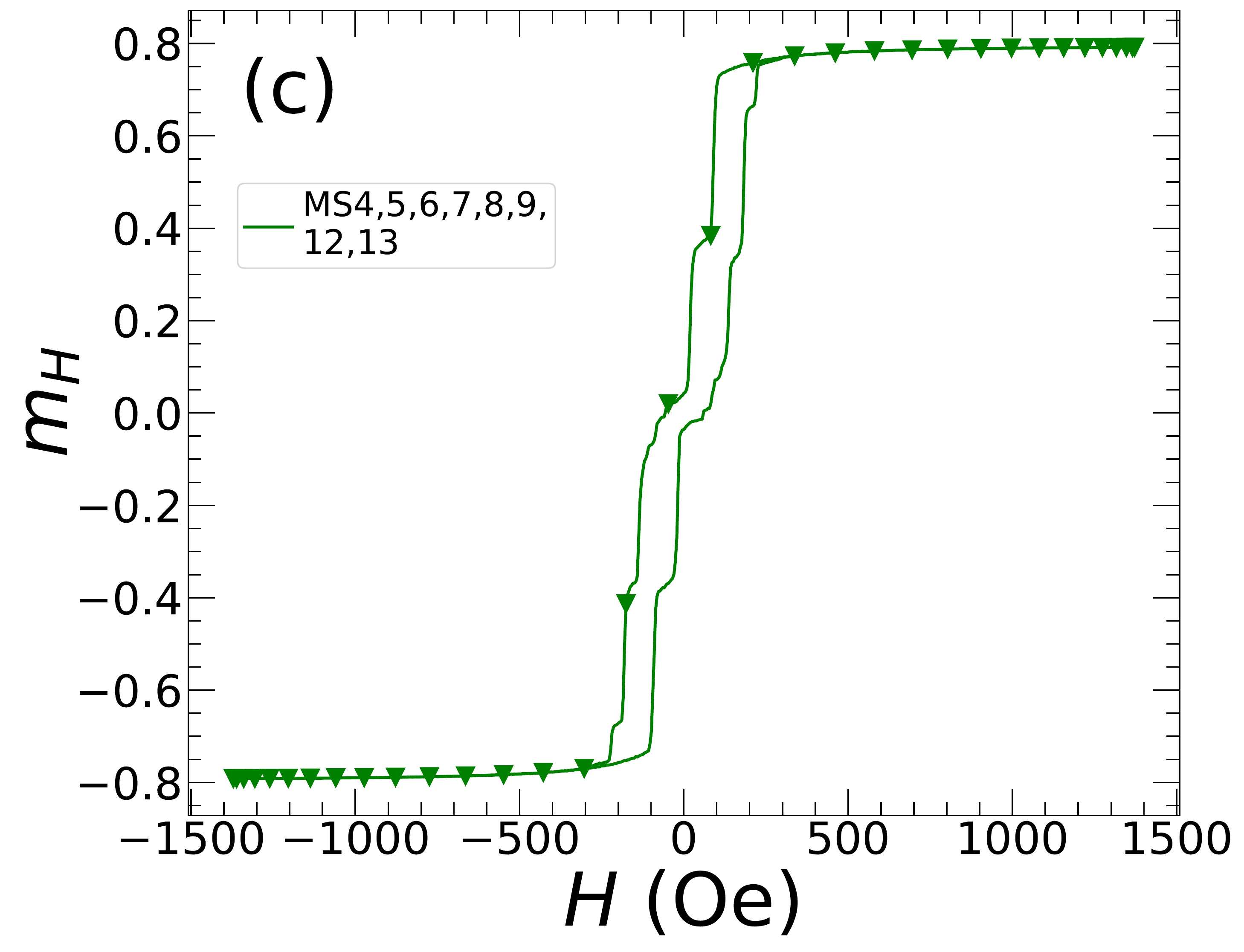}
    \includegraphics[width= 0.495\textwidth]{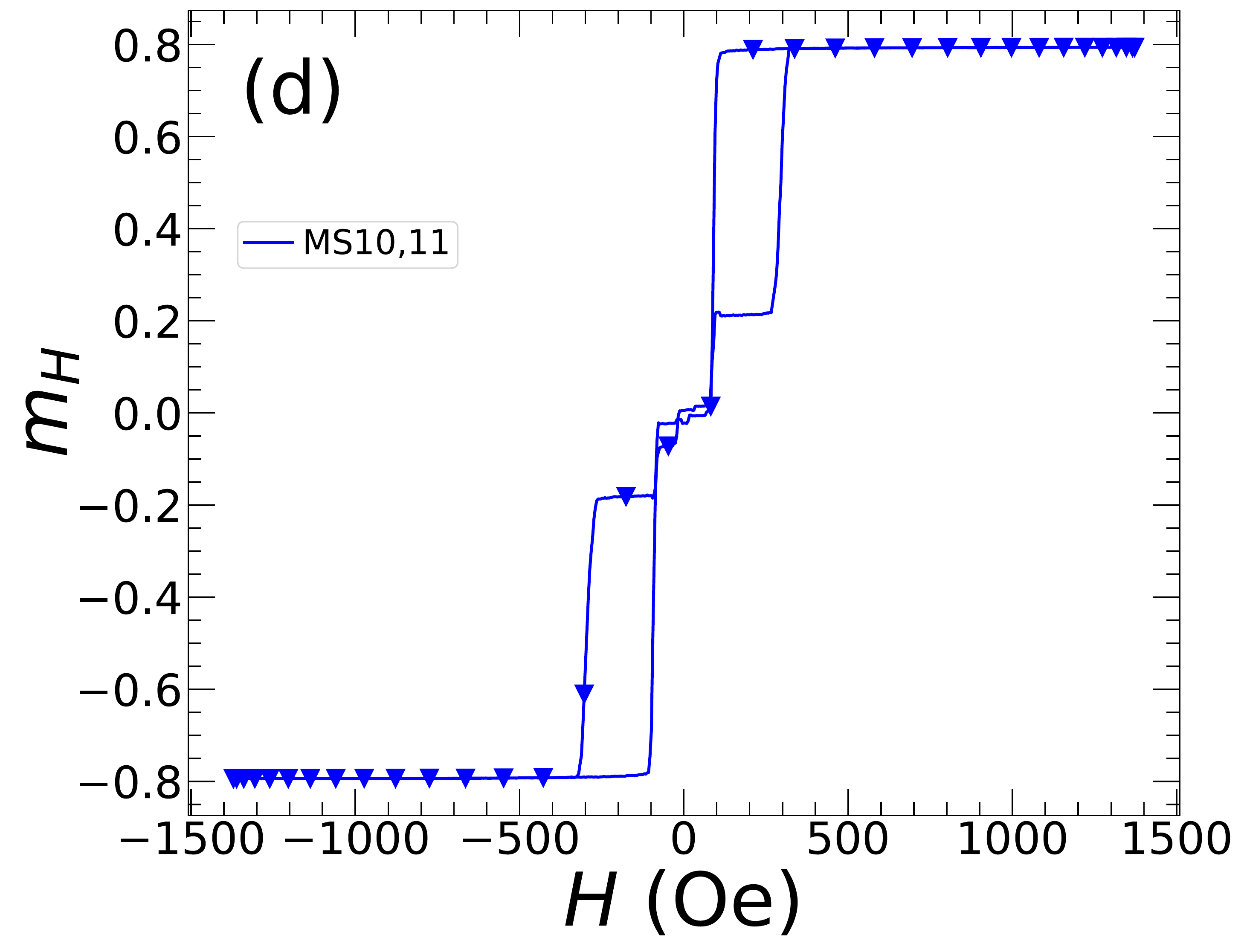}
    \caption{fcc structure made of 13 MSs, and their global and local hysteresis loops when neighboring particles are separated by distance $r=1.5d$. Closed markers distinguish the first ($H_{\rm max}\rightarrow -H_{\rm max}$) half of a cycle. Inset of panel (a) shows the labels on MSs in the fcc arrangement having uniaxial anisotropy along the applied field. a) Global hysteresis loop for 13 MSs shown in black and the local loop for the central MS (labeled 1 in the inset) in {\blue red}. Local hysteresis loops for b) MS2 and MS3, c) MS4, MS5, MS6, MS7, MS8, MS9, MS12, MS13 d) MS10 and MS11. Each loop is calculated via averaging over 1500 independent field cycles}
    \label{fig:fcc_Global_Localr1_5d}
\end{figure*}

The different dynamics we see for individual MSs in a triangle encourage us to consider the local loops for MSs in the fcc cluster. A labeled diagram of MSs and their local loops for nearest neighbour distance $r=d$ and $1.5d$ are shown in Figs.~\ref{fig:fcc_Global_Local_r1d} and \ref{fig:fcc_Global_Localr1_5d}, respectively, with the MSs' uniaxial anisotropy and applied field both along the $z$ axis. 


Owing to the symmetries of the structure, some sites should have identical loops, which are plotted in the same panel in Figs.~\ref{fig:fcc_Global_Local_r1d} and \ref{fig:fcc_Global_Localr1_5d}. As in Fig.~\ref{fig:localLoop_3MS}, the half cycle of the hysteresis curve where the external field is decreasing is indicated with symbols.  There are no inverted loops, but many portions of the curve are closed ($m_H$ is the same for increasing and decreasing field).
Each loop is calculated via averaging over 1500 independent field cycles and averaging over equivalent sites.

As shown in Fig.~\ref{fig:fcc_Global_Local_r1d}a, the central particle MS1 is the symmetry center of the structure with the same distance from 12 neighbouring MSs. When $r=d$, the effect of neighbouring dipoles on MS1 results in a sudden magnetization flip at a smaller $H_c$ than for an independent particle (red curve). The average of all individual loops  (black curve) results in a global hysteresis {\blue loop with small area, small $M_r$, and the same small $H_c$ as MS1}. As shown in the labeled model, 
MS2 and MS3 have equivalent positions and exhibit similar dynamics. Their hysteresis loop has the same $H_c$ as MS1 but smaller $M_r$, as shown in Fig.~\ref{fig:fcc_Global_Local_r1d}b. 
Given the symmetry of the cluster, there are four pairs of MSs that are all equivalent: MS4--MS6, MS5--MS7, MS8--MS9 and MS12--MS13.
In a given field cycle, each MS in a pair exhibits the same magnetization dynamics, similar to the two MSs forming the base of the triangle discussed above.  However, the neighbouring pair has opposite dynamics in a given loop.  For example, if MS4--MS6 shows a {\blue mostly} normal loop, then MS8--MS9 has a {\blue mostly} inverted loop and vice versa. The same applies for MS5--MS7 with respect to MS12--MS13.  Averaging over these eight loops results in a mostly closed loop with three tiny open areas as shown in Fig.~\ref{fig:fcc_Global_Local_r1d}c.  An individual pair will exhibit a normal or  inverted loop with equal probability.  
However, small displacements of MSs from their ideal fcc sites can result in the deviation in the inverted-to-normal loop ratio of 1 to 1 for a given pair, resulting in some pairs exhibiting averaged loops that are normal and some pairs exhibiting inverted loops.
{\blue MS10 and MS11 are the last equivalent sites, and these two MSs have opposite magnetization dynamics and loop areas during a given loop cycle, resulting in a closed loop when the two loops are averaged; see Fig.~\ref{fig:fcc_Global_Local_r1d}d. The peculiarity of MS10 and MS11 is that, in contrast to the case of oppositely magnetizing pairs shown in Fig.~\ref{fig:fcc_Global_Local_r1d}, their magnetizations appear to cancel completely, despite the influence of the external field.  Perhaps the field produced by the alignment of MSs 2, 1 and 3, is strong enough to counteract the symmetry-breaking effect of the external field.  
Furthermore, their individual loops are quite square (not shown).
It may also be noted that MS10 and MS11 are special in that they each have unique $x$ and $y$ coordinates, unlike all the other positions within the cluster.
Introducing disorder in the orientation of MSs (Fig.~\ref{fig:fccMS_randAnis}, Appendix~\ref{sec:additional}), decreases both the global loop and local loops for all MSs except MS10 and MS11, the loop areas of which increase.
} 
Fig.~\ref{fig:fcc_Global_Localr1_5d} shows that at $r=1.5d$, local and global loops have bigger areas, 
{\blue behavior consistent with weaker dipole interactions, particularly for MS10 and MS11.}
{\blue The behavior of individual NPs in an fcc cluster, shown in Fig.~\ref{fig:fccNP} for $r=d$ in Appendix~\ref{sec:additional}, is less distinct than in the MS case; only NP10 and NP11 (in the same positions as MS10 and MS11), and to a lesser extent the central NP, have loops with significant areas.  The loops of NP10 and NP11 do not cancel.}

\begin{figure*}
    \centering
    \includegraphics[width= 0.95\textwidth]{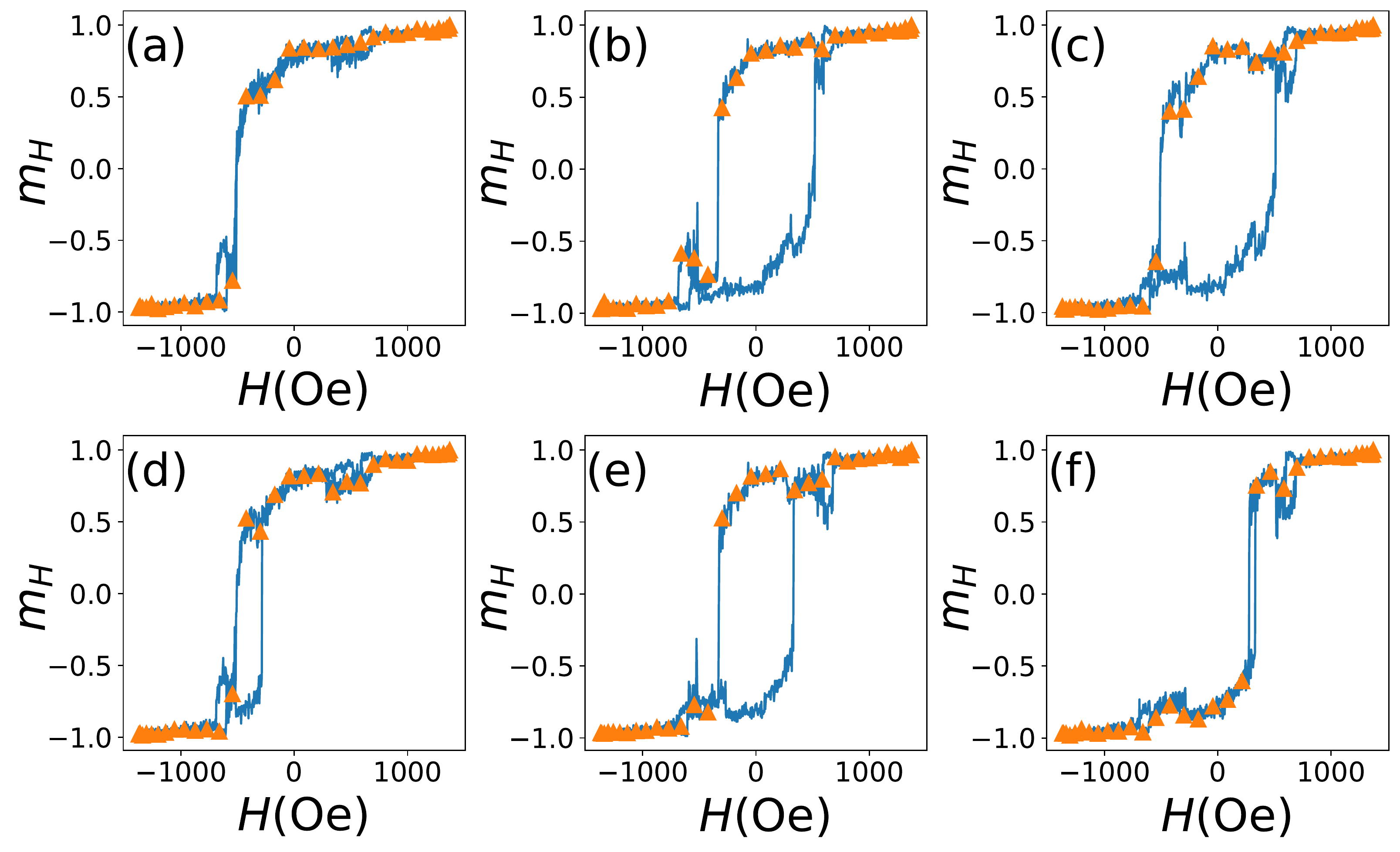}
    \caption{Six different types of loops for MS8 in the fcc structure. 
    {\blue While all loops shown are normal, inverted versions occur as well.   The asymmetry of loops in panels a, b, d, and f can occur at positive or negative values of the field.   Triangles label the part of the loop with decreasing field.}}\label{fig:MS8DiffLoops}
\end{figure*}

As expected, due to the thermal fluctuations, the local loops of a typical MS are not identical in different runs and the results shown in Fig~\ref{fig:fcc_Global_Local_r1d} are averaged over 1500 runs. To better understand the loop in Fig~\ref{fig:fcc_Global_Local_r1d}c, which is an average of eight equivalent sites (MS4--6, MS5--7, MS8--9, MS12--13), we show the different possible loops for MS8 in Fig~\ref{fig:MS8DiffLoops}. As before, the symbols on the loops identify the decreasing field half-cycle, and for all the loops shown the inverted  counterparts  occur.
We observe that the magnetization flips at certain values of the field, resulting in a discrete set of possible loops. 

With these results and the description of local loops from Munoz et al.~\cite{munoz2020disentangling} in mind, we conclude that although the global heating is less for $r=1d$ compared to $r=1.5d$, inverted local loops may mean high local heating near particular MSs.  It would be interesting to apply the dynamic framework of Munoz et al. for calculating local heating to this configuration of MSs,
given that for a system of interacting particles, the loops for individual particles calculated with respect  to the external field do not account for all transfers of energy. 

\section{Conclusions}\label{sec:conclusions}

The present study reports on an application of the RG-based coarse-graining method~\cite{grinstein2003coarse},  implemented, modified and extended in I and II, to simulate larger collections of magnetic nanoparticles that would otherwise not be computationally feasible, and examines the validity of the macrospin 
approximation, hitherto the exclusive approach to simulating clusters. We apply this method to simulate dynamic hysteresis of clusters of three and thirteen complex magnetite nanoparticles, made of nanorod building blocks, at $T=310$~K. 

Most simulation studies of magnetic nanoparticles are based on the MS model. We explore and compare the MS model, with both a single and two uniaxial anisotropies, to the complex NP model for three of many possible aggregations of magnetic particles.
For simulations of chains of three NPs and MSs, we find that the loop areas in both cases vary almost identically with interparticle distance, decaying as $1/r^3$ as one expects for dipoles. MSs and complex NPs further apart than three NP diameters act effectively as non-interacting particles.
In contrast, for triangular order, the nontrivial effect of the interactions between and within complex NPs, compared to the simpler dipolar interactions between equivalent MSs, results in dissimilar hysteresis loops for all
interparticle distances below the independence limit. In the triangular arrangement, MSs interact more strongly than NPs, becoming independent only by $r\approx5d$ compared to approximately $3d$ for NPs. 
Strikingly, for the triangular arrangement, interactions between NPs increase loop area compared to independent particles, while dipolar interactions between MSs reduce loop area.
For the fcc structure, the MSs become independent by $7d$,  while, based on mild extrapolation of the apparent trend, complex NPs do so by $3.5d$.
Thus, the dipole approximation within the MS model in non-chain geometries overemphasizes the effect of interparticle magnetostatic interactions, compared to more detailed modeling.  At closer distances, the MS model can give qualitatively different results.

We examine the local hysteresis loops of individual NPs and MSs in different clusters and compare their magnetization dynamics in terms of the dipole interactions they experience due to their location in a cluster. The appearance of inverted loops, while not directly yielding the degree of local cooling or heating, signifies significant work being done by MSs on each other, which can mean that local heating around MSs is uneven. 
{\blue Local hysteresis loops, calculated with respect to the external field, do not represent local heating in interacting systems like ours [12, 13].  However, we report them here for two reasons.  First, they provide a simplified description of local dynamics, and hence a way of distinguishing the responses of clustered NPs and MSs.  Second, considering such loops may provide a starting point for developing a loop-based thermodynamic description of local heating consistent with a dynamical approach as presented, for example, in Ref.~\cite{munoz2020disentangling}.  Inverted loops, which naively would imply cooling rather than heating, provide a particularly interesting case.}


In Appendix~\ref{sec:Eff_K}, we examine the use of two anisotropy axes in the MS description of a NP, and see {\blue some improvement at the level of a single particle}. 
We study the effect (at finite $T$) of drawing anisotropy directions for micromagnetic cells from a distribution, finding an approximately linear decrease in loop area with distribution width, amounting to a 10\% reduction in loop area when the standard deviation of the distribution is 50$^\circ$. {\blue We also find that cubic magnetocrystalline anisotropy in magnetite NPs decreases SLP by approximately 10\%.}


In Appendix~\ref{sec:SR_alpha}, the SR and $\alpha$ scaling technique for multiple particles is validated for both OOMMF and Vinamax, two micromagnetic software packages that we use here, in which faster simulations performed with higher SRs give equivalent loops to those performed with slower SRs so long as  SR$/\alpha$ is held fixed. However, the scaling has a wider range of validity with OOMMF, which is likely due to the type of solver employed. 

{\blue Finally, we show in  Appendix~\ref{sec:additional} that global and local loops have different shapes for $10z$ NPs in a triangle, but global loop area is unchanged from the $6z4y$ case.  Adding anisotropy of 10~kJ/m$^3$ to the $10z$ NP increases area, and hence SLP, by 28\%.  Orientational disorder for the 13-MS fcc cluster reduces heating efficiency. Also, individual NPs within the $r=d$ fcc cluster have similar local loops, in contrast to the case of MSs, where local loops are quite distinct between non-equivalent positions.}

\section*{Acknowledgment}
We thank Jonathan Leliaert for his guidance on employing the updated version of Vinamax. 
R.B. thanks Mikko Karttunen for hosting her stay at Western University.  We acknowledge the financial support from the Natural Sciences and Engineering Research Council (Canada).  Computational resources were provided by ACENET and Compute Canada.

\appendix
\section{Effective anisotropy in complex NPs}\label{sec:Eff_K}

{\blue In II, we found appropriate values of $K_u$ and $M_{\rm s}$ for an MS so that it exhibits a hysteresis loop with the same $H_c$ and $M_r$ as a same-volume $6z4y$ magnetite NP under a rotationally averaged field. However, when the field is applied parallel to the anisotropy axis, a value of $K_u$ approximately 25\% smaller is needed to produce an equivalent loop.}
{\blue Here, we continue our exploration to find a suitable anisotropy for MSs that provide equivalent hysteresis loops for different directions of the applied field for a $6z4y$ NP made of magnetite (for which $K_{u0}=0$) and a hypothetical material with uniaxial magnetocrystalline anisotropy $K_{u0}=10$~kJ/m$^3$ but otherwise having properties of magnetite.  Such a material might be formed by including a high-anisotropy element such as cobalt~\cite{fantechi_influence_2015,hasegawa_stabilisation_2019,aldaoud_magneto-thermal_2022}, and here we simply refer to it as anisotropic magnetite (anisotropic Fe$_3$O$_4$).} 

\begin{figure*}
    \centering
    \includegraphics[width=0.495 \textwidth]{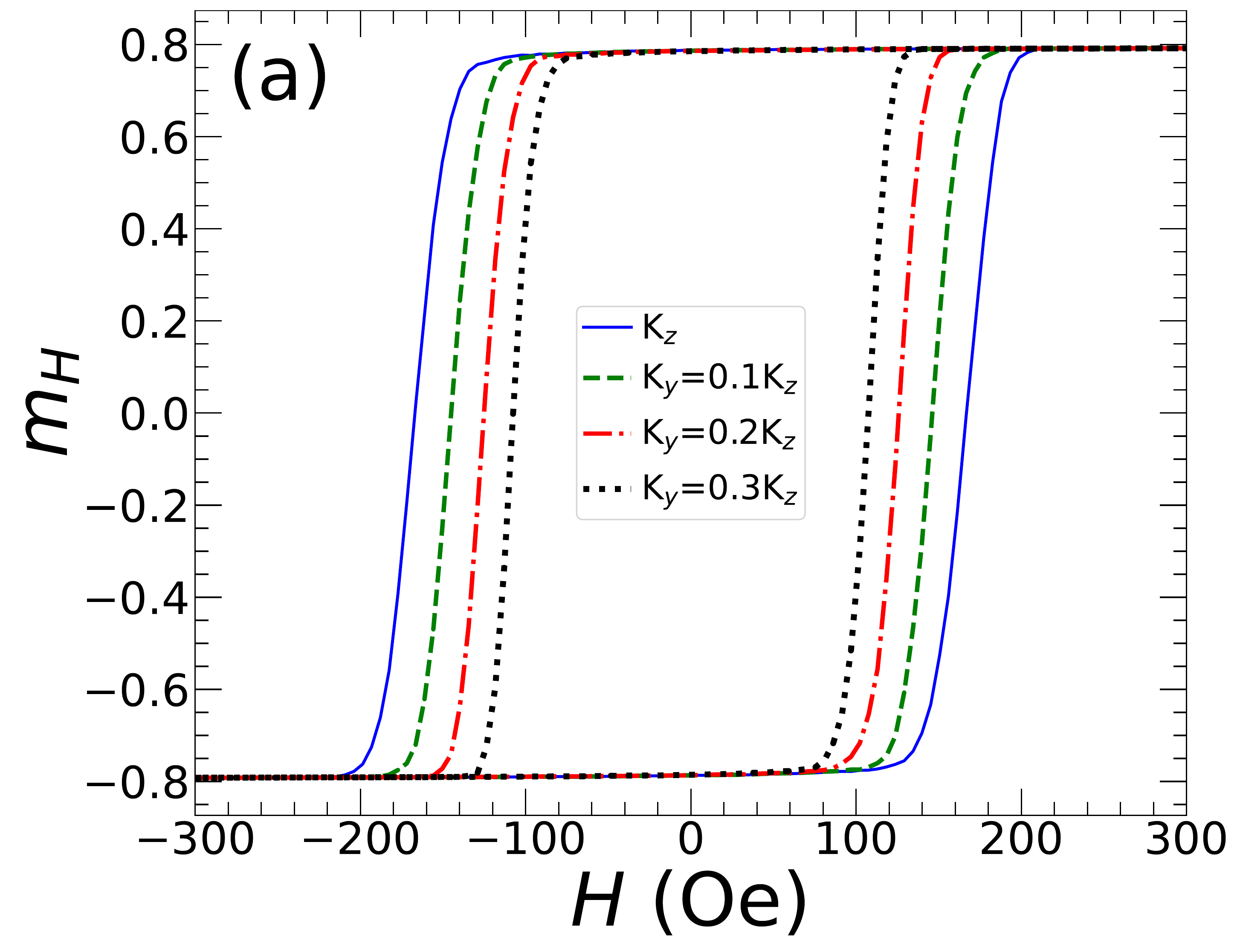}
    \includegraphics[width=0.495 \textwidth]{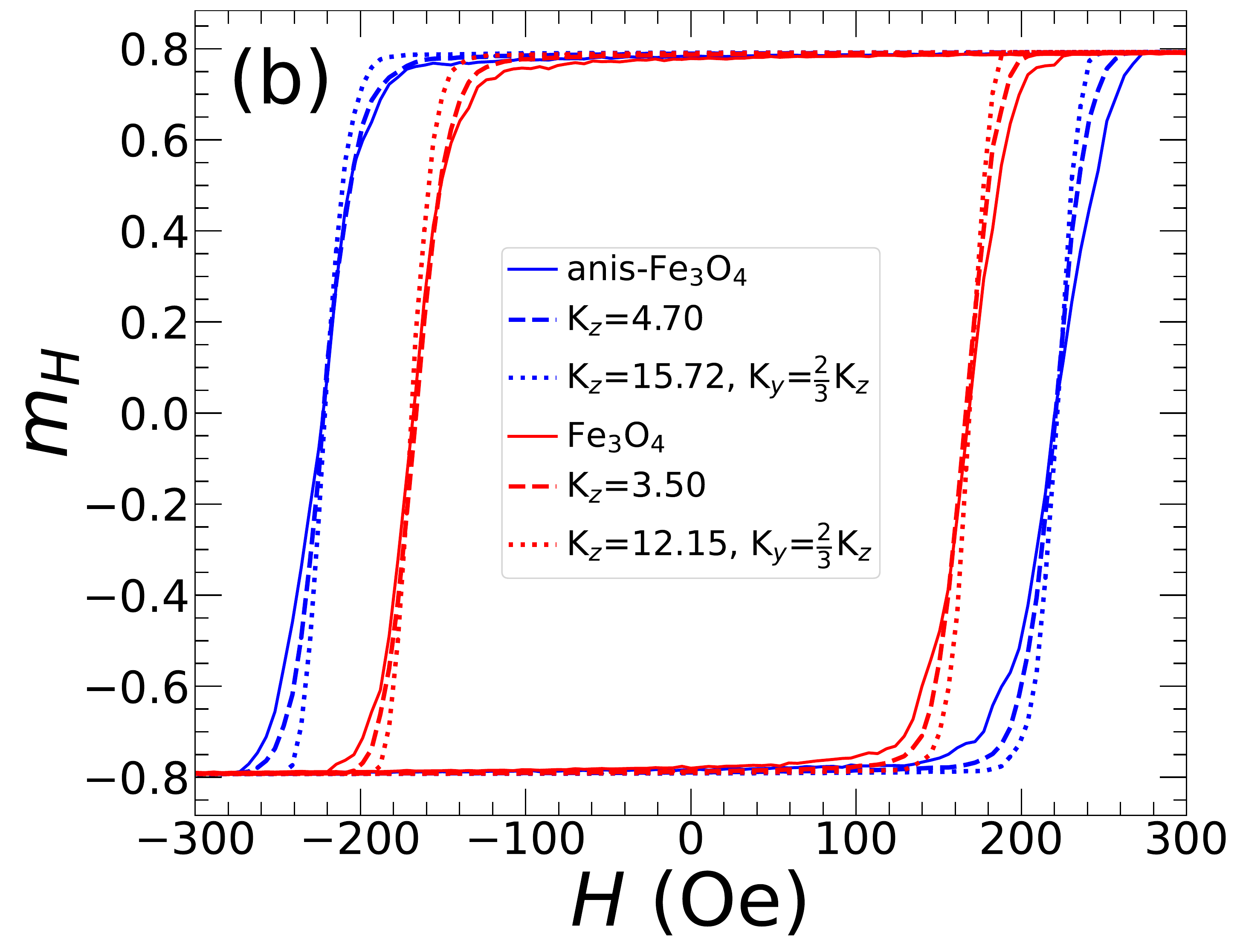}
    \caption[Impact of two uniaxial anisotropies on the hysteresis loop of the MS model]{a) Changes to the hysteresis loop when a second anisotropy axis along the $y$ direction with small strength $K_y$ is added to a MS with {\blue $K_z$=3.50 kJ/m$^3$} and $M_{\rm s}$=381~kA/m.  
    b) Hysteresis loops for $6z4y$ NPs of Fe$_3$O$_4$ and {\blue anisotropic  Fe$_3$O$_4$ ($K_{u0}$=10~kJ/m$^3$)} and their equivalent MSs with single (dashed line) or double uniaxial anisotropies (dotted lines).
    The external field is along $z$ for both panels.}
    \label{fig:KzKy}
\end{figure*}

The internal structure of the $6z4y$ NP means that nanorod anisotropies are found to lie in both the $z$ and $y$ directions; modelling the MS with two anisotropy axes, along $z$ and along $y$, may provide a route to finding a better quantitative match between MS and NP loops.
As shown in Fig.~\ref{fig:KzKy}a, simply adding $K_y$ to the original $K_z$, even with a very small value, shrinks the hysteresis loop, which is not desired. Hence, for the assumption of having two perpendicular anisotropies with $K_z$ and $K_y=2K_z/3$ energy densities, analogous to ordering six nanorods along the $z$ and four along the $y$ axis, new parameters need to be used. 
As shown in Fig.~\ref{fig:KzKy}b, the hysteresis loop of a $6z4y$ {\blue anisotropic Fe$_3$O$_4$} NP exposed to a field along the $z$ axis ($H_z$), can be reproduced with a MS having either a single anisotropy {\blue $K_{z}$=4.70 kJ/m$^3$ (dashed lines), or two perpendicular anisotropies with $K_{z}$=15.72 kJ/m$^3$}, $K_{y}=2K_{z}/3$ (dotted line). Similarly, for a magnetite $6z4y$ NP,
the best fit with only a $z$-axis anisotropy is with {\blue $K_z$=3.50~kJ/m$^3$, whereas if $K_y=2K_z/3$ is included, a value $K_z$=12.15~kJ/m$^3$} works well.

\begin{figure*}
    \centering
    \includegraphics[width=0.495 \textwidth]{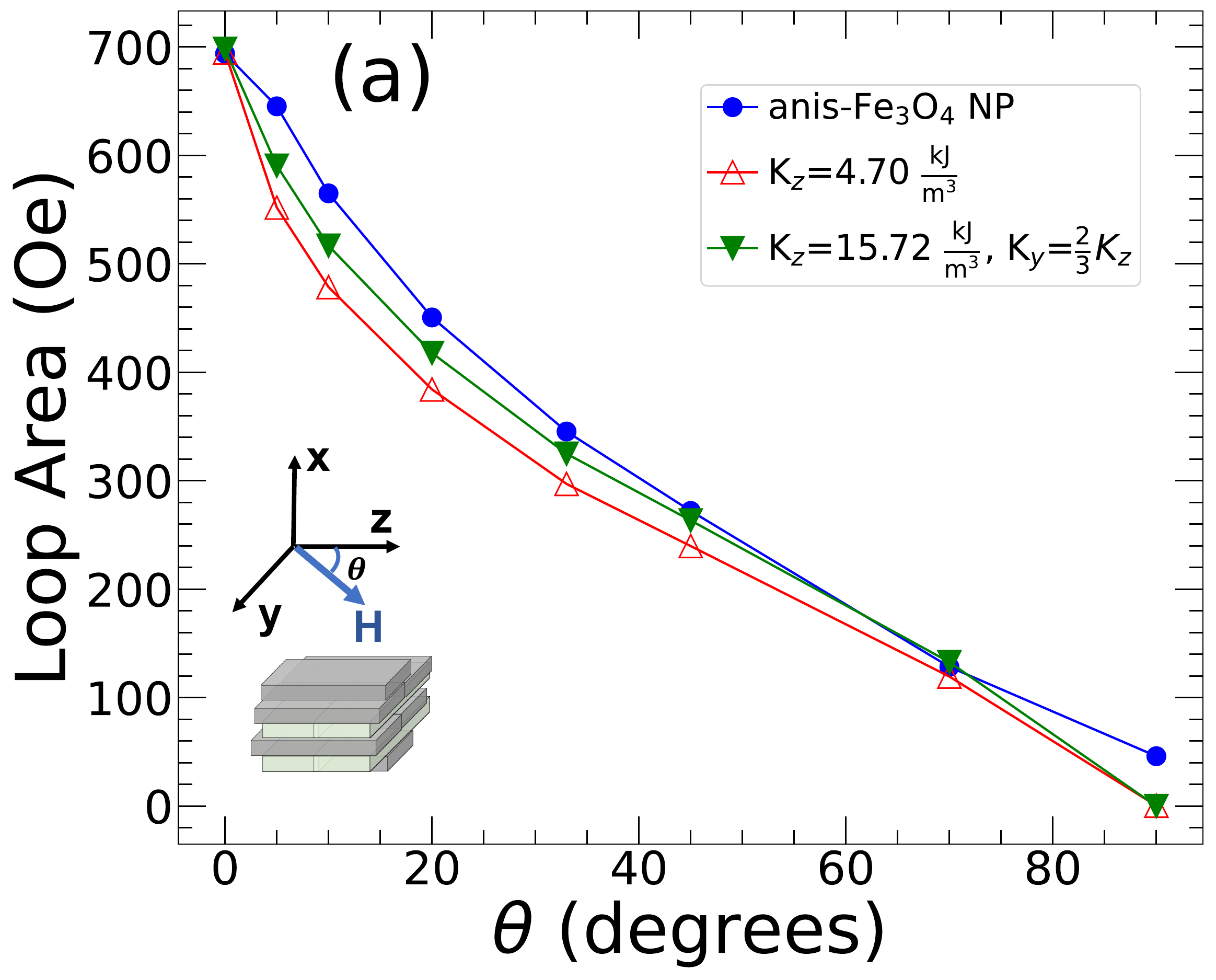}
    \includegraphics[width=0.495 \textwidth]{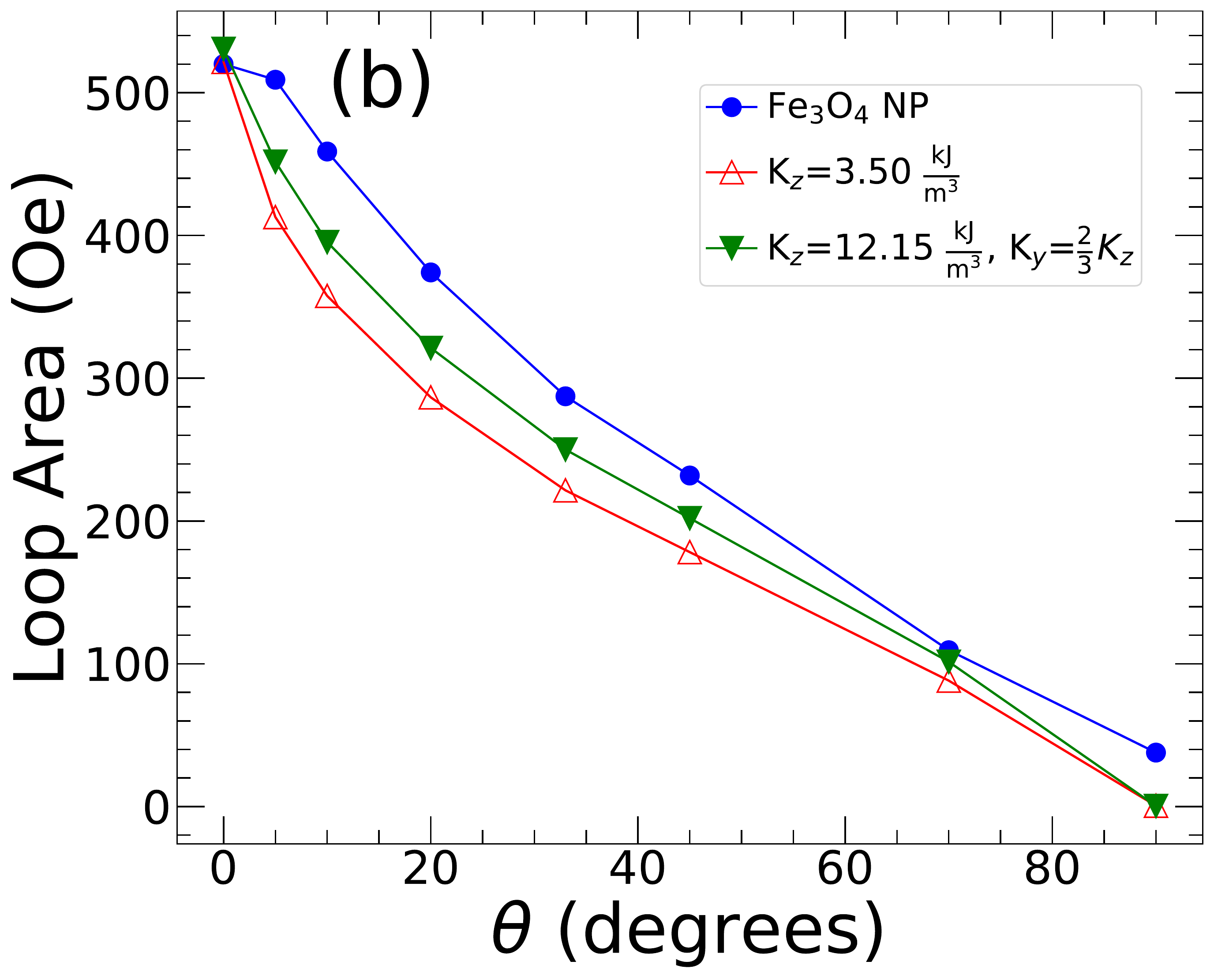}
    \caption[Loop area comparison for NPs and equivalent MSs with single and double uniaxial anisotropies]{Hysteresis loop area as a function of applied field angle $\theta$, see inset of panel (a), on $6z4y$ NPs and equivalent MSs with single and double uniaxial anisotropies for a) {\blue anisotropic Fe$_3$O$_4$ ($K_{u0}$=10~kJ/m$^3$)} and b) Fe$_3$O$_4$ ($K_0=0$) NPs.}
    \label{fig:NP_MS_area_theta}
\end{figure*}

To get better insight on the accuracy of replacing complex NPs with MSs using these anisotropy values, we compare the MS and NP loop areas upon tilting the field away from the $z$ axis in the $z-y$ plane by an angle $\theta$, as shown in the inset of Fig.~\ref{fig:NP_MS_area_theta}a. Comparing the loop areas for {\blue anisotropic Fe$_3$O$_4$} (Fig.~\ref{fig:NP_MS_area_theta}a) and magnetite (Fig.~\ref{fig:NP_MS_area_theta}b) NPs with respect to their equivalent MSs reveals that the presence of the magnetocrystalline anisotropy in {\blue anisotropic Fe$_3$O$_4$} encourages a closer match between a complex NP and its equivalent MS. Also, adding the second anisotropy to a MS, slightly improves the loop area agreement for 
{\blue both  Fe$_3$O$_4$ and anisotropic Fe$_3$O$_4$.} 
The relatively large value of $K_z$ when two anisotropy axes are used is closer to that of the MS model of a single nanorod, and thus physically appealing.
However, this larger value of $K_z$ also likely causes the increased ``squareness'' in the shoulder area of the loops seen in Fig.~\ref{fig:KzKy} for the two-axes cases.  Thus, further investigation into using two axes may prove to be fruitful, but it appears that there will be unavoidable trade-offs.  

\begin{figure*}
    \centering
    \includegraphics[width=0.505\textwidth]{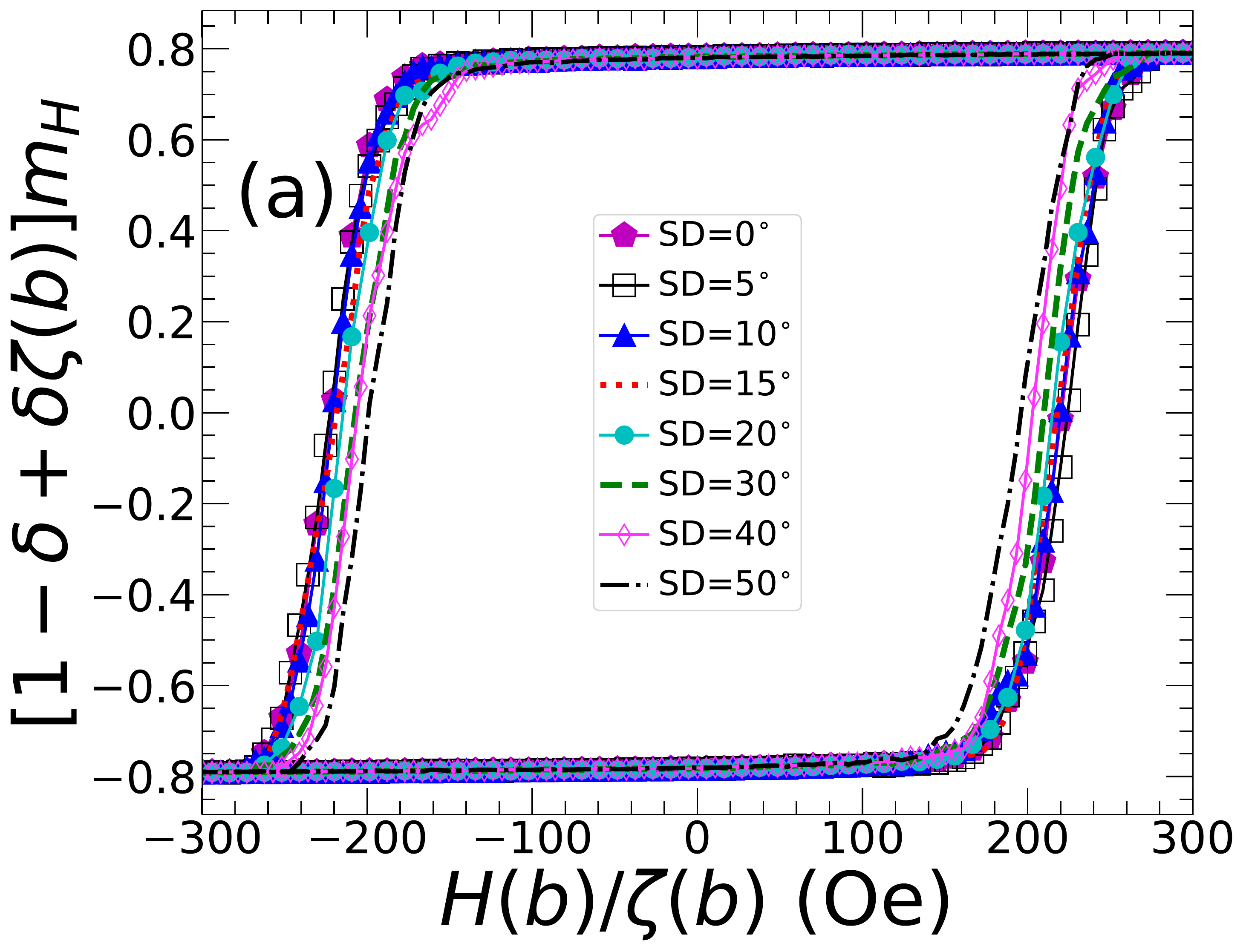}
    \includegraphics[width=0.485\textwidth]{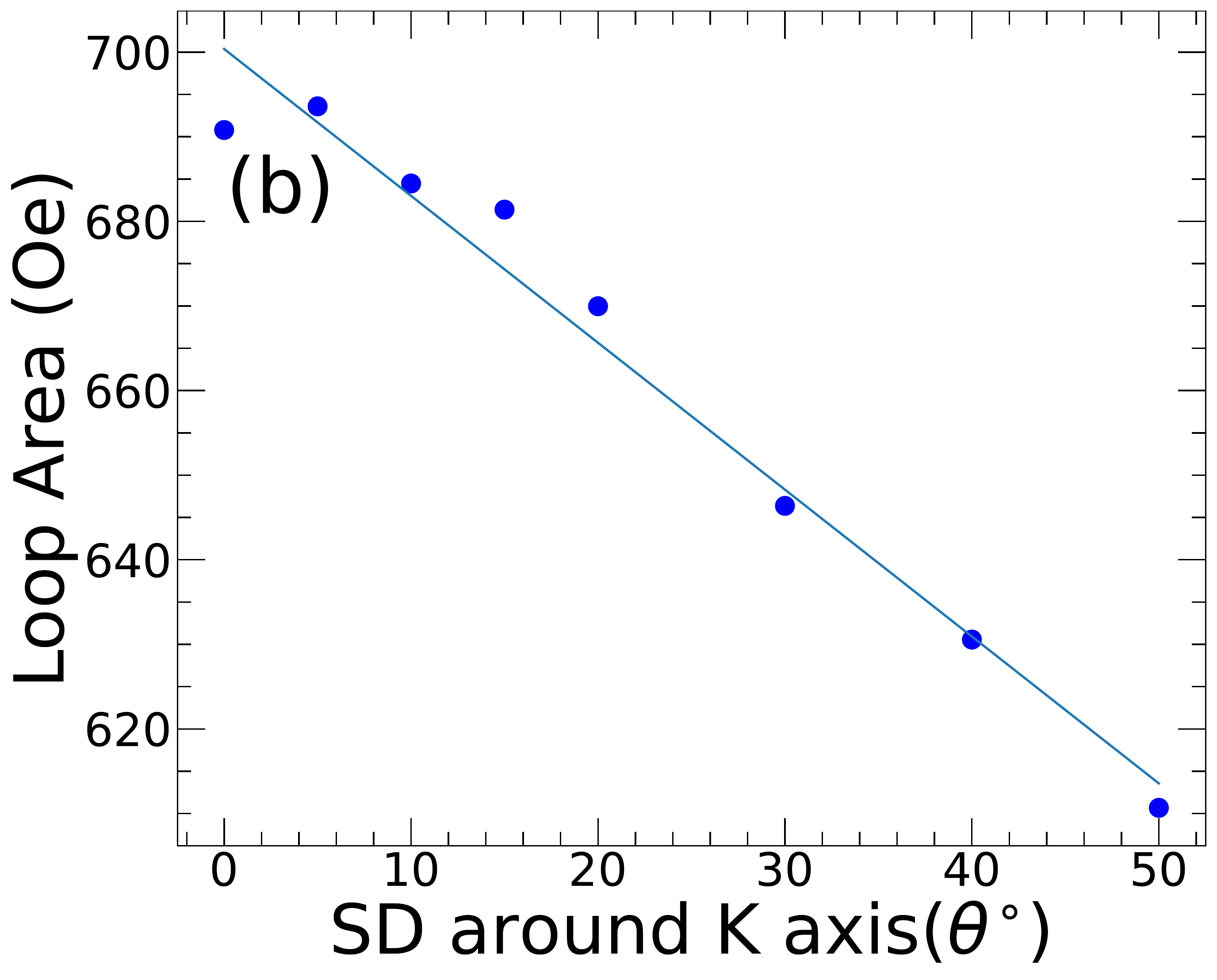}
    \caption[Effect of anisotropy axis distribution on a NP hysteresis loop]{a) Hysteresis loops of $6z4y$ {\blue anisotropic Fe$_3$O$_4$ ($K_{u0}$=10)} nanoparticles when the field is applied along the $z$ axis, and the anisotropy axis of each cell has a random angular deviation from the long axis of the rod of which it is a part.  The random angles are drawn from a normal distribution with zero mean and standard deviation (SD) ranging from 0 to 50$^{\circ}$.  b) Loop area versus standard deviation of the anisotropy direction.}
    \label{fig:vs}
\end{figure*}

Real materials will have properties that vary depending on structural defects, chemical impurities, size polydispersity, and other forms of disorder. One way to model such effects in simulations is to introduce distributions in micromagnetic cell properties~\cite{plumer2010micromagnetic}.
While the relationships between the standard deviations of such distributions and the degree of various forms of disorder are often difficult to quantify, it is not uncommon to encounter variations of parameters in the range of 0 to 20\%~\cite{serantes2014multiplying, hergt2005magnetic}.
Here, we simulate a 6$z$4$y$ {\blue anisotropic Fe$_3$O$_4$} NP when each cell's anisotropy axis is chosen from a normal distribution around the direction given by the nanorod's longest edge.  
We vary the standard deviation (SD) of the distribution between 0 and 50 degrees and plot the resulting hysteresis loops. 
As shown in Fig.~\ref{fig:vs}b, the loop area changes approximately linearly with the SD of the anisotropy axis direction with a slope of -1.74~Oe/degree and an intercept of 700~Oe.  With SD = 10 degrees, the loop area is reduced by 1\%, with SD = 20 it decreases by 3\%, and with SD = 50 degrees it decreases by almost 12\%.  Thus we see that, for this system at 310~K, the effect of varying anisotropy directions is not very large.


\begin{figure*}
    \centering
    \includegraphics[width = 0.495 \textwidth]{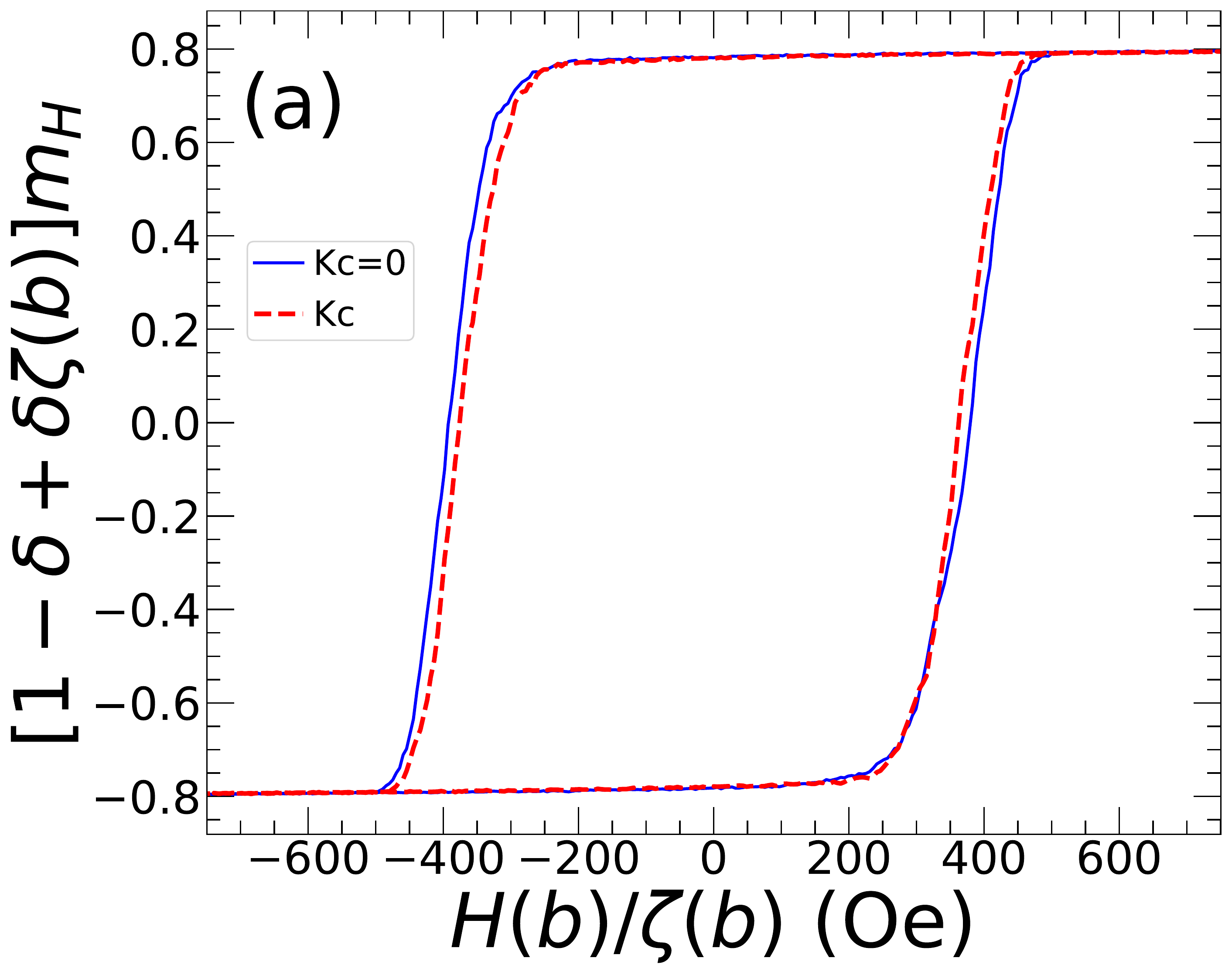}
    \includegraphics[width = 0.495 \textwidth]{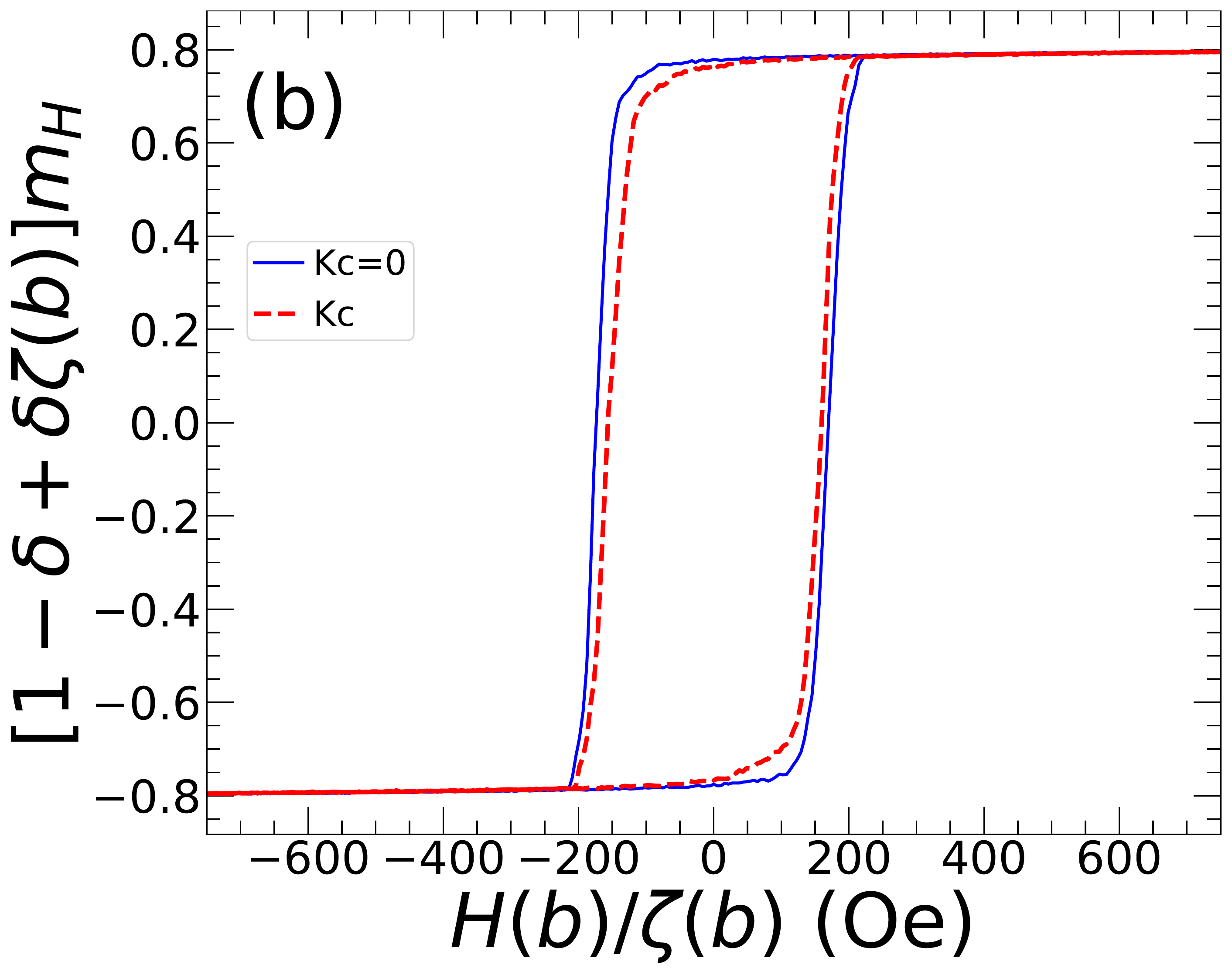}
    \caption{
    {\blue Including cubic anisotropy (a) for a single rod shrinks the loop area 3.7\% from 1188 to 1144~Oe and (b) for a $6z4y$ NP shrinks the loop area 10.5\% from 530 to 474~Oe. For rod and NP, simulations are done at ${\rm SR}^{\rm sim}=2.5$ and 25~Oe/ns, and $\alpha^{\rm sim}=1$ and 10, respectively, equivalent to the laboratory parameters ${\rm SR}=0.25/\zeta(4)=0.34$~Oe/ns and $\alpha=0.1$.
    }}
    \label{fig:cubicanisotropy}
\end{figure*}

{\blue Finally, we report in Fig.~\ref{fig:cubicanisotropy} the effects on loop areas of including magnetite's cubic anisotropy with coefficient $K_{c0}=-10$~kJ/m$^3$ for a single nanorod (Fig.~\ref{fig:cubicanisotropy}a) and a $6z4y$ NP (Fig.~\ref{fig:cubicanisotropy}b).  For the nanorod case, the calculated loop is 3.7\% smaller when cubic anisotropy is added, while for the NP, the loop is 10.5\% smaller and has more rounded corners.
}


\section{SR-scaling for multiple MSs and NPs}\label{sec:SR_alpha}

\begin{figure*}
    \centering
    \includegraphics[width = 0.495\textwidth]{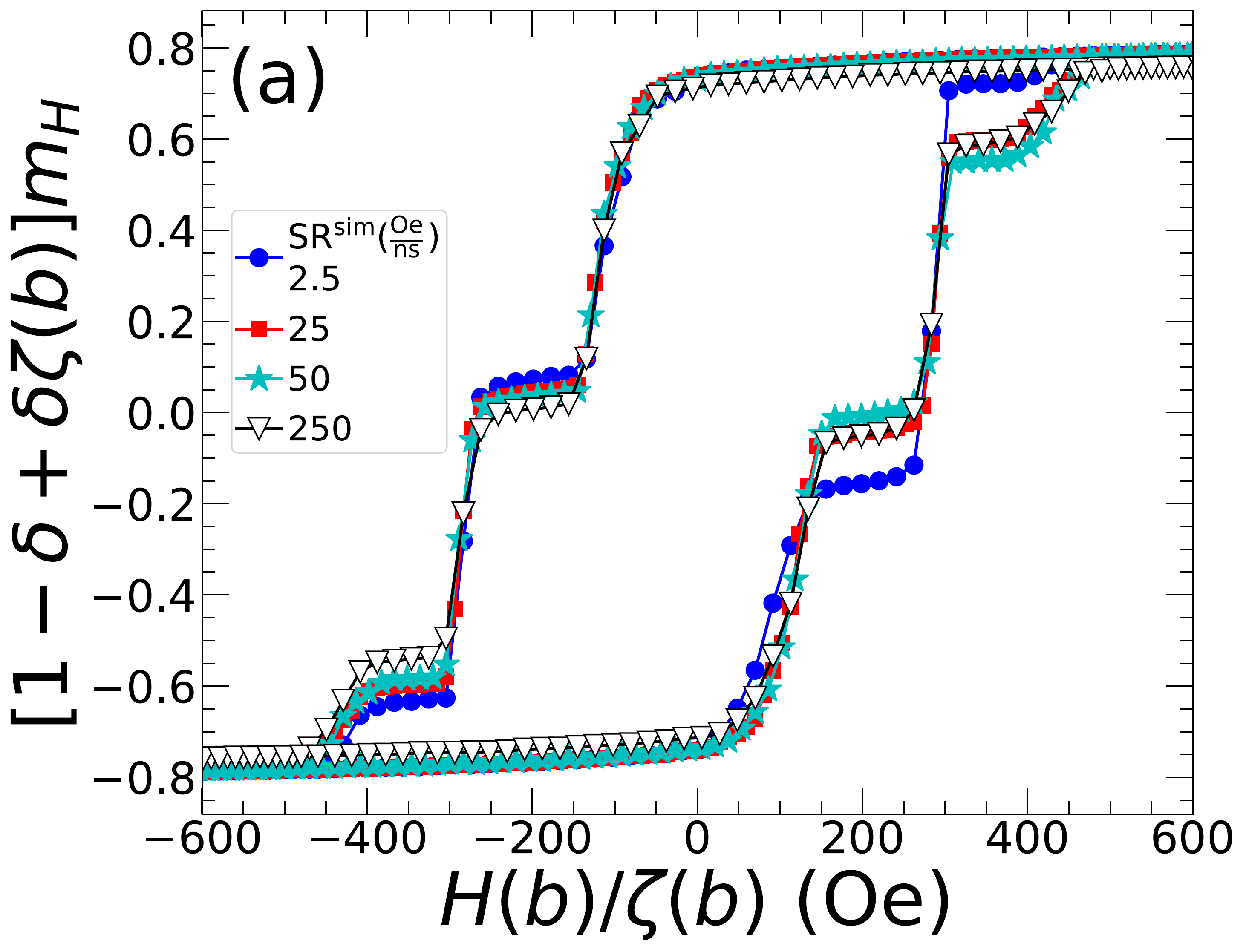}
    \includegraphics[width = 0.495\textwidth]{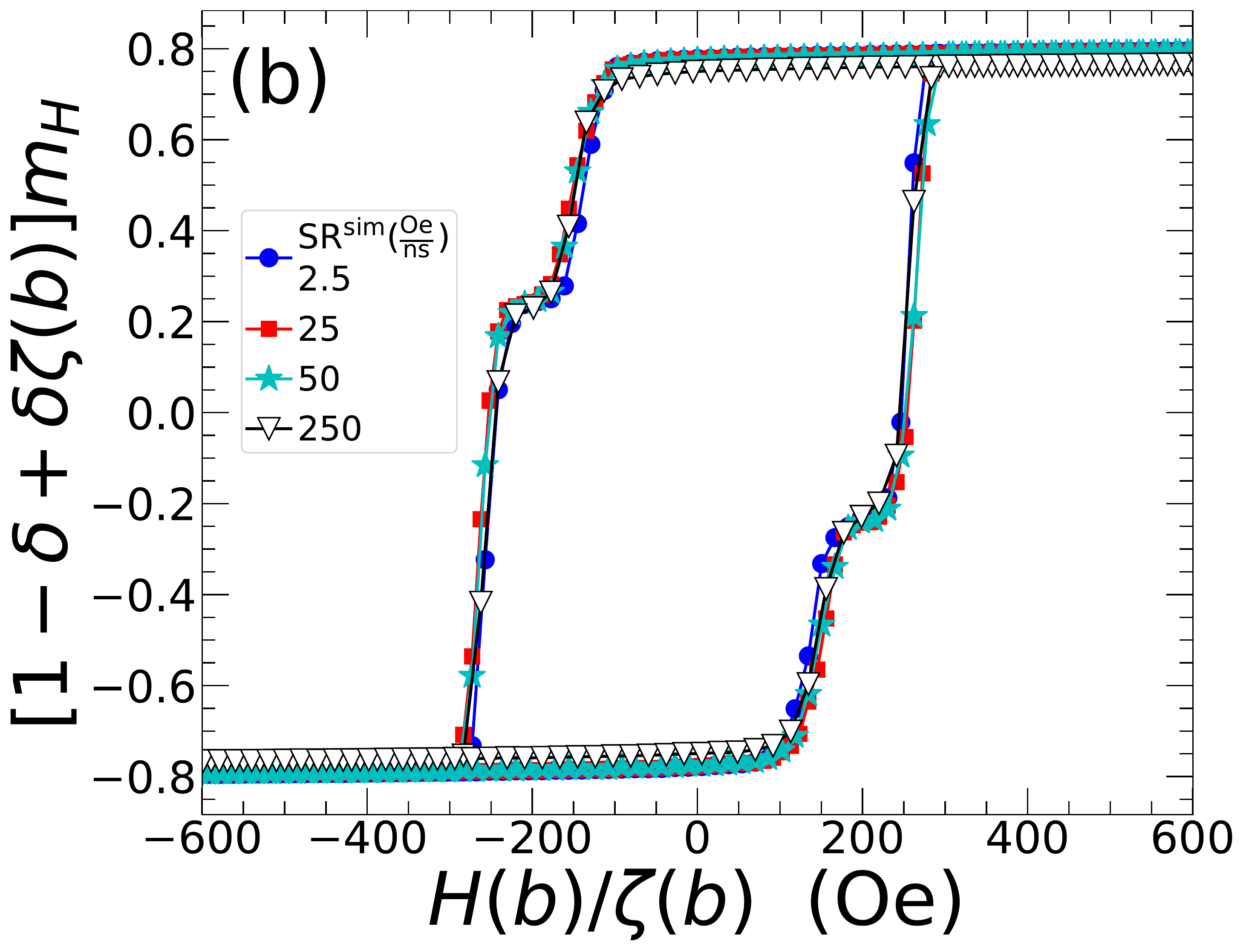}
    \caption[Test of ${\rm SR}/\alpha$ scaling for NP simulations in OOMMF]{Test of SR-scaling for three NPs (simulated in OOMMF) in triangular order with different separation a) $r=d$, b) $r =1.5d$, where $d$ is a NP diameter.}
    \label{fig:SRalpha3NP}
\end{figure*}

In addition to coarse-graining and using an MS model, a useful technique for decreasing the calculation time is to simulate the magnetic system with a faster SR but keeping the ratio ${\rm SR}/\alpha$ constant.
A more detailed explanation of this equivalence can be found in I and here we test this technique for assemblies of NPs and MSs using OOMMF and Vinamax, respectively. We start our investigation with NPs at the closest distance ($r=d$) in a triangular array, {\blue with $H_{\rm max}^{\rm sim}=1000$~Oe. As shown in Fig.~\ref{fig:SRalpha3NP}a, except for slight mismatches at the shoulders of the loop, we still get an acceptable loop agreement for loop areas with values 641, 663, 662, 637~Oe for ${\rm SR}^{\rm sim}=2.5$ ($\alpha^{\rm sim}=1$), 25, 50 and 250~Oe/ns ($\alpha^{\rm sim}=100$), respectively.} At higher particle separation, $r=1.5d$, the loop agreement is even better over the two orders of magnitude of SR simulated; see Fig.~\ref{fig:SRalpha3NP}b.   

\begin{figure*}
    \centering
    \includegraphics[width = 0.495 \textwidth]{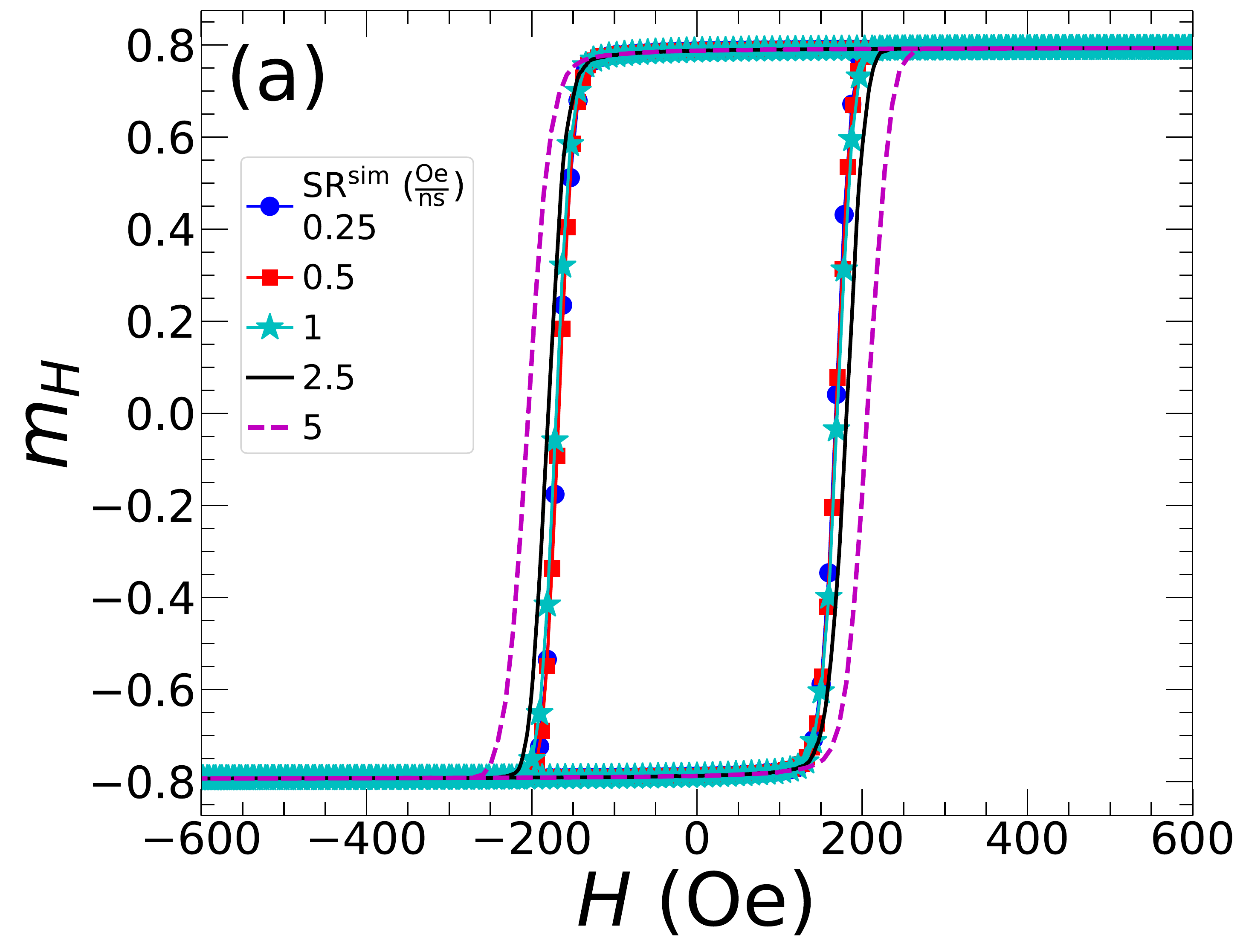}
    \includegraphics[width = 0.495 \textwidth]{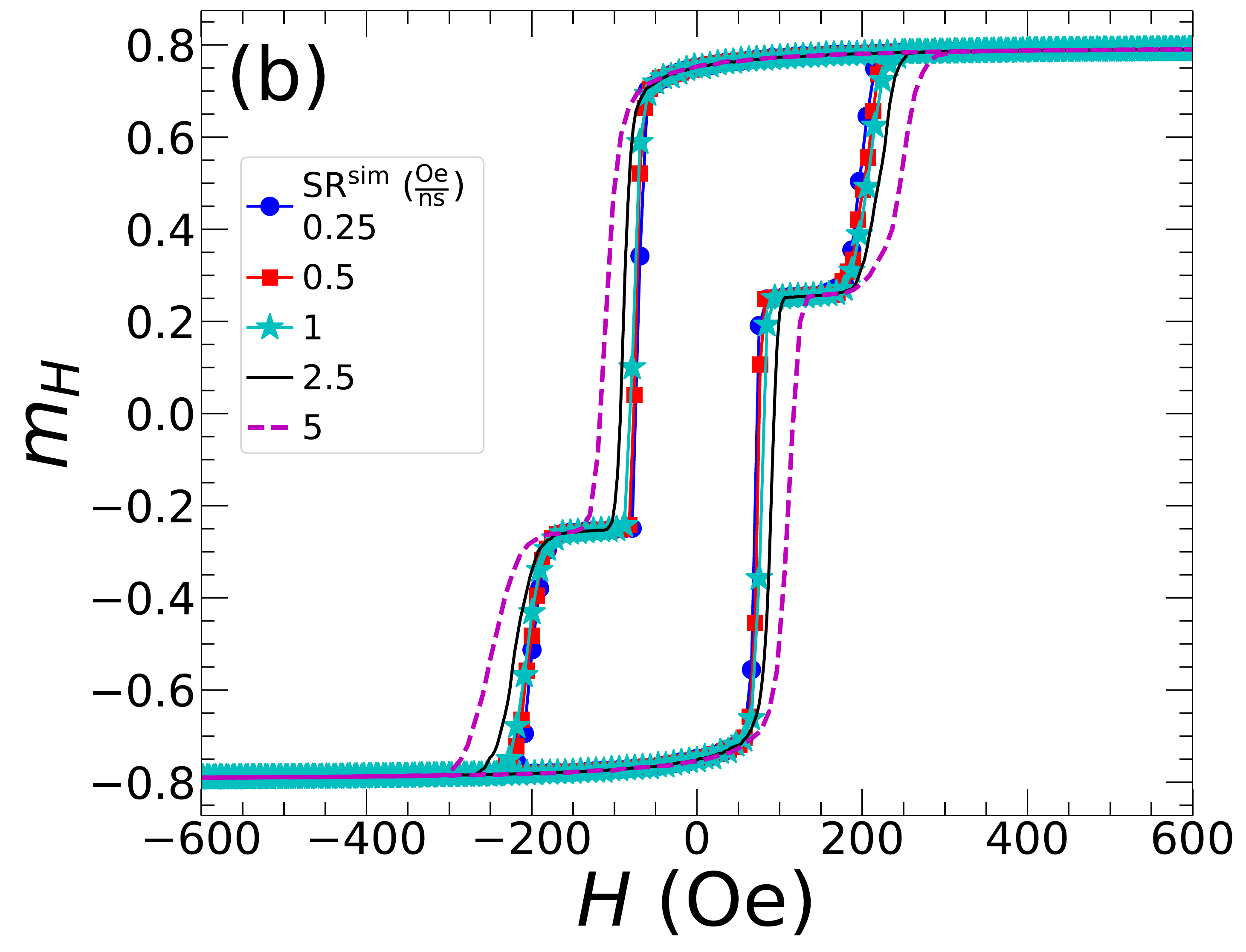}
    \caption{Test of SR-scaling for simulations in Vinamax for a) a single MS, b) three MSs in triangular order, $r=1.5d$.}
    \label{fig:SRalphaVMX}
\end{figure*}

Testing Vinamax for validity of this technique, we first compare a single MS hysteresis loop over a range of SR and $\alpha$ values, and as shown in Fig.~\ref{fig:SRalphaVMX}a, the limit of validity of this scaling appears to be ${\rm SR}=1$~Oe/ns, which is only four times faster than ${\rm SR}=0.25$~Oe/ns. 
This perhaps can be attributed to the numerical approach that Vinamax is based upon, the Dormand-Prince solver (an embedded Runge-Kutta method)~\cite{dormand1980family}, versus Euler for OOMMF.   

Moreover, a combination of three MSs in a triangular order simulated with Vinamax confirms the results that ${\rm SR}=1$~Oe/ns is the threshold of validity of the scaling using this software.  The difference in the limits of applicability for the SR-scaling technique for different numerical solvers is a matter of future investigation.

\section{Additional results for clusters}\label{sec:additional}

\begin{figure*}
    \centering
    \includegraphics[width = 0.495 \textwidth]{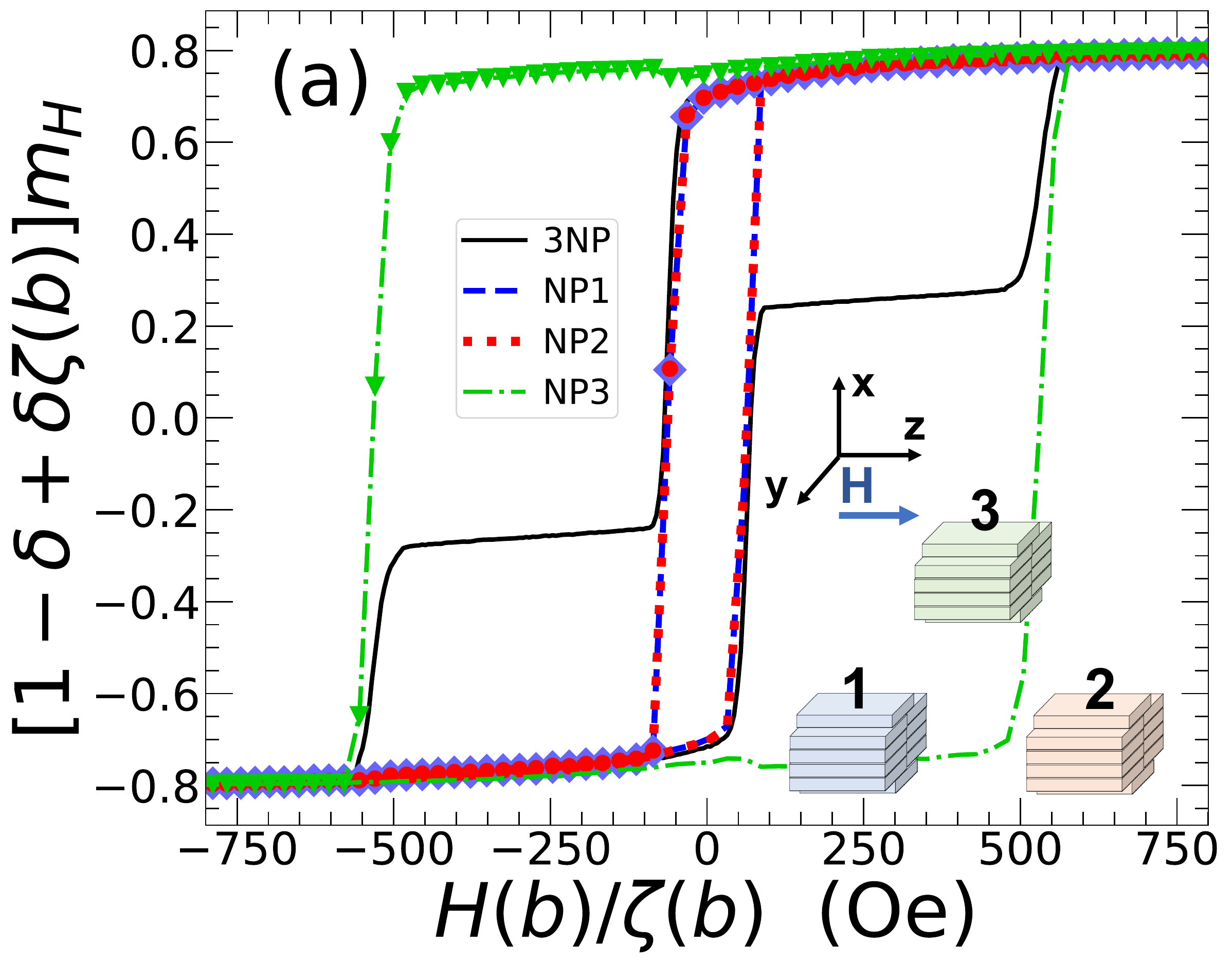}
    \includegraphics[width = 0.495 \textwidth]{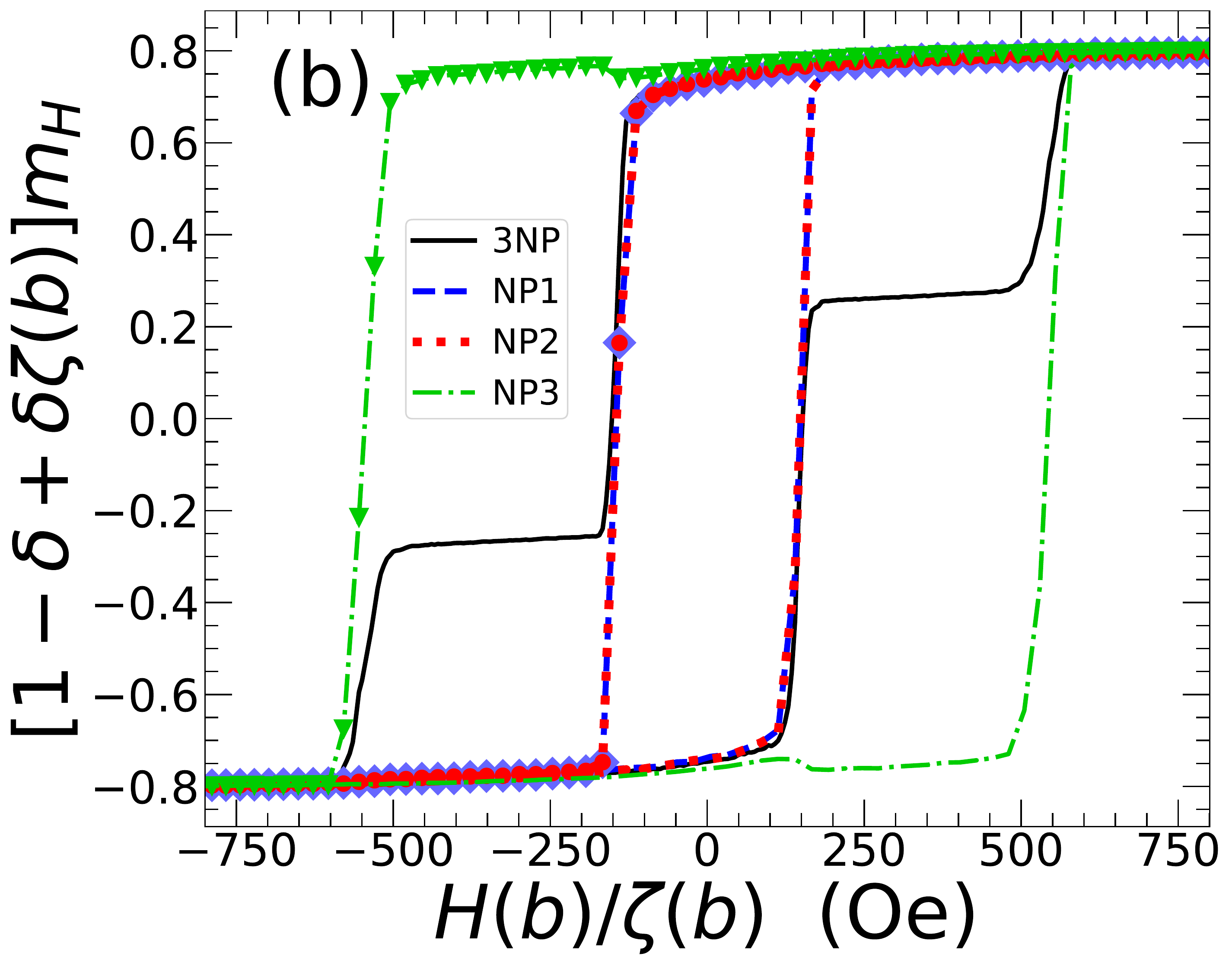}
    \caption{
    {\blue Three NPs in a triangular arrangement when the internal structure are a) 10z, $K_u=0$, and b) $10z$, $K_{u0}=10$~kJ/m$^3$. 
    Including uniaxial anisotropy result in 28\% global loop area increase (from 665 in panel a to 840~Oe in panel b). Unlike the different local dynamics for NPs with 10z (panel a) compared to 6z4y (Fig.~\ref{fig:localLoop_3MS}c), and despite the different global loop shapes, their global loop areas are similar: 655 and 663~Oe for $10z$ and $6z4y$ magnetite NPs, respectively.}}
    \label{fig:Triangle_10z}
\end{figure*}

\begin{figure*}
    \centering
    \includegraphics[width = 0.495 \textwidth]{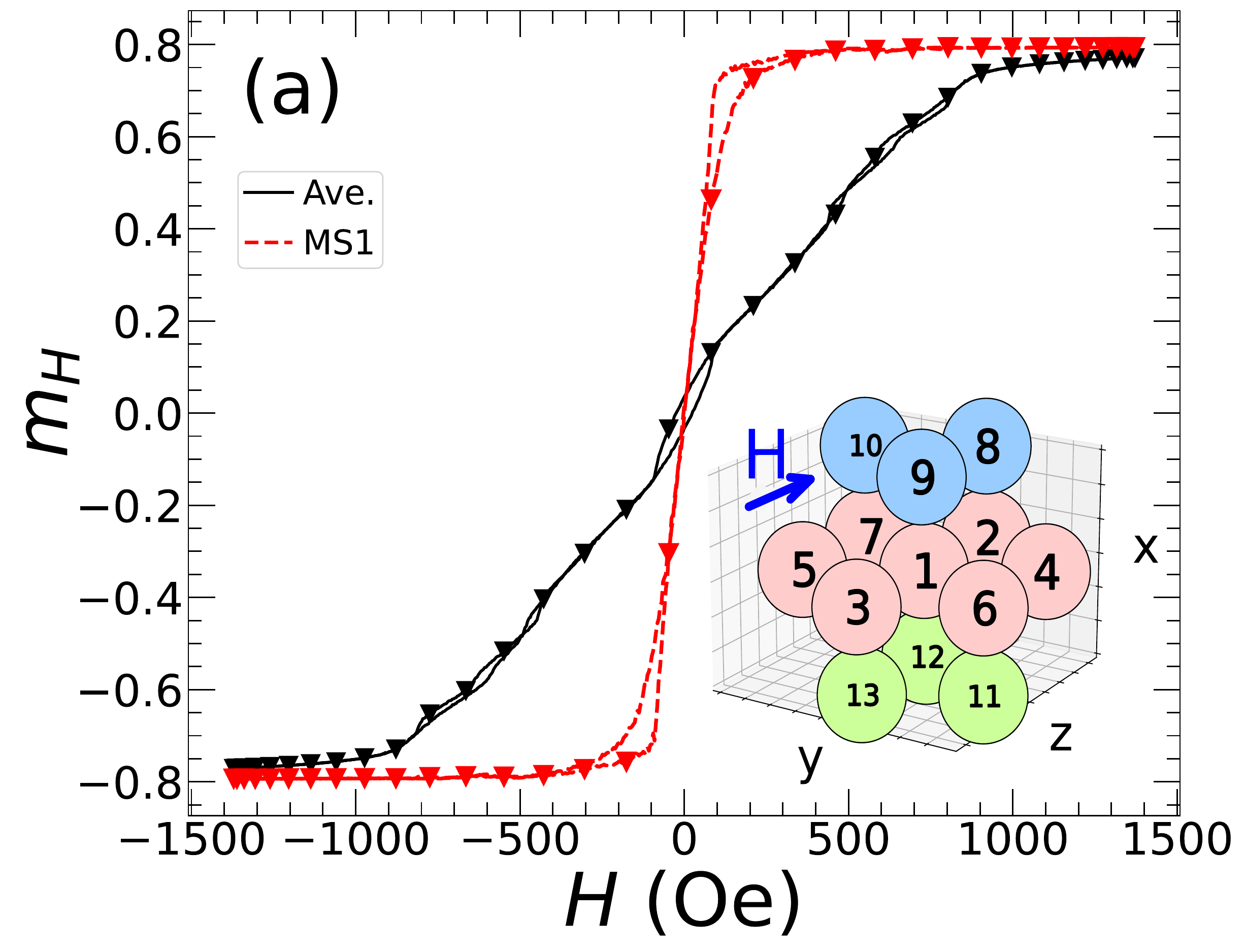}
    \includegraphics[width = 0.495 \textwidth]{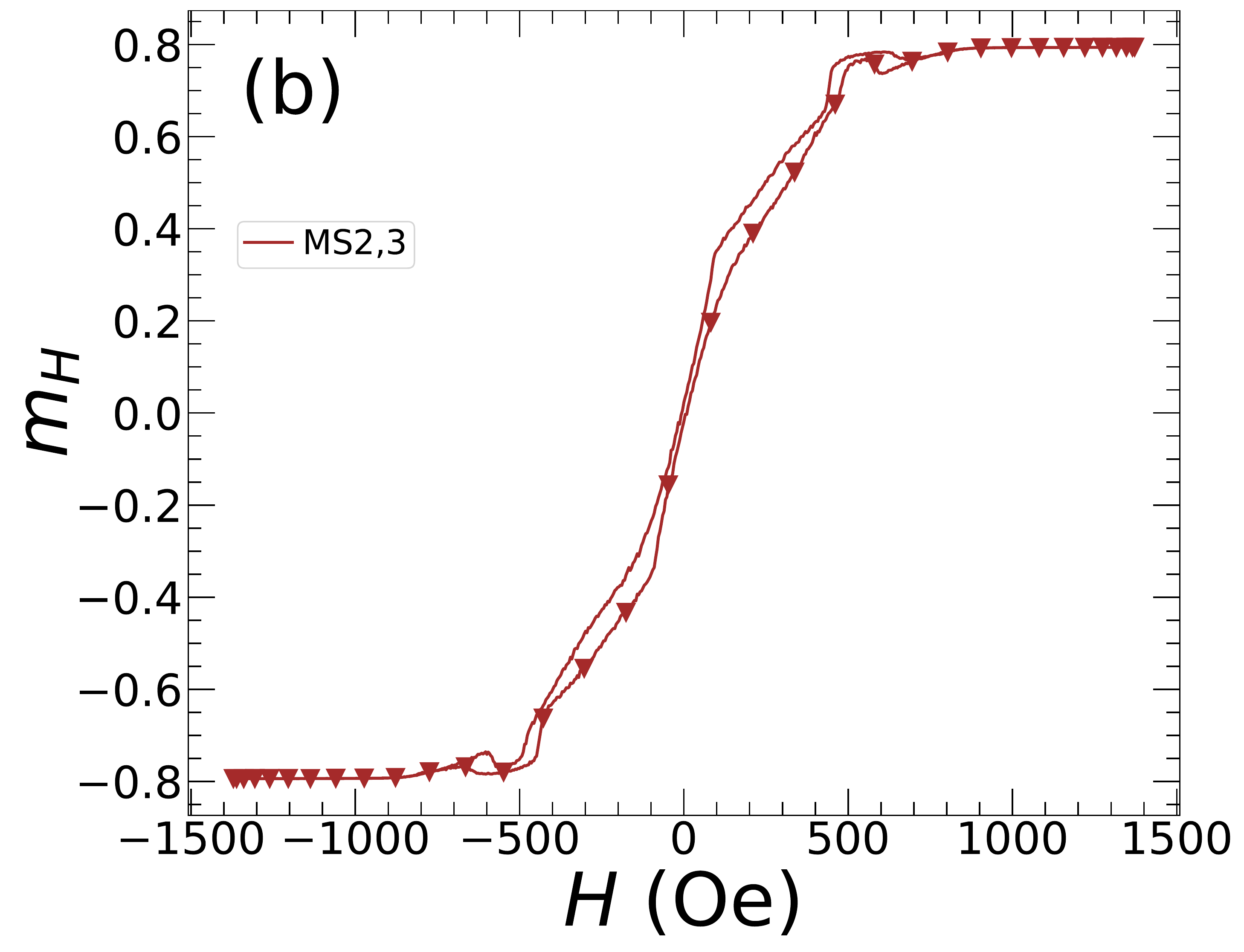}
    \includegraphics[width = 0.495 \textwidth]{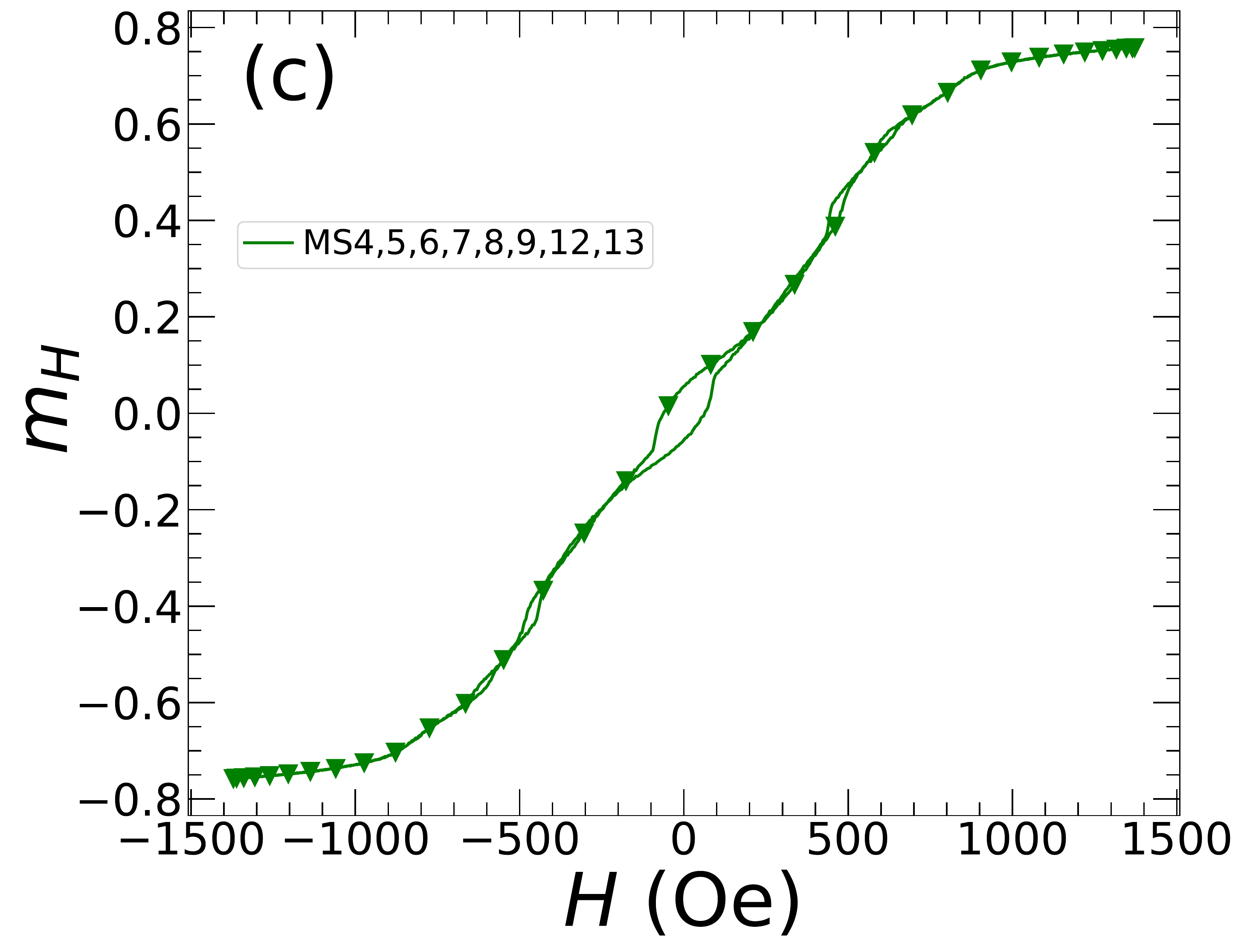}
    \includegraphics[width = 0.495 \textwidth]{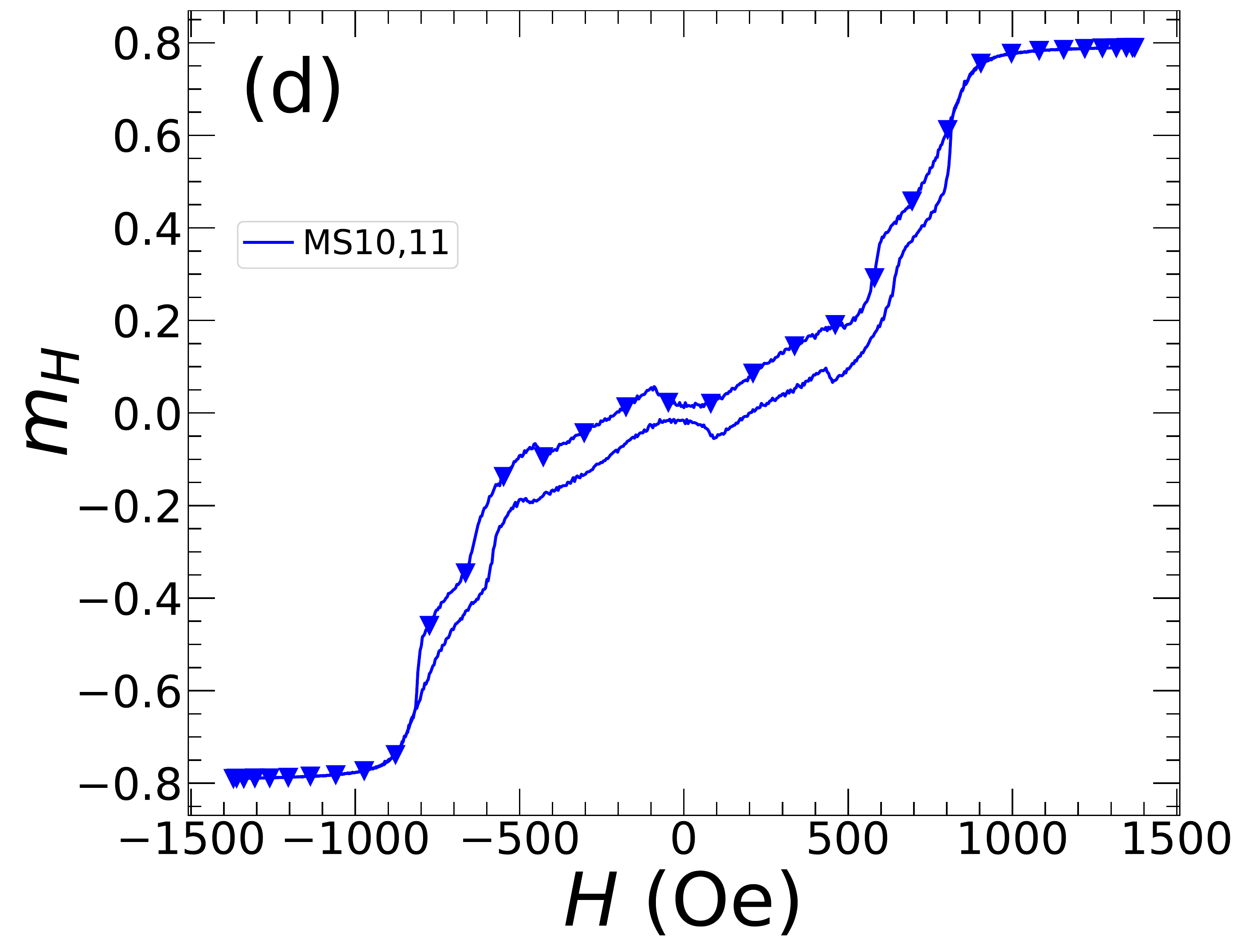}
    \caption{
    {\blue fcc structure made of 13 MSs, and their global and local hysteresis loops when $r=d$, and uniaxial anisotropy axes of MSs are not aligned and chosen as u1=[0 1 0], u2=[0 0 1], u3=[1 0 0], u4=[0 1 1], u5=[1 0 1], u6=[1 1 0], u7=[1 0 -1], u8=[-1 1 0], u9=[0 -1 1], u10=[1 1 1], u11=[-1 1 1], u12=[-1 -1 1], u13=[1 -1 1] for the labeled MSs in the inset of panel (a) and field is applied along the $z$ axis. a) Global hysteresis loop for 13 MS is shown in black and the local loops for the central MS in dashed red line. Local hysteresis loops for b) MS2 and MS3, c) MS4, MS5, MS6, MS7, MS8, MS9, MS12, MS13 d) MS10 and MS11. Each loop is calculated via averaging over 500 independent field cycles.}}
    \label{fig:fccMS_randAnis}
\end{figure*}

\begin{figure*}
    \centering
    \includegraphics[width = 0.495 \textwidth]{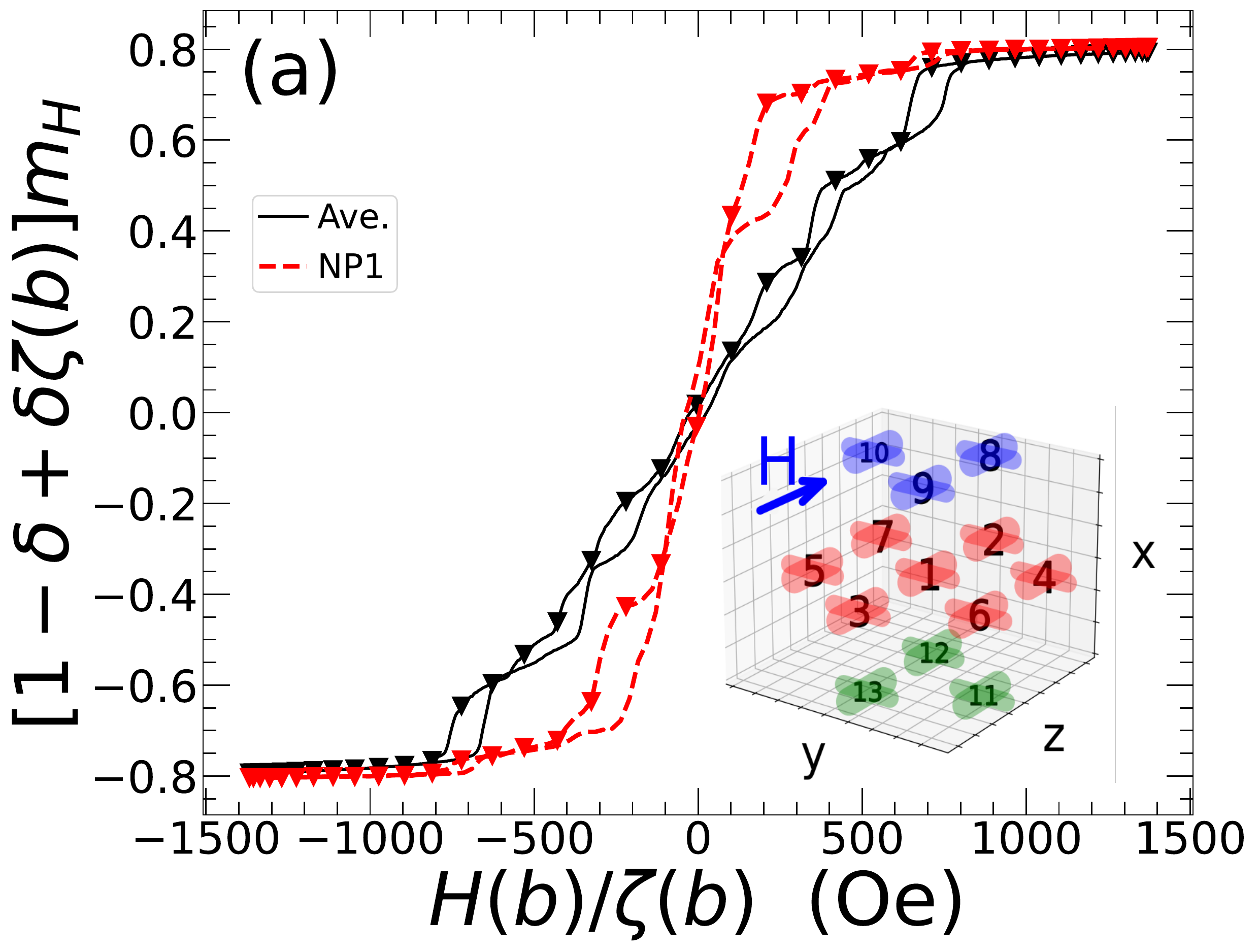}
    \includegraphics[width = 0.495 \textwidth]{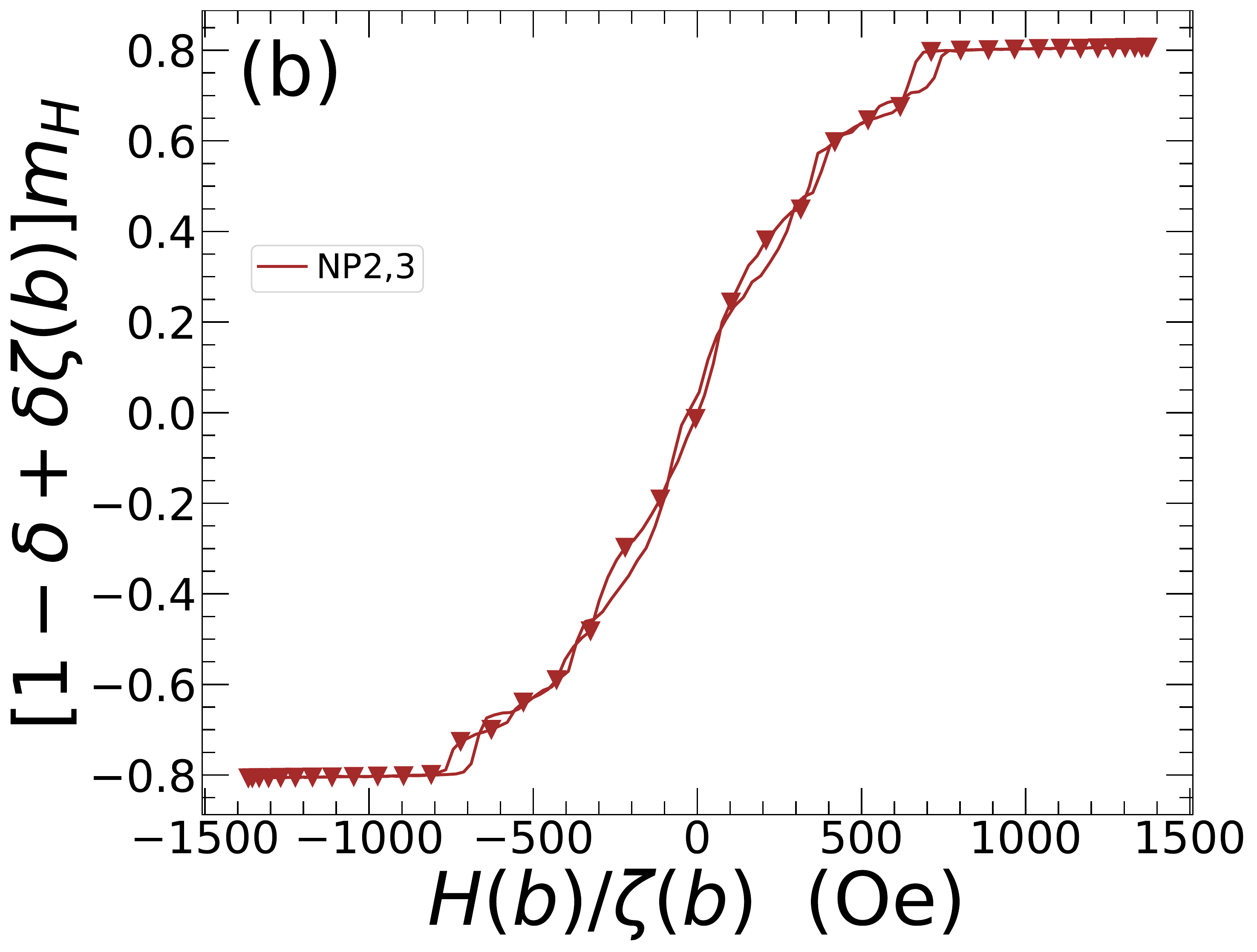}
    \includegraphics[width = 0.495 \textwidth]{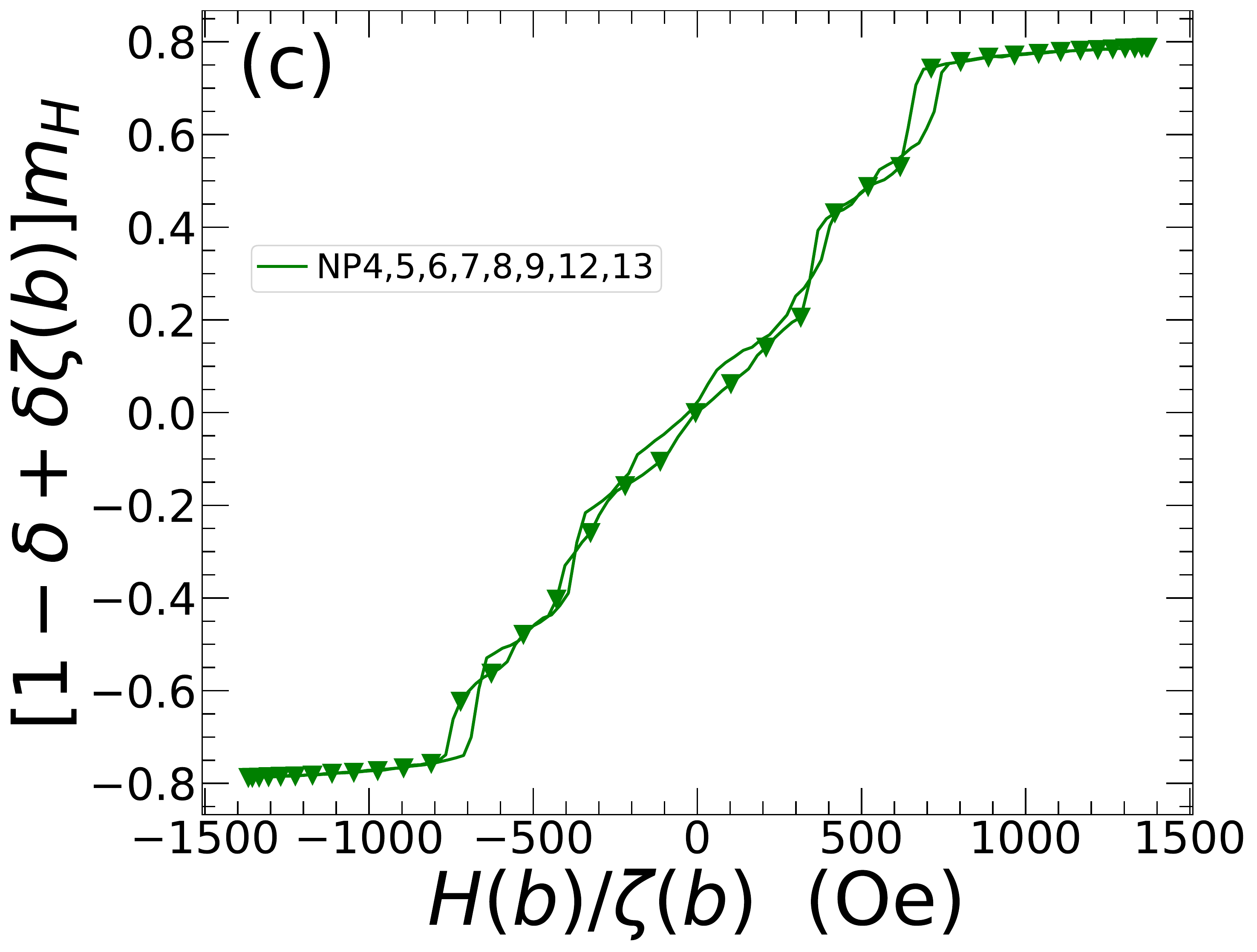}
    \includegraphics[width = 0.495 \textwidth]{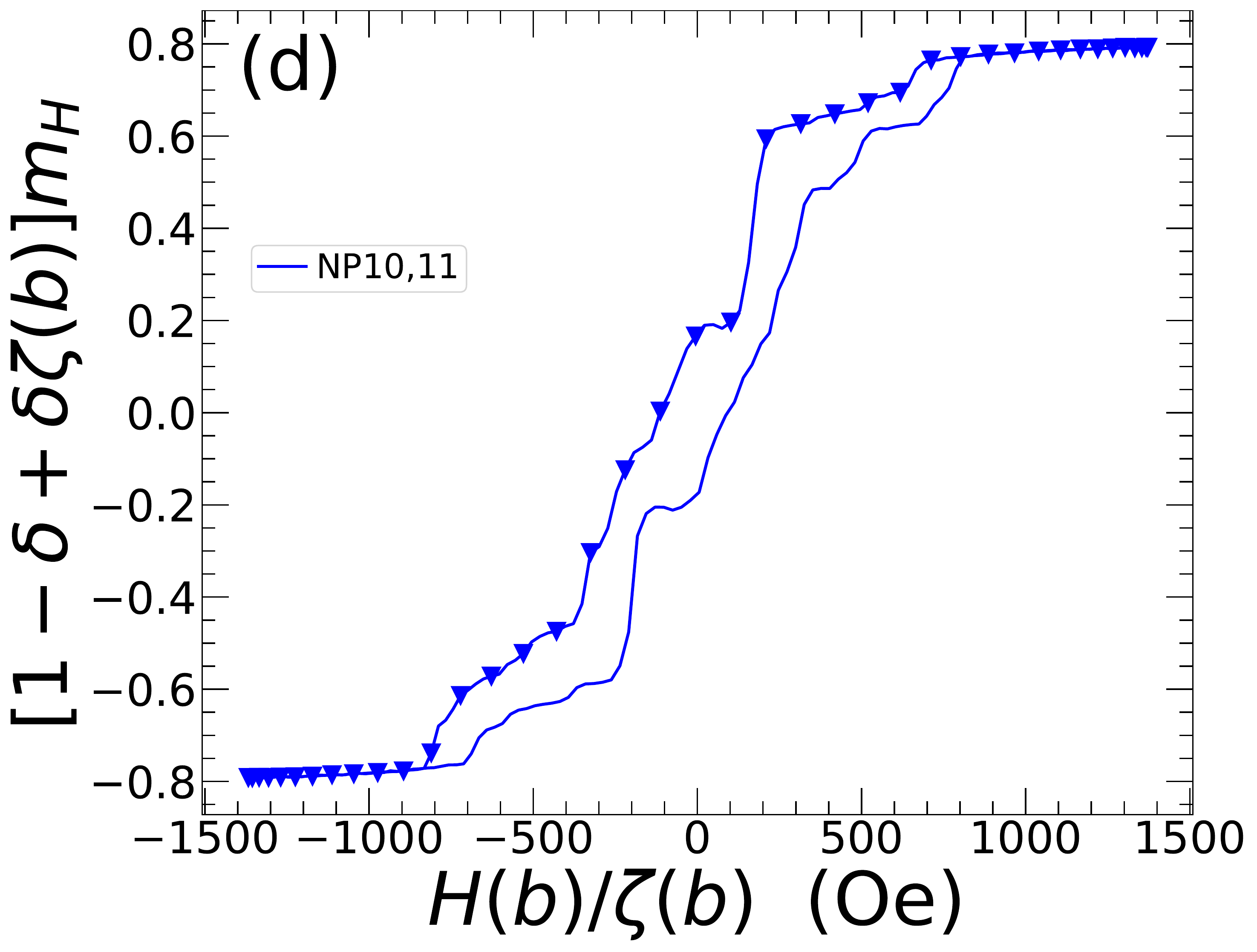}
    \caption{
    {\blue fcc structure made of 13 6z4y magnetite NPs, and their global and local hysteresis loops when $r=d$ and the external field is applied along the $z$ axis. a) Global hysteresis loop for 13 NPs is shown in black and the local loops for the central NP in dashed red line. Local hysteresis loops for b) NP2 and NP3, c) NP4, NP5, NP6, NP7, NP8, NP9, NP12, NP13 d) NP10 and NP11. Each loop is calculated via averaging over 50 independent field cycles.}}
    \label{fig:fccNP}
\end{figure*}

{\blue 
Fig.~\ref{fig:Triangle_10z} shows the effects for a triangular arrangement of NPs of changing the internal structure of the NPs to $10z$ from $6z4y$ (Fig.~\ref{fig:Triangle_10z}a), where all nanorods are aligned along the field, as well as additionally adding uniaxial magnetocrystalline anisotropy artificially to  magnetite (Fig.~\ref{fig:Triangle_10z}b).  
Adding anisotropy increases the loop area by 28\% from 665 to 840~Oe.  The global loop and individual loops for NP1 and NP2 are wider, but the loop for NP3 is approximately unchanged.
Comparing the loops of $10z$ NPs ($K_{u0}=0$) in Fig.~\ref{fig:Triangle_10z}a with those for 
$6z4y$ NPs shown in Fig.~\ref{fig:localLoop_3MS}c, the global and local loops have very different shapes, but the areas of the global loops are approximately the same, surprisingly.

Fig.~\ref{fig:fccMS_randAnis} shows the effect of randomizing the anisotropy directions in 13-MS fcc cluster with $r=d$ on global and individual loops.  The directions of the anisotropy axes of the MSs are all different and chosen from variations of the [100], [110] and [111] directions.  Directions for the MSs are specified in the caption of Fig.~\ref{fig:fccMS_randAnis}.
In comparison with the anisotropy-aligned 13-MS fcc cluster (Fig.~\ref{fig:fcc_Global_Local_r1d}) the near-zero global loop area is significantly smaller, the reduction most evident for MS1, MS2 and MS3.  The loops for MS10 and MS11 do not cancel, and have significant areas.

Fig.~\ref{fig:fccNP} shows the global and local loops for an fcc cluster of 13 $6z4y$ NPs with nearest neighbor separation set to $r=d$.  In comparison to the corresponding MS cluster (Fig.~\ref{fig:fcc_Global_Local_r1d}), the global loop is significantly smaller, as is the local loop for the central NP.  The local loops for NP10 and NP11 are significant, and differ greatly from MS10 and MS11 in the equivalent MS cluster. The loops of MS10 and MS11, and in fact those of all of the other MSs, are more similar to those in the randomized-anisotropy MS fcc cluster in Fig.~\ref{fig:fccMS_randAnis}. 
}




\newpage

%

\end{document}